\documentclass[a4paper,11pt]{article}
\pdfoutput=1 
\usepackage{jcappub} 
\makeatletter
\renewcommand{\@fpheader}{%
  Prepared for submission to JCAP%
  \hfill INR-TH-2026-006%
}
\makeatother
\usepackage{amsfonts,amsmath,graphics,graphicx,color,diagbox,float}
\usepackage{rotating}
\usepackage{pdflscape}
\usepackage{booktabs}
\usepackage{longtable}
\usepackage{array}
\usepackage{amsmath}
\usepackage{tabularx}
\graphicspath{{images/}}

\newcommand{\redd}{\color{black}}
\newcommand{\red}{\color{black}}
\newcommand{\black}{\color{black}}
\usepackage{hyperref}

\title{
Isochrone fitting of Galactic globular clusters -- IX. Tip of the red-giant branch for 27 clusters and constraints on new particle physics}

\author[a,b]{G.A.\ Gontcharov,}
\author[c,b]{A.M.\ Kudashov,}
\author[d,b]{O.S.\ Ryutina,}   
\author[b,c,1]{and S.V.\ Troitsky\note{Corresponding author.}
}
\affiliation[a]{Central (Pulkovo) Astronomical Observatory of the Russian Academy of Sciences,\\ Pulkovskoye Chauss\'ee 65/1, St. Petersburg 196140, Russia}
\affiliation[b]{Institute for Nuclear Research of the Russian Academy of Sciences,\\ 60th October Anniversary Prospect 7a, Moscow 117312, Russia}
\affiliation[c]{Faculty of Physics, M.V. Lomonosov Moscow State University,\\ 1-2 Leninskie Gory,  Moscow, 119991, Russia}
\affiliation[d]{Institute of Applied Astronomy of the Russia Academy of Sciences, Kutuzova quay, 10, Saint-Petersburg 191187, Russia}

\emailAdd{st@ms2.inr.ac.ru}

\abstract{Red-giant evolution is sensitive to non-standard channels of energy loss in stellar cores and therefore provides a probe of new physics. Such losses increase the luminosity of the tip of the red-giant branch (TRGB), the brightest point reached by a red giant in the color--magnitude diagram. Previous analyses of TRGB luminosities in globular clusters placed stringent bounds on the axion--electron coupling $g_{ae}$, but were limited by observational and theoretical uncertainties. To obtain a homogeneous and systematically controlled constraint, we use new parameters and TRGB estimates for 27 globular clusters, derived from Stetson, {\it Hubble Space Telescope}, and {\it Gaia} photometry, \red  together with \black cluster-specific MESA calculations. Our primary likelihood uses 24 clusters after excluding three hot-flasher systems and \red explicitly accounts for stellar variability in identifying the brightest RGB stars.\black\ Finite sampling of the upper RGB is described by the full conditional distribution, and shared observational and stellar-physics uncertainties are retained as correlated nuisance parameters. \red The observed TRGB is fainter than the standard-cooling prediction by $0.078^{+0.036}_{-0.037}$~mag; to avoid an artificially strong bound from this downward fluctuation, we use the power-constrained profile likelihood method and obtain  \black $g_{ae}<6.4\times10^{-14}$ \red (95\% CL). Sensitivity estimates for the neutrino magnetic dipole moment and for millicharged particles are also presented.\black}


\begin{document}
\maketitle

\flushbottom

\section{Introduction}
\label{sec:intro}
Many extensions of the Standard Model of particle physics predict light, weakly interacting particles. Of particular interest are axions \cite{Peccei:1977hh,Peccei:1977ur,Weinberg:1977ma,Wilczek:1977pj}, which provide a possible solution to the strong charge--parity (CP) problem, and other light pseudoscalars arising in models with two symmetry-breaking scales, commonly referred to as axion-like particles (ALPs) \cite{Jaeckel:2010ni}. Light particles coupled to photons, electrons, or nuclei can be produced abundantly in hot and dense astrophysical plasmas, in particular in stellar interiors \cite{Raffelt:1996Stars,Raffelt:1999tx,Caputo:2024oqc}. If their interactions are sufficiently weak, most of the produced particles escape from the star, thereby providing an additional non-standard channel of energy loss. This energy loss may become sufficiently large to modify observable characteristics of stellar evolution.

One approach to testing such effects is based on population studies of Galactic globular clusters. In particular, additional energy losses delay helium ignition in red giants, allowing the degenerate helium core to grow to a larger mass and making the tip of the red-giant branch (TRGB) more luminous \cite{Raffelt:1990yz,Viaux:2013lha}. Previous studies have compared observed and predicted TRGB luminosities to constrain ALPs, millicharged particles, hidden photons, non-standard neutrino interactions, and other new-physics scenarios \cite{Giannotti:2015kwo}. Their sensitivity has been limited, however, by uncertainties associated with the selection of cluster RGB members, the precision of cluster distances, photometric measurements, and theoretical stellar modeling. Nevertheless, TRGB constraints on the electron coupling of light ALPs, as well as on several other new-physics models, remain among the strongest available in the literature \cite{Caputo:2024oqc}.

 In the present work, we develop this approach with a homogeneous cluster sample and an explicit treatment of correlated systematic uncertainties:
\begin{itemize}
\item
we consider a new homogeneous sample of 27 Galactic globular clusters \cite{gcP8}, constructed using the latest available photometric data and combining Gaia Data Release~3 (DR3), Hubble Space Telescope (HST), and ground-based observations;
\item
we use state-of-the-art distance determinations and Gaia DR3 proper motions, improving both the precision of the luminosity measurements and the reliability of cluster-membership selection;
\item
we determine both a non-variable TRGB estimator and an estimator that uses variable stars only when sufficiently well-sampled photometric time series are available.   The variable-inclusive estimator is adopted for the primary particle-physics likelihood, while the non-variable estimator is retained as a population-selection control;
\item
we use stellar-population parameters determined homogeneously through isochrone fitting as inputs to theoretical simulations with Modules for Experiments in Stellar Astrophysics (MESA) \cite{Paxton:2011,Paxton:2013,Paxton:2015,Paxton:2018,Paxton:2019,Jermyn:2023}, performed separately for each cluster;
\item
 we determine observational and theoretical uncertainties cluster by cluster using dedicated calculations, explicitly distinguish cluster-specific contributions from systematic effects shared by the sample, and combine the measurements by profiling over the corresponding common nuisance parameters.  Three hot-flasher clusters are excluded from the primary axion fit and restored as a stability test.
\end{itemize}

In Sec.~\ref{sec:obs}, we describe the observational data and the procedures used to select the brightest red giants, determine their bolometric luminosities and associated uncertainties, and infer the TRGB luminosity from these measurements. Section~\ref{sec:sim} presents the theoretical part of the analysis, including our MESA simulations with ALP--electron interactions and a detailed discussion of the various sources of theoretical uncertainty. In Sec.~\ref{sec:comparison}, we compare the standard-cooling predictions directly with the observed TRGB magnitudes, examine their dependence on cluster parameters, and compare our results with previous TRGB studies. \red Section~\ref{sec:constraints} then introduces the likelihood method, derives the ALP--electron constraint, compares it with previous results, and gives indicative sensitivity estimates for other cooling mechanisms. We conclude in Sec.~\ref{sec:concl}.\black

\section{Observed TRGB luminosities in 27 globular clusters}
\label{sec:obs}

\subsection{Cluster sample}
\label{sec:obs:sample}
In Ref.~\cite{gcP8}, we derived a homogeneous set of metallicity $[$Fe$/$H$]$, age, distance from the Sun $D$, and reddening for 27 Galactic globular clusters from fitting of their colour--magnitude diagrams (CMDs) with theoretical isochrones derived from stellar evolution models by Dartmouth Stellar Evolution Database (DSED, \cite{dotter2008})\footnote{\url{http://stellar.dartmouth.edu/models/}} and a Bag of Stellar Tracks and Isochrones (BaSTI, \cite{BaSTIalpha2021})\footnote{\url{http://basti-iac.oa-abruzzo.inaf.it/index.html}} for $\alpha$--enrichment $[\alpha/$Fe$]=+0.4$, which is typical for these clusters.

For robust fitting, we used helium mass fraction estimates $Y_{\rm 1G}$ and $Y_{\rm 2G}$ for the primordial (1G) and enriched (2G) generations of cluster members from Ref.~\cite{milone2018} to consider only clusters with a mild helium mass fraction difference between the generations.

We pay special attention to identify and analyze variable stars among the cluster members. Accordingly, we put off several clusters, for which we have not analyzed their variable stars.

To cover the entire cluster areas, we consider only clusters, for which observations are available in all of the three following data sets:
(i) {\it Hubble Space Telescope (HST)} Wide Field Channel of the Advanced Camera for Surveys (ACS) \cite{nardiello2018} (hereafter {\it HST} data set)\footnote{\url{http://groups.dfa.unipd.it/ESPG/treasury.php}} which covers only a few central arcminutes of the areas, 
(ii) {\it Gaia} DR3 \cite{gaiadr3} (hereafter {\it Gaia} data set) covering mainly periphery of the areas, and
(iii) the {\it Gaia} DR3 cluster members among the stars in the reduction of ground-based photometric observations by \cite{stetson2019} (hereafter Stetson data set)\footnote{\url{http://cdsarc.u-strasbg.fr/viz-bin/cat/J/MNRAS/485/3042}, its ongoing updates are presented at
\url{https://www.canfar.net/storage/vault/list/STETSON/homogeneous/Latest_photometry_for_targets_with_at_least_BVI}.}, which also covers the peripheries.
The latter is important as a large amount of independent photometric observations averaged for several decades and, hence, representing nearly average colours and magnitudes of variable stars.
The data sets are cross-identified to each other.

We use CMDs with the {\it Gaia} $G_\mathrm{BP}$ (effective wavelength $\lambda_\mathrm{eff}=505$ nm) and $G_\mathrm{RP}$ ($\lambda_\mathrm{eff}=770$ nm) filters,
the {\it HST} ACS $F606W$ ($\lambda_\mathrm{eff}=599$ nm) and $F814W$ ($\lambda_\mathrm{eff}=807$ nm) filters,
and the Stetson $B$ ($\lambda_\mathrm{eff}=452$ nm) and $I$ ($\lambda_\mathrm{eff}=807$ nm) filters.

For robust determination of the parameters, we consider only relatively close clusters, for which the three data sets cover their red giant branch (RGB), horizontal branch (HB), asymptotic giant branch (AGB), subgiant branch (SGB), turn-off (TO), and a brighter part of their main sequence (MS). This effectively limits the cluster distance to $\sim$20 kpc from the Sun.    
We consider only clusters, whose RGB, HB, and AGB stars are rather confidently separated in any CMD under consideration.

To select cluster members, we use astrometry from {\it HST} \cite{libralato2022} or {\it Gaia} DR3.

We analyze and take into account differential reddening in the areas of the clusters using the method of \cite{bonatto2013}.

Isochrones are fitted directly to the bulk of cluster members. For various sets of the derived parameters ($[$Fe$/$H$]$, distance from the Sun $D$, reddening, and age) in their 4-dimensional space, we find the best solution that is the set with the minimal sum of the squares of the residuals between the isochrones and the data points.
The contribution of different CMD domains is balanced by a weight of each data point, which is inversely proportional to the number of stars of a given magnitude. Some stars are excluded from the fitting:
the extremely blue HB (with effective temperature $T_\mathrm{eff}>9000$ K),
blue stragglers, RR~Lyrae stars, and some other variable stars including all variable red giants near the TRGB.

More details of our approach are presented in our previous papers \cite{gcP1,gcP2,gcP3,gcP4,gcP5,gcP6,gcP7,gcP8}.

Different origin and association of the clusters may impact the relations between their parameters.
Therefore, in our figures, we mark clusters of different origin and association by different colours, the same as in our previous paper \cite{gcP8}.
We consider clusters associated with the disk, bulge, low energy, Helmi stream, Gaia-Sausage-Enceladus progenitor, and Sequoia progenitor following Refs.~\cite{callingham2022,belokurov2024,boldrini2025,massari2025}.

\subsection{Selection of the most luminous red giants: non-variable stars}
\label{sec:obs:selection}
Our  first  analysis of each cluster is based on its non-variable star with the highest luminosity among all the stars in all three data sets under consideration. 
We have to use the luminosity of each candidate instead of its magnitude, since we consider six different filters of the three data sets, whose conversion into each other or into bolometric magnitude is poorly defined for very red and cool TRGB stars.
Our deliberate selection of clusters, whose RGB and AGB stars are rather confidently separated in any CMD under consideration, allows us to be sure about the RGB or AGB status of any candidate.
We consider only candidates positioned with respect to isochrone in CMD within a combination of all uncertainties, including those of photometry, model, and derived cluster parameters. Typically, this combined uncertainty is several hundredths of a magnitude for both colour (abscissa) and magnitude (ordinate) in CMD.
The luminosity of each candidate is derived as that of the isochrone point (interpolated if needed) closest to the candidate in the CMD under consideration.

Finally, we average up to six independent estimates of the luminosity for each candidate: up to three data sets (and three corresponding CMDs) by two isochrones/models (BaSTI or DSED). We expected some cases when the same pair of candidates is observed in two or more data sets and one candidate is more luminous in one data set, while the other is so in another data set. However, we have meet only a few such cases. For example, the stars Gaia DR3 4365633698594456576 and 4365647481152213632 in NGC\,6254 are slightly more luminous using the {\it Gaia} and Stetson data sets, respectively. Such a ``competition'' between candidates is rare primarily because the brighter part of the RGB is rather sparsely populated. Yet, our combination of the three data sets appears very important, since in some clusters the stars at the TRGB populate predominantly inner or outer parts of the cluster area, observed by {\it HST} and {\it Gaia}, respectively. Accordingly, some previous estimates of the TRGB luminosity or bolometric magnitude using only {\it HST}  or {\it Gaia}  data sets may be biased loosing the brightest RGB star at the cluster periphery.

The identification and exclusion of the variable stars at the TRGB is based on the information about variability in the database of Clement \cite{clement2017}\footnote{\url{https://www.astro.utoronto.ca/~cclement/cat/listngc.html}}, the parameter \verb"VarFlag=VARIABLE" from the {\it Gaia} data set, or the parameters \verb"Vary" (Welch-Stetson variability index) and \verb"Weight" (weight of the variability index) for stars in the Stetson data set.

\subsection{Analysis with variable red giants}
In the  first  analysis, we exclude the variable stars from the set of candidate TRGB tracers and define the observational TRGB from the brightest retained non-variable red giant, with a separate correction for the removal of the brightest $k$ stars, see Sec.~\ref{sec:obs:MCcorr}. Indeed, variable red giants experience changes in their luminosities and temperatures \cite{Ita2021,Nicholls2009,Lebzelter2014}, and an instantaneous position of a variable red giant on the CMD cannot be interpreted as straightforwardly as for a non-variable RGB star. Notably, Ref.~\cite{CapozziRaffelt2020} found that the brightest star previously used in M5 was in fact a semiregular long-period variable with a large amplitude and therefore dropped M5 from their TRGB calibration sample altogether \cite{CapozziRaffelt2020}.  More generally, Anderson et al.\ demonstrated that variability-related population selection near the TRGB can shift the inferred tip magnitude at the level comparable to the precision targeted in modern TRGB work \cite{Anderson2024}.

However, we perform an alternative analysis for ten clusters having at least one well-defined variable star at the TRGB with its luminosity higher than that of the brightest non-variable star. 
We consider variable star well-defined, if 
\begin{enumerate}
\item its coordinates, magnitude and amplitude are estimated in the database of Clement \cite{clement2017},
\item its amplitude does not exceed 0.5~mag,
\item its magnitude in the Stetson data is averaged over enough observations (namely, the amplitude divided to square root of the number of observations is less than all other uncertainties) is within the stated photometric uncertainties from the magnitude estimate in \cite{clement2017},
\item its deviation in the Stetson CMD from the best-fit isochrone is not larger than the total uncertainty (variable stars at the TRGB are not used in our fitting),
\item it is certainly (more than the sum of all uncertainties) more luminous than the most luminous non-variable star. For example, variable star {\it HST}~R0005537 in NGC\,362 has nearly the same luminosity as non-variable star {\it HST}~R0003183, hence, we ignore the former and use the latter.
\end{enumerate}
To derive the TRGB luminosity from the most luminous variable star, we use only its Stetson data suggesting that they cover the entire variability period. We ignore the {\it HST} and {\it Gaia} data for such a variable, since they are based on a few observations at a random variability phase and, hence, may provide biased magnitude and colour.

It is worth noting that many of the clusters have poorly defined variable stars which are more luminous than the most luminous non-variable star: they have too high or uncertain amplitude, large deviation from the best-fit isochrones in a CMD, etc.
We have to exclude such stars and, hence, to compensate it by Monte-Carlo correction.
 Therefore, we retain both estimators.  The well-defined variable stars in ten clusters reduce the residual scatter and restore the expected metallicity trend; the variable-inclusive estimator is consequently used in the primary particle-physics likelihood, while the estimator based only on non-variable stars is used as a population-selection control.  
Comparison of the  results of the two analyses  is presented in Sec.~\ref{sec:constraints:result}.  

The most luminous stars used for the TRGB luminosity estimate are presented in Table~\ref{tab:stars}.
Their derived bolometric luminosities are given in Table~\ref{tab:results}.
\begin{sidewaystable}[p]
\centering
\def\baselinestretch{1}\normalsize\normalsize
\setlength{\tabcolsep}{2.8pt}
\caption{The most luminous stars used for the TRGB luminosity estimate.
The stars are identified by the 19-digit {\it Gaia} DR3 Source identifier and/or {\it HST} number from \cite{nardiello2018} in the form R1234567.
Their equatorial coordinates R.A. and Dec J2000 are from {\it Gaia} DR3 or, if absent, from {\it HST}.}
\label{tab:stars}
\resizebox{\textheight}{!}{%
\begin{tabular}{rrrrrrrrr}
\toprule
        & \multicolumn{4}{c}{Non-variable star} & \multicolumn{4}{c}{Variable star} \\
Cluster & {\it Gaia} & {\it HST} & R.A. & Dec & {\it Gaia} & {\it HST} & R.A. & Dec \\
NGC288  & 2342904492467793280 &          &  $13.25473$ & $-26.60192$ & 2342907928440967296 & R0001617 &  $13.17140$ & $-26.55752$ \\
NGC362  &                     & R0003183 &  $15.80709$ & $-70.85389$ &                     &          &             &             \\
NGC1261 & 4733794691729785728 & R0006350 &  $48.06529$ & $-55.20389$ &                     &          &             &             \\
NGC5024 & 3938022459637112960 & R0007845 & $198.23188$ &  $18.18990$ & 3938022326492775040 &          & $198.18967$ & $18.19136$  \\
NGC5053 & 3938681999108164992 &          & $199.08589$ &  $17.71290$ &                     &          &             &             \\
NGC5272 &                     & R0007468 & $205.54455$ &  $28.38275$ &                     &          &             &             \\
NGC5466 & 1452626177948989440 &          & $211.34507$ &  $28.50382$ &                     &          &             &             \\
NGC5897 & 6252660878976844928 &          & $229.37684$ & $-21.08359$ & 6252678642961484928 &          & $229.31148$ & $-20.94938$ \\
NGC5904 & 4421574105733508096 &          & $229.76352$ &   $2.13424$ & 4421573379883544704 & R0014658 & $229.64244$ &   $2.10704$ \\
NGC6093 & 6050422821900009728 & R0004344 & $244.25342$ & $-22.98016$ &                     &          &             &              \\
NGC6101 & 5806468821225885312 &          & $246.63320$ & $-72.30973$ &                     &          &             &               \\
NGC6171 & 4330465028905802240 &          & $248.16271$ & $-13.04740$ & 4330463620156488320 &          & $248.13088$ & $-13.04973$  \\
NGC6205 & 1328057184182441472 & R0010259 & $250.44655$ &  $36.46216$ & 1328057871377540224 & R0007780 & $250.39796$ &  $36.45761$  \\
NGC6218 & 4379076469493227008 & R0003187 & $251.82753$ &  $-1.93790$ &                     &          &             &               \\
NGC6254 & 4365647481152213632 &          & $254.37301$ &  $-4.04583$ & 4365659300902222208 &          & $254.36409$ & $-4.02356$ \\
NGC6341 & 1360405430444910208 & R0004099 & $259.30051$ &  $43.13306$ &                     &          &             &               \\
NGC6352 & 5950114891490146048 &          & $261.50577$ & $-48.43926$ & 5950129219503048576 &          & $261.40632$ & $-48.36942$   \\
NGC6362 & 5813088293539416448 & R0001785 & $262.96315$ & $-67.04561$ &                     &          &             &               \\
NGC6366 & 4362226861812857344 &          & $261.93818$ &  $-5.17641$ &                     &          &             &               \\
NGC6397 & 5921747938171016320 &          & $265.08405$ & $-53.70051$ &                     &          &             &               \\
NGC6541 & 6724042095309908864 & R0007644 & $271.99192$ & $-43.70834$ &                     &          &             &              \\
NGC6723 & 6730905075065184640 & R0002036 & $284.89407$ & $-36.63852$ &                     &          &             &               \\
NGC6752 & 6638378251014612096 &          & $287.87581$ & $-59.90872$ &                     &          &             &               \\
NGC6779 & 2039206599066699136 &          & $289.16120$ &  $30.10906$ & 2039259921076958464 & R0005969 & $289.15761$ & $30.20941$     \\
NGC6809 & 6751389006757832832 & R0001085 & $295.01699$ & $-30.97240$ &                     &          &             &               \\
NGC6838 & 1821608643754363520 & R0001274 & $298.44778$ &  $18.78087$ & 1821608849912714368 &          & $298.47502$ & $18.78428$    \\
NGC7099 & 6816576031804765440 &          & $325.01424$ & $-23.19560$ &                     &          &             &               \\
\bottomrule
\end{tabular}
}
\end{sidewaystable}
\begin{sidewaystable}[p]
\centering
\def\baselinestretch{1.1}\normalsize\small
\setlength{\tabcolsep}{4pt}
\caption{The results for the clusters:
$M_{\rm bol,*}^{\rm non}$ and $M_{\rm bol,*}^{\rm var}$ are observed bolometric magnitudes of the most luminous non-variable and variable star, respectively;
the TRGB $M_{\rm bol}^{\rm non}$ and $M_{\rm bol}^{\rm var}$ are these magnitudes corrected for the removal of the brightest stars and some variable stars;
$N_{\rm obs}^{\rm non}$ and $N_{\rm obs}^{\rm var}$ are the number of the RGB stars observed within 2.5 magnitudes from the most luminous non-variable and variable star, respectively;
$k^{\rm non}$, and $k^{\rm var}$ are the number of the ignored variable stars, if the most luminous non-variable or variable star is adopted, respectively.
 The tabulated TRGB central values use the median finite-sampling correction, and the asymmetric errors summarize its 16th and 84th percentiles combined with the remaining one-star uncertainty. The analysis does not approximate this correction by these two errors, but uses the full conditional distribution described in Sec.~\ref{sec:obs:MCcorr}. 
The cluster association: disk, bulge, low energy (low-E), Helmi stream, Gaia-Sausage-Enceladus (GSE), or Sequoia.
}
\label{tab:results}
\resizebox{0.8\textheight}{!}{%

\begin{tabular}{rrcccccccc}
~\vspace{0.5mm}\\ \toprule
Cluster & Association & $M_{\rm bol,*}^{\rm non}$ & $N_{\rm obs}^{\rm non}$ & $k^{\rm non}$ & TRGB $M_{\rm bol}^{\rm non}$ & $M_{\rm bol,*}^{\rm var}$ & $N_{\rm obs}^{\rm var}$ & $k^{\rm var}$ & TRGB $M_{\rm bol}^{\rm var}$ \\
NGC\,288 & GSE & $-2.699{\pm0.078}$ & 64 & 1 & $-2.859_{{-0.214}}^{{+0.127}}$ & $-3.583{\pm0.083}$ & 34 & 0 & $-3.703_{{-0.270}}^{{+0.127}}$ \\
NGC\,362 & GSE & $-3.480{\pm0.116}$ & 109 & 3 & $-3.692_{{-0.205}}^{{+0.155}}$ & $-3.486{\pm0.116}$ & 116 & 2 & $-3.623_{{-0.170}}^{{+0.137}}$ \\
NGC\,1261 & GSE & $-3.388{\pm0.092}$ & 83 & 2 & $-3.590_{{-0.216}}^{{+0.142}}$ & $-3.388{\pm0.092}$ & 83 & 2 & $-3.590_{{-0.216}}^{{+0.142}}$ \\
NGC\,5024 & Helmi & $-3.402{\pm0.062}$ & 210 & 1 & $-3.443_{{-0.078}}^{{+0.067}}$ & $-3.461{\pm0.063}$ & 211 & 0 & $-3.476_{{-0.069}}^{{+0.064}}$ \\
NGC\,5053 & Helmi & $-2.325{\pm0.059}$ & 42 & 1 & $-2.585_{{-0.354}}^{{+0.174}}$ & $-2.325{\pm0.059}$ & 42 & 1 & $-2.585_{{-0.354}}^{{+0.174}}$ \\
NGC\,5272 & Helmi & $-3.359{\pm0.077}$ & 192 & 5 & $-3.540_{{-0.131}}^{{+0.106}}$ & $-3.359{\pm0.077}$ & 192 & 5 & $-3.540_{{-0.131}}^{{+0.106}}$ \\
NGC\,5466 & Sequoia & $-3.250{\pm0.087}$ & 34 & 0 & $-3.370_{{-0.271}}^{{+0.129}}$ & $-3.250{\pm0.087}$ & 34 & 0 & $-3.370_{{-0.271}}^{{+0.129}}$ \\
NGC\,5897 & GSE & $-3.346{\pm0.093}$ & 52 & 4 & $-4.008_{{-0.549}}^{{+0.320}}$ & $-3.403{\pm0.088}$ & 52 & 4 & $-4.065_{{-0.548}}^{{+0.319}}$ \\
NGC\,5904 & GSE & $-3.346{\pm0.082}$ & 380 & 2 & $-3.381_{{-0.087}}^{{+0.084}}$ & $-3.569{\pm0.088}$ & 380 & 1 & $-3.590_{{-0.092}}^{{+0.089}}$ \\
NGC\,6093 & Low-E & $-3.455{\pm0.123}$ & 116 & 3 & $-3.652_{{-0.197}}^{{+0.155}}$ & $-3.455{\pm0.123}$ & 116 & 3 & $-3.652_{{-0.197}}^{{+0.155}}$ \\
NGC\,6101 & GSE & $-3.349{\pm0.107}$ & 52 & 2 & $-3.701_{{-0.399}}^{{+0.221}}$ & $-3.349{\pm0.107}$ & 52 & 2 & $-3.701_{{-0.399}}^{{+0.221}}$ \\
NGC\,6171 & Bulge & $-2.959{\pm0.175}$ & 43 & 1 & $-3.212_{{-0.381}}^{{+0.237}}$ & $-3.343{\pm0.170}$ & 34 & 0 & $-3.463_{{-0.308}}^{{+0.195}}$ \\
NGC\,6205 & GSE & $-2.546{\pm0.082}$ & 303 & 14 & $-2.870_{{-0.146}}^{{+0.122}}$ & $-3.340{\pm0.086}$ & 160 & 0 & $-3.360_{{-0.096}}^{{+0.088}}$ \\
NGC\,6218 & Disk & $-3.209{\pm0.071}$ & 50 & 3 & $-3.740_{{-0.493}}^{{+0.275}}$ & $-3.480{\pm0.071}$ & 57 & 0 & $-3.546_{{-0.155}}^{{+0.088}}$ \\
NGC\,6254 & Low-E & $-3.276{\pm0.085}$ & 102 & 4 & $-3.584_{{-0.251}}^{{+0.162}}$ & $-3.470{\pm0.072}$ & 104 & 2 & $-3.625_{{-0.159}}^{{+0.110}}$ \\
NGC\,6341 & GSE & $-3.296{\pm0.078}$ & 119 & 0 & $-3.324_{{-0.097}}^{{+0.081}}$ & $-3.296{\pm0.078}$ & 119 & 0 & $-3.324_{{-0.097}}^{{+0.081}}$ \\
NGC\,6352 & Disk & $-3.013{\pm0.115}$ & 77 & 2 & $-3.234_{{-0.247}}^{{+0.166}}$ & $-3.013{\pm0.110}$ & 77 & 2 & $-3.234_{{-0.245}}^{{+0.162}}$ \\
NGC\,6362 & Disk & $-3.355{\pm0.091}$ & 60 & 0 & $-3.418_{{-0.158}}^{{+0.103}}$ & $-3.355{\pm0.091}$ & 60 & 0 & $-3.418_{{-0.158}}^{{+0.103}}$ \\
NGC\,6366 & Disk & $-2.988{\pm0.162}$ & 30 & 0 & $-3.125_{{-0.337}}^{{+0.195}}$ & $-2.988{\pm0.162}$ & 30 & 0 & $-3.125_{{-0.337}}^{{+0.195}}$ \\
NGC\,6397 & Low-E & $-3.398{\pm0.124}$ & 42 & 0 & $-3.492_{{-0.235}}^{{+0.145}}$ & $-3.398{\pm0.124}$ & 42 & 0 & $-3.492_{{-0.235}}^{{+0.145}}$ \\
NGC\,6541 & Low-E & $-3.450{\pm0.091}$ & 109 & 0 & $-3.481_{{-0.111}}^{{+0.094}}$ & $-3.450{\pm0.091}$ & 109 & 0 & $-3.481_{{-0.111}}^{{+0.094}}$ \\
NGC\,6723 & Bulge & $-3.053{\pm0.158}$ & 106 & 2 & $-3.205_{{-0.210}}^{{+0.178}}$ & $-3.053{\pm0.158}$ & 106 & 2 & $-3.205_{{-0.210}}^{{+0.178}}$ \\
NGC\,6752 & Disk & $-3.478{\pm0.122}$ & 97 & 1 & $-3.577_{{-0.168}}^{{+0.136}}$ & $-3.478{\pm0.122}$ & 97 & 1 & $-3.577_{{-0.168}}^{{+0.136}}$ \\
NGC\,6779 & GSE & $-3.434{\pm0.068}$ & 76 & 2 & $-3.659_{{-0.233}}^{{+0.139}}$ & $-3.444{\pm0.076}$ & 61 & 1 & $-3.613_{{-0.226}}^{{+0.131}}$ \\
NGC\,6809 & Low-E & $-3.439{\pm0.072}$ & 76 & 0 & $-3.487_{{-0.122}}^{{+0.081}}$ & $-3.439{\pm0.072}$ & 76 & 0 & $-3.487_{{-0.122}}^{{+0.081}}$ \\
NGC\,6838 & Disk & $-2.846{\pm0.103}$ & 33 & 3 & $-3.600_{{-0.698}}^{{+0.384}}$ & $-3.691{\pm0.100}$ & 34 & 1 & $-4.016_{{-0.448}}^{{+0.228}}$ \\
NGC\,7099 & GSE & $-3.421{\pm0.055}$ & 40 & 1 & $-3.695_{{-0.373}}^{{+0.181}}$ & $-3.421{\pm0.055}$ & 40 & 1 & $-3.695_{{-0.373}}^{{+0.181}}$ \\
\bottomrule
\end{tabular}
}
\end{sidewaystable}
\begin{figure*}
   \centering
   \includegraphics[width=14cm, angle=0]{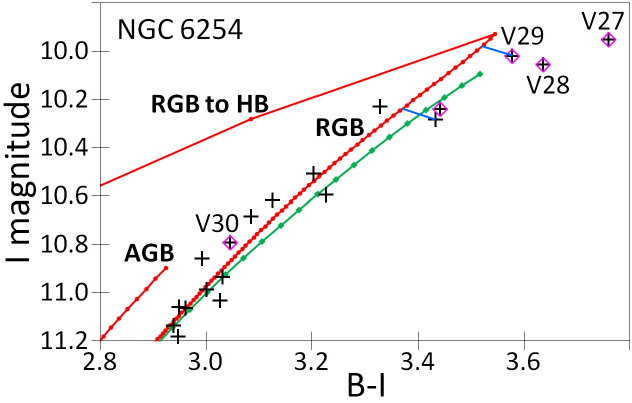}
   \caption{The TRGB domain of the Stetson CMD for NGC\,6254. The cluster members -- black crosses, variable stars -- magenta diamonds with known labels. The isochrone points for $Y\approx0.25$ from the best-fitting BaSTI and DSED isochrones are red and green symbols, respectively, with the interpolated lines between them. 
   The RGB, AGB and RGB-to-HB parts of the BaSTI isochrone are labeled. 
The blue lines connect the most luminous non-variable and variable stars from Table~\ref{tab:stars} to the corresponding points on the isochrones (the variable star is more luminous than any DSED point).
}
\label{ngc6254rgb}
\end{figure*}
Fig.~\ref{ngc6254rgb} presents an example of the TRGB domain of the Stetson CMD for NGC\,6254 where luminosity of the most luminous variable and non-variable RGB stars is determined.
It is worth noting that both the stars are far from the cluster centre (6.0 and 6.5 arcmin for the non-variable and variable stars, respectively) and, hence, never observed by {\it HST}. This emphasizes the importance of a combination of different data sets covering the entire cluster area in order to estimate $M_{\rm bol}$ correctly.
Also Fig.~\ref{ngc6254rgb} shows that 
(i) the variable star is much brighter than the non-variable one,
(ii) both the stars are certainly belong to the RGB, not to the AGB,
(iii) we ignore two variable stars brighter than the adopted one (V29): the reddest star (V27) has too high amplitude of 0.68 mag and it is far from the isochrones, while V28 provides the same $M_{\rm bol}^{\rm var}$, but it is farther from the isochrones.

Despite considerable difference between magnitudes of the variable and non-variable stars, the corrected bolometric luminosities of TRGB derived in 
 their  
analyses coincide within the uncertainties, see Table~\ref{tab:results}.

\subsection{Brightest-red-giant luminosities and their uncertainties}
\label{sec:obs:Mbol}
The average luminosity of the most luminous star of a cluster is corrected for the removal of the brightest $k$ stars and some variable stars and then converted into bolometric absolute magnitude $M_{\rm bol}=4.74-2.5\log(L/L_\odot)$, where $L$ denotes luminosity and the subscript $\odot$ denotes the solar value.

For each cluster, we assign an uncertainty to the adopted bolometric magnitude of the brightest non-variable RGB star, $M_{\rm bol}^{\rm obs}$.  The goal is to include not only the formal observational errors, but also the dominant uncertainties associated with the conversion from multi-band photometry and cluster parameters to an absolute bolometric magnitude. 

 The adopted $M_{\rm bol}^{\rm obs}$ values use the two $\alpha$-enhanced isochrone families fitted independently to each cluster, BaSTI and DSED.  For a non-variable TRGB candidate, each available Gaia, HST, and Stetson CMD provides one value from each family; the adopted value is the average of the resulting two to six determinations. For an accepted variable star, snapshot Gaia and HST photometry cannot replace a phase-averaged magnitude, so only Stetson-photometry determinations (for BaSTI and DSED) enter.  Direct SED integration is retained as an external check for the 22 non-variable stars having infrared photometry.

We decompose the optical method spread as follows.  For cluster $i$, let $K_i=1,2,3$ be the number of available photometric systems, $\bar m_{iB}$ and $\bar m_{iD}$ the family means, and $\bar m_{ip}$ the mean of BaSTI and DSED in system $p$.  The signed isochrone-family response is
\begin{equation}
 s_{i,{\rm iso}}=\frac{\bar m_{iB}-\bar m_{iD}}{2}.
 \label{eq:method-isochrone}
\end{equation}
Writing $r_{ip}=\bar m_{ip}-K_i^{-1}\sum_p\bar m_{ip}$ for an available system and setting it to zero otherwise, the two orthogonal photometric-system responses are
\begin{equation}
 s_{i,1}=\frac{r_{iG}-r_{iH}}{\sqrt{2K_i}},\qquad
 s_{i,2}=\frac{r_{iG}+r_{iH}-2r_{iS}}{\sqrt{6K_i}}.
 \label{eq:method-photometry}
\end{equation}
Their quadratic sum is the population variance of the available system means.  The small family--system interaction and any residual difference between the recorded average and the raw determinations are retained as cluster-specific terms $u_{i,{\rm int}}$ and $u_{i,{\rm avg}}$.

For the 22 non-variable stars with infrared coverage, direct SED integration provides a useful external check on the adopted bolometric magnitudes.  The SED--baseline differences have median $+0.0216$~mag and robust scatter $0.0441$~mag.  However, each of these SED estimates is based on only a few broad-band photometric points, so we do not regard either the sample-wide SED offset or its scatter as a sufficiently robust calibration of the bolometric zero point or of an additional noise term.  No SED-derived common nuisance parameter and no SED-derived residual uncertainty enter the likelihood; the SED comparison is used only as an external consistency check.

The adopted optical-only method uncertainty is therefore
\begin{equation}
 \delta_{{\rm method},i}^{2}=s_{i,{\rm iso}}^2+s_{i,1}^2+s_{i,2}^2
 +u_{i,{\rm int}}^2+u_{i,{\rm avg}}^2.
 \label{eq:method-decomposition}
\end{equation}
The first three terms are implemented as shared signed nuisance modes; the last two are cluster-specific.  The SED availability is shown in Table~\ref{tab:method-systematics} and is purely diagnostic.

Reddening is one of the largest external inputs.  We propagated two related effects separately.  First, we estimated the sensitivity to the adopted extinction law by repeating the analysis with $R_V=3.1$ and $R_V=3.3$, using the Cardelli--Clayton--Mathis law with the O'Donnell optical update~\cite{Cardelli1989,ODonnell1994}.  Second, we propagated the uncertainty in the reddening value itself.  The statistical reddening term was obtained by recomputing the BaSTI solution at $E(B-V) \pm\sigma_{E(B-V)}$.  The reddening-scale systematic amplitude was calibrated from comparisons with independent reddening determinations, including Ref.~\cite{Legnardi2023}; for the correlated analysis we additionally retain the sign of the corresponding change of $M_{\rm bol}$.

 The statistical distance term is the quoted uncertainty of our baseline isochrone-fit distance.  For the distance-scale systematic in Table~\ref{tab:mbol_error_budget_full}, we convert the uncertainty of the comparison scale of Ref.~\cite{BaumgardtVasiliev2021} to magnitudes.  Its sign in the correlated analysis follows the direction of the central $M_{\rm bol}$ shift between the comparison and baseline distances.

We treated metallicity, age, and helium abundance in the same way.  For metallicity, the statistical term was obtained by shifting the baseline $[\mathrm{Fe/H}]$ by its quoted uncertainty, while the systematic term was estimated by replacing the baseline metallicity scale with the comparison scale of \cite{Carretta2009}.  For age and helium content $Y$, statistical terms were computed from the response to the adopted uncertainties in age and $Y$, and systematic terms were estimated from the shifts, using the age compilation of Ref.~\cite{WagnerKaiser2017} and the helium estimates of Ref.~\cite{Valcin2025}.  In all cases, the response was evaluated by recomputing the bolometric magnitude with the modified input parameter.  For the shared-systematic treatment, the relative signs of the metallicity, age, and helium responses are fixed by the direction of the corresponding parameter shift and by the monotonic response of the bolometric TRGB in the relevant old-cluster regime.

Table~\ref{tab:mbol_error_budget_full} gives the revised one-cluster budget for the non-variable estimator, and Table~\ref{tab:method-systematics} lists the method terms for both estimators.  We denote by $s_{a,i}$ the signed response of cluster $i$ to a shared systematic source $a$.  Here $a$ labels $R_V$, $E(B-V)$, distance, $[\mathrm{Fe/H}]$, age, $Y$, the isochrone family, and the two photometric-system modes.  The absolute values $|s_{a,i}|$ are the systematic amplitudes contributing to the one-cluster uncertainty budget, while the signs specify the coherent direction of the response. In the joint likelihood these terms are not treated as independent errors of the clusters. Instead, a single nuisance amplitude $\eta_a$ is assigned to each source and constrained by a unit Gaussian penalty,
\begin{equation}
M_{{\rm bol},i}^{\rm obs}\rightarrow M_{{\rm bol},i}^{\rm obs}+\sum_a s_{a,i}\eta_a .
\label{eq:obs-common-nuisance}
\end{equation}
The remaining statistical terms and the residual $u_i$ are cluster-specific. The columns $\sigma_{\rm stat}$, $\delta_{\rm syst}$, and $\sigma_{\rm total}$ in Table~\ref{tab:mbol_error_budget_full} are therefore useful summaries of the stand-alone uncertainty budget for each cluster, but the joint likelihood uses the signed response vectors and profiles over the corresponding common nuisance amplitudes rather than adding these systematic terms independently for each cluster. The corresponding total uncertainties for individual clusters are typically at the level of $\simeq0.1$ mag.

\IfFileExists{mbol_error_budget_rotated.tex}{%
  \begin{sidewaystable}[p]
\centering
\scriptsize
\setlength{\tabcolsep}{4pt}
\caption{Error budget for the non-variable bolometric TRGB absolute magnitudes. Statistical and combined uncertainties are denoted by $\sigma$.  The six columns denoted by $s_a$ give signed responses to external systematic sources shared by the cluster sample.  The displayed $\delta_{\rm method}$ is the optical-only value of Eq.~\eqref{eq:method-decomposition}.  The estimator-dependent method terms used for the variable-inclusive likelihood are listed in Table~\ref{tab:method-systematics}.  In the last row, signed-response columns are averaged in absolute value.  See the text for more details.}
\label{tab:mbol_error_budget_full}
\resizebox{\textheight}{!}{%
\begin{tabular}{lrrrrrrrrrrrrrrrr}
\toprule
Cluster &  $\sigma_{\rm M}$ & $\delta_{\rm method}$ & $s_{\rm RV}$ & $\sigma_{\rm EBV}$ & $s_{\rm EBV}$ & $\sigma_{\rm dist}$ & $s_{\rm dist}$ & $\sigma_{\rm Fe/H}$ & $s_{\rm Fe/H}$ & $\sigma_{\rm age}$ & $s_{\rm age}$ & $\sigma_{\rm Y}$ & $s_{\rm Y}$ & $\sigma_{\rm stat}$ & $\delta_{\rm syst}$ & $\sigma_{\rm total}$ \\
\midrule
NGC288 & 0.0025 &  0.0409 & 0.0000 & 0.0340 &  -0.0225 & 0.0440 &  -0.0214 & 0.0045 & 0.0140 & 0.0091 & 0.0071 & 0.0007 & 0.0042 & 0.0566 &  0.0538 &  0.0781 \\
NGC362 & 0.0001 &  0.0075 & 0.0000 & 0.0834 &  -0.0512 & 0.0244 & 0.0236 & 0.0368 & 0.0339 & 0.0010 &  -0.0035 & 0.0016 & 0.0157 & 0.0944 &  0.0682 &  0.1165 \\
NGC1261 & 0.0001 &  0.0422 & 0.0316 & 0.0465 & 0.0000 & 0.0263 & 0.0253 & 0.0304 &  -0.0308 & 0.0019 &  -0.0039 & 0.0013 & 0.0163 & 0.0615 &  0.0682 &  0.0918 \\
NGC5024 & 0.0001 &  0.0291 & 0.0000 & 0.0317 & 0.0106 & 0.0295 &  -0.0216 & 0.0163 & 0.0166 & 0.0001 &  -0.0002 & 0.0001 & 0.0003 & 0.0463 &  0.0413 &  0.0621 \\
NGC5053 & 0.0020 &  0.0122 &  -0.0042 & 0.0122 & 0.0122 & 0.0438 &  -0.0289 & 0.0070 & 0.0097 & 0.0072 &  -0.0032 & 0.0009 & 0.0090 & 0.0466 &  0.0365 &  0.0592 \\
NGC5272 & 0.0001 &  0.0521 & 0.0000 & 0.0268 & 0.0133 & 0.0321 &  -0.0174 & 0.0200 &  -0.0214 & 0.0003 &  -0.0006 & 0.0015 & 0.0094 & 0.0464 &  0.0611 &  0.0768 \\
NGC5466 & 0.0018 &  0.0257 & 0.0000 & 0.0265 & 0.0264 & 0.0510 &  -0.0220 & 0.0169 & 0.0391 & 0.0011 &  -0.0100 & 0.0014 & 0.0221 & 0.0600 &  0.0629 &  0.0869 \\
NGC5897 & 0.0008 &  0.0309 & 0.0000 & 0.0300 & 0.0594 & 0.0311 & 0.0408 & 0.0165 &  -0.0010 & 0.0008 &  -0.0073 & 0.0025 & 0.0181 & 0.0463 &  0.0809 &  0.0931 \\
NGC5904 & 0.0026 &  0.0196 & 0.0000 & 0.0310 &  -0.0310 & 0.0628 &  -0.0174 & 0.0070 & 0.0019 & 0.0016 &  -0.0027 & 0.0003 & 0.0061 & 0.0705 &  0.0412 &  0.0816 \\
NGC6093 & 0.0001 &  0.0461 & 0.0249 & 0.0000 & 0.0000 & 0.0311 & 0.0243 & 0.0103 & 0.0103 & 0.0010 &  -0.0033 & 0.0042 & 0.1030 & 0.0330 &  0.1186 &  0.1232 \\
NGC6101 & 0.0031 &  0.0216 & 0.0192 & 0.0291 & 0.0759 & 0.0535 &  -0.0279 & 0.0132 & 0.0000 & 0.0008 &  -0.0021 & 0.0006 &  -0.0082 & 0.0624 &  0.0863 &  0.1064 \\
NGC6171 & 0.0030 &  0.0442 & 0.0305 & 0.0634 & 0.1410 & 0.0159 &  -0.0295 & 0.0422 & 0.0012 & 0.0204 &  -0.0220 & 0.0002 &  -0.0028 & 0.0805 &  0.1553 &  0.1750 \\
NGC6205 & 0.0001 &  0.0276 & 0.0000 & 0.0521 & 0.0317 & 0.0088 &  -0.0221 & 0.0263 &  -0.0040 & 0.0090 &  -0.0103 & 0.0004 & 0.0268 & 0.0597 &  0.0556 &  0.0816 \\
NGC6218 & 0.0001 &  0.0292 & 0.0000 & 0.0163 & 0.0000 & 0.0569 &  -0.0206 & 0.0084 & 0.0112 & 0.0026 &  -0.0032 & 0.0003 & 0.0040 & 0.0598 &  0.0378 &  0.0707 \\
NGC6254 & 0.0019 &  0.0464 & 0.0251 & 0.0254 & 0.0252 & 0.0382 & 0.0272 & 0.0254 &  -0.0004 & 0.0026 & 0.0051 & 0.0027 & 0.0147 & 0.0526 &  0.0662 &  0.0847 \\
NGC6341 & 0.0001 &  0.0533 & 0.0000 & 0.0164 & 0.0164 & 0.0383 &  -0.0180 & 0.0105 & 0.0033 & 0.0006 &  -0.0002 & 0.0005 & 0.0281 & 0.0430 &  0.0651 &  0.0779 \\
NGC6352 & 0.0066 &  0.0342 & 0.0000 & 0.0666 & 0.0480 & 0.0369 &  -0.0284 & 0.0432 &  -0.0243 & 0.0104 &  -0.0202 & 0.0012 & 0.0048 & 0.0884 &  0.0729 &  0.1145 \\
NGC6362 & 0.0001 &  0.0025 & 0.0000 & 0.0593 & 0.0000 & 0.0393 & 0.0189 & 0.0513 & 0.0046 & 0.0023 &  -0.0137 & 0.0000 &  -0.0002 & 0.0877 &  0.0240 &  0.0910 \\
NGC6366 & 0.0049 &  0.1353 & 0.0360 & 0.0262 & 0.0241 & 0.0395 &  -0.0318 & 0.0248 &  -0.0103 & 0.0210 &  -0.0326 & 0.0028 & 0.0244 & 0.0577 &  0.1515 &  0.1621 \\
NGC6397 & 0.0055 &  0.0550 & 0.0209 & 0.0210 & 0.0210 & 0.0798 &  -0.0166 & 0.0125 & 0.0080 & 0.0025 &  -0.0021 & 0.0032 & 0.0633 & 0.0837 &  0.0909 &  0.1236 \\
NGC6541 & 0.0001 &  0.0303 & 0.0000 & 0.0323 &  -0.0184 & 0.0201 &  -0.0290 & 0.0203 & 0.0189 & 0.0006 &  -0.0007 & 0.0030 & 0.0626 & 0.0432 &  0.0798 &  0.0908 \\
NGC6723 & 0.0001 &  0.0391 & 0.0000 & 0.0489 & 0.1304 & 0.0322 &  -0.0261 & 0.0455 & 0.0091 & 0.0040 &  -0.0129 & 0.0019 &  -0.0084 & 0.0743 &  0.1398 &  0.1583 \\
NGC6752 & 0.0054 &  0.0125 & 0.0000 & 0.0729 & 0.0432 & 0.0485 &  -0.0216 & 0.0617 & 0.0012 & 0.0174 & 0.0008 & 0.0032 & 0.0209 & 0.1087 &  0.0542 &  0.1214 \\
NGC6779 & 0.0054 &  0.0138 & 0.0196 & 0.0197 & 0.0000 & 0.0507 & 0.0298 & 0.0093 & 0.0000 & 0.0006 &  -0.0002 & 0.0002 & 0.0054 & 0.0554 &  0.0386 &  0.0675 \\
NGC6809 & 0.0001 &  0.0225 & 0.0000 & 0.0335 & 0.0446 & 0.0287 &  -0.0209 & 0.0165 & 0.0064 & 0.0010 &  -0.0006 & 0.0002 & 0.0003 & 0.0471 &  0.0545 &  0.0721 \\
NGC6838 & 0.0001 &  0.0488 & 0.0000 & 0.0578 & 0.0000 & 0.0218 &  -0.0271 & 0.0449 & 0.0257 & 0.0145 &  -0.0260 & 0.0016 & 0.0036 & 0.0777 &  0.0668 &  0.1025 \\
NGC7099 & 0.0054 &  0.0288 & 0.0181 & 0.0181 & 0.0180 & 0.0182 &  -0.0230 & 0.0166 & 0.0033 & 0.0012 &  -0.0018 & 0.0010 & 0.0058 & 0.0311 &  0.0453 &  0.0550 \\
\midrule
 Mean abs. & 0.0019 & 0.0352 & 0.0085 & 0.0363 & 0.0320 & 0.0372 & 0.0245 & 0.0236 & 0.0115 & 0.0050 & 0.0073 & 0.0014 & 0.0181 & 0.0620 & 0.0710 & 0.0972 \\
\bottomrule
\end{tabular}%
}
\end{sidewaystable}
}{%
  \fbox{\parbox{0.8\textwidth}{Missing rotated table file: \texttt{mbol\_error\_budget\_rotated.tex}}}
}
\begin{table}[tbp]
\centering
\small
\renewcommand{\arraystretch}{1.08}
\setlength{\tabcolsep}{7pt}

\caption{Construction of the method uncertainty for the brightest red giant bolometric magnitude. $N_{\rm non}$ is the number of BaSTI/DSED determinations from the available Gaia, HST, and Stetson CMDs for the  non-variable star. The SED column records availability of the independent optical--infrared consistency check. For a selected variable, $N_{\rm var}=2$ because only the time-averaged Stetson magnitudes can be used; otherwise the variable-inclusive estimator reuses the non-variable star.}
\label{tab:method-systematics}
\begin{tabular}{lcrrcrr}
\toprule
Cluster & $N_{\rm non}$ & SED & $\delta_{\rm method}^{\rm non}$ & Variable estimator & $N_{\rm var}$ & $\delta_{\rm method}^{\rm var}$ \\
\midrule
NGC~288 & 4 & yes & 0.0409 & variable & 2 & 0.0500 \\
NGC~362 & 2 & no & 0.0075 & non-variable & 2 & 0.0075 \\
NGC~1261 & 6 & no & 0.0422 & non-variable & 6 & 0.0422 \\
NGC~5024 & 6 & yes & 0.0291 & variable & 2 & 0.0288 \\
NGC~5053 & 4 & no & 0.0122 & non-variable & 4 & 0.0122 \\
NGC~5272 & 6 & yes & 0.0521 & non-variable & 6 & 0.0521 \\
NGC~5466 & 4 & yes & 0.0257 & non-variable & 4 & 0.0257 \\
NGC~5897 & 4 & yes & 0.0309 & variable & 2 & 0.0025 \\
NGC~5904 & 4 & yes & 0.0196 & variable & 2 & 0.0388 \\
NGC~6093 & 4 & yes & 0.0461 & non-variable & 4 & 0.0461 \\
NGC~6101 & 4 & yes & 0.0216 & non-variable & 4 & 0.0216 \\
NGC~6171 & 4 & yes & 0.0442 & variable & 2 & 0.0050 \\
NGC~6205 & 6 & no & 0.0276 & variable & 2 & 0.0375 \\
NGC~6218 & 4 & no & 0.0292 & non-variable & 4 & 0.0292 \\
NGC~6254 & 4 & yes & 0.0464 & variable & 2 & 0.0100 \\
NGC~6341 & 6 & yes & 0.0533 & non-variable & 6 & 0.0533 \\
NGC~6352 & 4 & yes & 0.0342 & variable & 2 & 0.0063 \\
NGC~6362 & 2 & yes & 0.0025 & non-variable & 2 & 0.0025 \\
NGC~6366 & 4 & yes & 0.1353 & non-variable & 4 & 0.1353 \\
NGC~6397 & 2 & yes & 0.0550 & non-variable & 2 & 0.0550 \\
NGC~6541 & 6 & yes & 0.0303 & non-variable & 6 & 0.0303 \\
NGC~6723 & 4 & yes & 0.0391 & non-variable & 4 & 0.0391 \\
NGC~6752 & 2 & yes & 0.0125 & non-variable & 2 & 0.0125 \\
NGC~6779 & 2 & yes & 0.0138 & variable & 2 & 0.0238 \\
NGC~6809 & 6 & yes & 0.0225 & non-variable & 6 & 0.0225 \\
NGC~6838 & 2 & yes & 0.0488 & variable & 2 & 0.0363 \\
NGC~7099 & 2 & yes & 0.0288 & non-variable & 2 & 0.0288 \\
\bottomrule
\end{tabular}

\end{table}

\subsection{From the brightest red giant to the TRGB: corrections for discreteness and variable-star removal}
\label{sec:obs:MCcorr}
The brightest retained RGB star in a finite stellar sample is, in general, fainter than the true TRGB.  This discreteness bias is especially important for globular clusters, where the number of stars in the upper RGB is modest and differs from cluster to cluster.  In addition, our observational TRGB candidate is defined after removing variables. We therefore correct the luminosity of the brightest retained red giant to an estimate of the underlying TRGB, and propagate the corresponding asymmetric uncertainty. We calculate the correction and its uncertainties by means of Monte-Carlo simulations inspired by Ref.~\cite{Straniero2020}, though with several important updates.

Let $M_{\rm tip}$ denote the true bolometric TRGB magnitude, and let $M_{(k+1)}$ be the bolometric magnitude of the brightest retained RGB star after the removal of the $k$ brighter variables.  The integer $k$ counts only stars that are otherwise plausible cluster RGB members; AGB stars and non-members are not included in $k$.  We define
\begin{equation}
    \delta_k \equiv M_{(k+1)} - M_{\rm tip} \ge 0 ,
\end{equation}
so that the corrected TRGB magnitude is brighter than the observed retained star by $\delta_k$.

The correction depends primarily on the number of upper-RGB stars available in the cluster \cite{Straniero2020}; our tests confirm that variations of the cluster parameters within the present sample have a much smaller effect. We construct a reference upper-RGB luminosity function from an $\alpha$-enhanced BaSTI evolutionary track \cite{BaSTIalpha2021,BaSTIWeb} with $M=0.8\,M_\odot$, $Z=0.000886$, and $Y=0.248146$. Adjacent track segments are weighted by their evolutionary time intervals. The parent distribution is tabulated down to 4~mag below the tip, while the hidden upper-RGB richness $N_{\rm tip}$ refers to the true 2.5-mag interval below the TRGB.

The observable count is $N_{\rm obs}$, the number of retained RGB stars with $M\le M_{(k+1)}+2.5$~mag. Since this window is displaced by $\delta_k$ relative to the true window, $N_{\rm obs}$ is not identical to the hidden $N_{\rm tip}$. We sum over the latter over the broad working range $1 \le N_{\rm tip} \le 550$, assigning equal weight to each allowed integer value.

While Ref.~\cite{Straniero2020} used Monte-Carlo simulations to determine the correction, it can be evaluated deterministically. Let \(F(\delta)\) and \(f(\delta)\) be, respectively, the CDF and PDF of the magnitude offset \(\delta\) for a randomly selected RGB star within the true 2.5-mag interval below the TRGB. We denote by \(F_{\rm full}\) the corresponding CDF of the parent luminosity function continued to fainter magnitudes beyond this interval.

The number of stars actually belonging to the true 2.5-mag upper RGB interval is not known a priori. We therefore introduce the latent integer \(n\), which we call the \emph{hidden richness}: it is the total number of RGB stars that would populate this interval before the \(k\) brightest stars are excluded and before the finite observational window is shifted relative to the true TRGB. For a specified \(n\), the probability distribution of the offset \(\delta\) of the \((k+1)\)-th brightest star is the standard order-statistics distribution,
\begin{equation}
f_{k+1:n}(\delta)=
\frac{n!}{k!(n-k-1)!}
F(\delta)^k[1-F(\delta)]^{n-k-1}f(\delta).
\label{eq:order-stat-discreteness}
\end{equation}

If the \((k+1)\)-th brightest star lies at an offset \(\delta\) below the true TRGB, then the 2.5-mag observational window is displaced by the same amount toward fainter magnitudes. As a result, additional stars may enter the sample at its faint end. For a population of hidden richness \(n\), their expected number is \(nR(\delta)\), with
\begin{equation}
R(\delta)=
\frac{F_{\rm full}(2.5+\delta)-F_{\rm full}(2.5)}
{F_{\rm full}(2.5)}.
\end{equation}
We model the number of these additional stars as Poisson distributed. Since, for fixed \(n\), the sample would contain \(n-k\) stars after exclusion of the \(k\) brighter objects, the observed richness \(N_{\rm obs}\) requires
$
N_{\rm obs}-(n-k)=N_{\rm obs}-n+k
$
additional stars to have entered through the faint boundary. We denote the Poisson probability by
\(
\mathcal{P}(m\mid\lambda)\equiv e^{-\lambda}\lambda^m/m!,
\)
where \(m\) is the observed count and \(\lambda\) is its expectation value. Summing over the unknown hidden richness therefore gives the conditional finite-sampling distribution used below,
\begin{equation}
p(\delta_k\mid N_{\rm obs},k)\propto
\sum_{n=k+1}^{\min(550,N_{\rm obs}+k)}
f_{k+1:n}(\delta_k) \,
\mathcal{P}\!\left(N_{\rm obs}-n+k\mid nR(\delta_k)\right).
\label{eq:exact-discreteness}
\end{equation}
The sign of this correction, \(
 M_{\rm tip}=M_{(k+1)}-\delta_k, \)
implies that the long positive tail of $p(\delta_k)$ maps to the brighter, more negative side of the TRGB-magnitude distribution.  We retain the complete density of Eq.~\eqref{eq:exact-discreteness} for every $(N_{\rm obs},k)$ used in the analysis and explicitly use this non-Gaussian distribution in the description of uncertainty. The median and central 68\% interval are reported in Table~\ref{tab:results} only for illustration.

\begin{figure*}
   \centering
   \includegraphics[width=0.65\textwidth, angle=0]{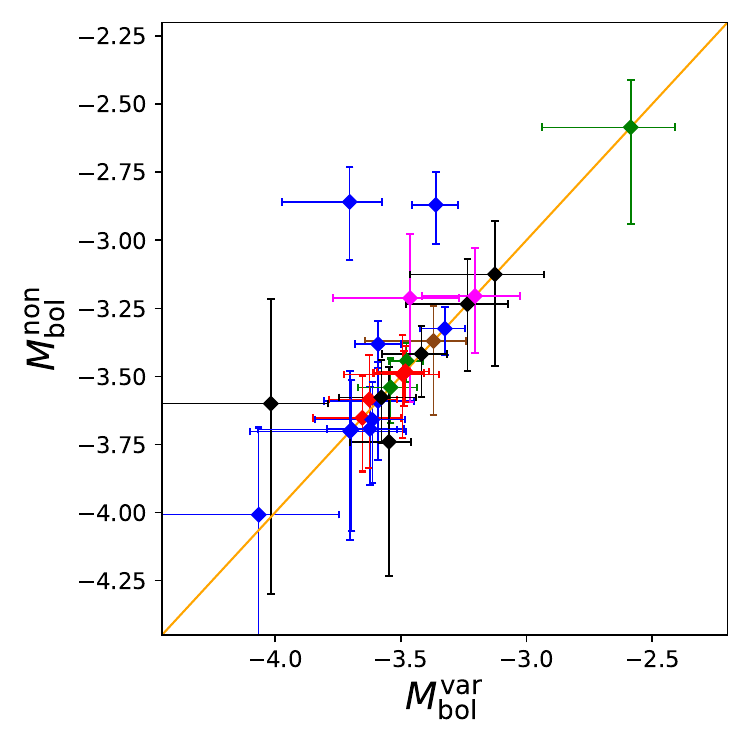}
   \caption{Comparison of the observed TRGB bolometric magnitudes $M_{\rm bol}^{\rm var}$ and $M_{\rm bol}^{\rm non}$ derived from the most luminous variable and non-variable RGB stars of the same cluster, respectively.
The disk and bulge clusters are marked by the black and magenta symbols, respectively;
clusters from the low energy subsample are marked by red symbols;
accreted (ex-situ) clusters associated with the Helmi stream, Gaia-Sausage-Enceladus progenitor, and Sequoia progenitor are marked by green, blue, and brown symbols, respectively.
The orange line shows one-to-one relation.
}
\label{mbolvarnon}
\end{figure*}

The corrected observed TRGB bolometric absolute magnitudes $M_{\rm bol}^{\rm var}$ and $M_{\rm bol}^{\rm non}$, derived, respectively, from the most luminous variable and non-variable RGB stars of the same cluster, are compared in Fig.~\ref{mbolvarnon}. 
Naturally, only ten clusters with the most luminous well-defined variable TRGB star do not lie on the bisector. $M_{\rm bol}^{\rm var}$ and $M_{\rm bol}^{\rm non}$ agree within their uncertainties for eight of them, while disagree only for NGC\,288 and NGC\,6205. These two outliers have no non-variable stars at the brightest part of their RGB. This may be related to the fact that a loose and low-mass cluster NGC\,288 must have a high mass loss at the RGB and high loss of low-mass stars during its evolution. Hence, it has a sparsely populated RGB providing a poor statistics to the correction for the removal of the brightest stars. For NGC\,6205, the explanation may be related to its highest fraction, about 80\%, of the 2G stars with their very high helium mass fraction, $Y=0.268$, making them variable.

\section{Simulated TRGB luminosities in 27 globular clusters}
\label{sec:sim}
Obtaining a robust constraint on the axion--electron coupling, \(g_{ae}\), requires a reliable prediction of the TRGB luminosity and a careful assessment of the theoretical uncertainties associated with stellar-evolution modeling. In this section, we describe our stellar-evolution simulations and discuss the corresponding sources of uncertainty.

We use the open-source stellar-evolution code MESA \cite{Paxton:2011,Paxton:2013,Paxton:2015}. Later releases are documented in Refs.~\cite{Paxton:2018,Paxton:2019,Jermyn:2023}. Our goal is to model the TRGB for the sample of 27 Galactic globular clusters. 

\subsection{Baseline MESA simulations}
\label{sec:sim:MESA}
\subsubsection{Cluster-specific composition and mass.}
\label{sec:cluster_specific_parameters}
In this subsection, we describe the stellar-physics assumptions and the
cluster-dependent input parameters adopted in our baseline simulations.
We begin with the initial stellar parameters. For each cluster, the
initial stellar mass, $M_{\rm init}$, is chosen such that the model
reaches the TRGB at an age consistent with the adopted age of the
cluster. Further details of the initial-mass determination and its
associated uncertainty are given in
Section~\ref{sec:theory_other_uncertainties}.

As described in Section~\ref{sec:obs:sample}, many of the adopted
cluster parameters, including the cluster ages, were obtained from
isochrones calculated at a fixed $\alpha$-enhancement of
$[\alpha/{\rm Fe}]=+0.4$. To maintain consistency between the
observational and theoretical inputs, we adopt the same
$\alpha$-enhancement in our MESA models. As discussed in
Ref.~\cite{gcP8}, allowing for variations in
$[\alpha/{\rm Fe}]$ changes the inferred cluster parameters only
insignificantly and within the quoted uncertainties. We likewise adopt
the $[{\rm Fe}/{\rm H}]$ values obtained from the same isochrone
analysis.

To achieve analogous consistency with the isochrone fitting, we
represent the helium content of each cluster by the population-weighted
abundance
\begin{equation}
Y_{\rm mix}
=
f_{\rm 1G}Y_{\rm 1G}
+
f_{\rm 2G}Y_{\rm 2G},
\qquad
f_{\rm 1G}+f_{\rm 2G}=1,
\label{eq:Ymix}
\end{equation}
where $Y_{\rm 1G}$ and $Y_{\rm 2G}$ are the initial helium abundances
of the first and second stellar generations, respectively, and
$f_{\rm 1G}$ and $f_{\rm 2G}$ are their relative fractions.

As discussed in Section~\ref{sec:obs:sample}, the clusters considered
here are characterized by two dominant stellar populations with
mild differences between their mean helium abundances. The
prescription in Eq.~\eqref{eq:Ymix} provides a single-composition
approximation to this stellar mixture. If $T(Y)$ denotes a mean
theoretical TRGB observable that depends smoothly on the initial
helium abundance, then its value for the population is
\begin{equation}
\left\langle T \right\rangle_{\rm pop}
\simeq
f_{\rm 1G}T(Y_{\rm 1G})
+
f_{\rm 2G}T(Y_{\rm 2G})
=
T(Y_{\rm mix})
+
\mathcal{O}\!\left((\Delta Y)^2\right),
\qquad
\Delta Y
\equiv
Y_{\rm 2G}-Y_{\rm 1G}.
\label{eq:Ymix_approximation}
\end{equation}
The correction linear in $\Delta Y$ vanishes because
\begin{equation}
f_{\rm 1G}
\left(Y_{\rm 1G}-Y_{\rm mix}\right)
+
f_{\rm 2G}
\left(Y_{\rm 2G}-Y_{\rm mix}\right)
=0.
\end{equation}
Therefore, evaluating the stellar model at $Y_{\rm mix}$ reproduces
the population-weighted theoretical prediction to first order in the
internal helium spread. This approximation does not imply that an
individual star has $Y=Y_{\rm mix}$. Instead, $Y_{\rm mix}$ is an
effective composition used to replace the unresolved population
mixture with a single stellar model.

Eq.~\eqref{eq:Ymix_approximation} applies only approximately to the
observational TRGB estimator, even when the helium difference between
the populations is small, because the probability
that the \emph{brightest} red giant belongs to the first generation need not be exactly equal to $f_{\rm 1G}$. We however expect that this difference is negligible. 

\subsubsection{Input physics} \label{sec:theory_input_physics} 

The principal physical ingredients adopted for the baseline MESA calculations are summarized in Table~\ref{tab:baseline_input_physics}. Some aspects of the setup require additional comments.

Convection is treated using the Ledoux stability criterion and the Cox mixing-length theory, with $\alpha_{\rm MLT}=1.82$ and $\alpha_{\rm sc}=0.01$. The adopted mixing-length parameter is motivated by the solar-calibrated value used in the MIST models \cite{Choi:2016}. We note, however, that the numerical value of
$\alpha_{\rm MLT}$ obtained from a solar calibration depends on
the overall physical setup of the stellar model, including the
specific implementation of mixing-length theory. We therefore consider the variation of $\alpha_{\rm MLT}$ as one of sources of uncertainty.

The mixing length of metal-poor globular-cluster stars is not known uniquely. Valcin et al.\ \cite{Valcin:2021} found best-fitting values near $\alpha_{\rm MLT}=1.9$ and adopted $\sigma_{\alpha_{\rm MLT}}=0.15$ as a conservative estimate. We therefore explored the relevant convective parameters over broad
ranges and also tested different values of the convective-overshooting
parameter, as described in Section~\ref{sec:theory_convection}.

Element diffusion is also included in our baseline calculations, as indicated in Table~\ref{tab:baseline_input_physics}. When available, we adopt cluster-specific RGB mass losses inferred from observations \cite{Tailo:2020}. For clusters without such estimates, the Reimers prescription is used with $\eta_{\rm R}=0.3$ \cite{Reimers:1975}. Further details are given in Section~\ref{sec:theory_other_uncertainties}. The variations of the physical prescriptions entering the theoretical uncertainty budget are discussed in Section~\ref{sec:theory_uncertainties}.

\begin{table}[t]
\centering
\caption{Physical ingredients adopted for the baseline stellar-evolution
calculations. Cluster-dependent composition and initial-mass inputs are
specified in Section~\ref{sec:cluster_specific_parameters}.}
\label{tab:baseline_input_physics}

\small
\renewcommand{\arraystretch}{1.08}
\setlength{\tabcolsep}{5pt}

\begin{tabularx}{\textwidth}{
    @{}
    >{\raggedright\arraybackslash}p{3.15cm}
    >{\raggedright\arraybackslash}X
    @{}
}
\toprule
Physical ingredient
&
Baseline prescription
\\
\midrule

Composition and initial mass
&
GS98 solar abundance mixture \cite{GrevesseSauval:1998} with
$[\alpha/{\rm Fe}]=+0.4$; cluster-dependent
$[{\rm Fe}/{\rm H}]$, $Y_{\rm mix}$, and $M_{\rm init}$ from
Section~\ref{sec:cluster_specific_parameters}.
\\

Equation of state
&
Blend of Skye, PC, and FreeEOS
\cite{Jermyn:2021,Potekhin:2010,Irwin:2004}.
\\

Radiative opacity
&
$\alpha$-enhanced OPAL radiative-opacity tables
\cite{Iglesias:1993,Iglesias:1996}, supplemented by the
Ferguson et al.\ low-temperature tables
\cite{Ferguson:2005}.
\\

Convection and mixing
&
Ledoux criterion; Cox MLT \cite{Cox:1968} with $\alpha_{\rm MLT}=1.82$;
semiconvection with $\alpha_{\rm sc}=0.01$ \cite{Langer:1983}.
\\

Element diffusion
&
Multicomponent element diffusion with the updated diffusion
coefficients described in Ref.~\cite[Sec.~9]{Jermyn:2023}.
\\

Nuclear burning and screening
&
\texttt{pp\_cno\_extras\_o18\_ne22.net};
JINA REACLIB reaction rates \cite{Cyburt:2010}, including the
LUNA ${}^{14}{\rm N}(p,\gamma){}^{15}{\rm O}$ rate
\cite{Imbriani2005} and the Fynbo et al.\ triple-$\alpha$ rate
\cite{Fynbo:2005}; Chugunov screening
\cite{Chugunov:2007}.
\\

Neutrino losses
&
Thermal-neutrino energy-loss rates from Itoh et al.\
\cite{Itoh:1996}.
\\
Conductive transport
&

Cassisi electron-conduction opacities with the undamped Blouin correction
\cite{Cassisi:2007,Blouin:2020}.

\\

Atmosphere
&
Grey Eddington $T(\tau)$ relation with
\texttt{atm\_T\_tau\_opacity = 'iterated'}
\cite[Sec.~6]{Jermyn:2023}.
\\

RGB mass loss
&
Reimers wind \cite{Reimers:1975}.
\\

\bottomrule

\end{tabularx}
\end{table}

\subsubsection{Hot-flasher clusters}
\label{sec:hot_flasher_clusters}

Among the clusters considered in this work, NGC\,6352, NGC\,6366 and NGC\,6752 form a particularly interesting
subset. For the adopted values of $[{\rm Fe}/{\rm H}]$, age,
and $\eta_{\rm R}$, the representative stellar model of each
of these clusters follows a hot-flasher (hereafter HF) evolutionary path.
The evolution of these stars differs from the canonical
evolution of low-mass stars. Their mass loss is sufficiently
strong to leave only a very thin hydrogen-rich envelope,
which substantially reduces the contribution of the
hydrogen-burning shell and may lead to its almost complete
extinction. Consequently, a HF star leaves the
red-giant branch before the onset of the core helium flash.

HF are commonly divided into early and late
hot flashers according to the stage of post-RGB evolution
at which core-helium ignition occurs \cite{Cassisi:2003}.
Early HF ignite helium while evolving from the
RGB towards the white-dwarf cooling sequence, whereas
late HF ignite helium only after reaching the
cooling sequence, when the hydrogen-burning shell has
already weakened substantially. For example, in the
late HF sequence calculated by Cassisi et al.\
\cite{Cassisi:2003}, the hydrogen-burning luminosity at
the onset of the helium flash was only
\(L_{\rm H}\simeq0.35\,L_\odot\), and the mass of the
residual hydrogen-rich envelope was
\(M_{\rm env}\simeq5.5\times10^{-4}\,M_\odot\).
Modeling the subsequent evolution of such stars is
challenging. After helium ignition, the flash-driven
convective region may reach the hydrogen-rich envelope and
carry protons into the hot helium-burning layers, where
hydrogen is rapidly consumed. The timescales of proton-capture
reactions and convective mixing then become comparable.
Consequently, convective mixing must be treated
time-dependently and solved simultaneously with the nuclear
reaction network \cite{Cassisi:2003}.

In the present work, however, we follow the evolution of
these stars only up to the TRGB, before the onset of the helium flash.
We therefore do not encounter the numerical difficulties
associated with flash-induced hydrogen mixing and adopt the
same physical ingredients and numerical treatment for the
HF clusters as for the remaining clusters in our
sample. See Sec.~\ref{sec:hf-energy-loss} for a discussion of additional energy losses in hot-flasher clusters.

\subsection{Axion-electron interactions and their implementation in MESA}
\label{sec:sim:axion}
Throughout RGB evolution up to the onset of helium ignition,
axions and axion-like particles with the masses and coupling
strengths considered in this work are in the free-streaming
regime \cite{DiLuzio:2020wdo,Raffelt:1994ry}. We therefore implemented
axion energy losses as an additional contribution to the
neutrino energy-loss rate evaluated in the MESA \texttt{neu} module.

In particle-physics units with $\hbar =c=k_B=1$ \cite{Raffelt:1996Stars}, the interaction of ALP, $a$, with electron, $e$, is described by
the Lagrangian
\begin{equation}
\mathcal{L}_{ae}
=
\frac{g_{ae}}{2m_e}
(\partial_\mu a)\,
\overline{e}\gamma^\mu\gamma^5e,
\end{equation}
where $m_e$ is the electron mass, and
$g_{ae}$ is the dimensionless axion--electron coupling. For
convenience, we introduce the rescaled coupling
\(g_{13}\equiv g_{ae}/10^{-13}\).

The emissivities adopted below are evaluated in the
light-axion limit, $m_a \ll T$, and therefore neglect
finite-mass suppression of axion production.

The dominant axion-production mechanisms induced by the
axion--electron coupling are Compton conversion,
$\gamma+e^-\rightarrow a+e^-$, and bremsstrahlung in
electron--ion collisions,
$e^-+(Z,A)\rightarrow e^-+(Z,A)+a$. Axions may also be
produced through electron--positron annihilation,
$e^+e^-\rightarrow\gamma+a$ \cite{Pantziris1986}, but this
process is less important under the stellar conditions considered
here.

Under the high-density conditions of the degenerate helium core,
bremsstrahlung is the dominant axion-production mechanism. The
specific energy-loss rates in the non-degenerate and degenerate
limits are, respectively \cite{DiLuzio:2020wdo,Raffelt:1994ry},
\begin{align}
\epsilon_{\rm B,ND}
&\simeq
47\,g_{ae}^{\,2}T^{5/2}
\frac{\rho}{\mu_e}
\sum_j
\frac{X_jZ_j}{A_j}
\left(
Z_j+\frac{1}{\sqrt{2}}
\right)
\ {\rm erg\,g^{-1}\,s^{-1}},
\\
\epsilon_{\rm B,D}
&\simeq
8.6\times10^{-7}\,
g_{ae}^{\,2}T^4
\left(
\sum_j\frac{X_jZ_j^2}{A_j}
\right)
F(\kappa,\beta_{\rm F})
\ {\rm erg\,g^{-1}\,s^{-1}}.
\end{align}
Here $T$ and $\rho$ are expressed in K and
${\rm g\,cm^{-3}}$, respectively, and
\begin{equation}
\mu_e
=
\left(
\sum_j\frac{X_jZ_j}{A_j}
\right)^{-1}
\end{equation}
is the mean molecular weight per electron. The quantities $X_j$,
$Z_j$, and $A_j$ are the mass fraction, charge number, and mass
number of ion species $j$, respectively. In the non-degenerate
rate, the terms proportional to $Z_j$ and $1/\sqrt{2}$ describe
the electron--ion and electron--electron contributions,
respectively.

In the degenerate limit, the mild dependence of the
bremsstrahlung rate on density is contained in the dimensionless
function
\begin{align}
F(\kappa,\beta_{\rm F})
={}&
\frac{2}{3}
\ln\left(
\frac{2+\kappa^2}{\kappa^2}
\right)
\\
&+
\left[
\frac{2+5\kappa^2}{15}
\ln\left(
\frac{2+\kappa^2}{\kappa^2}
\right)
-\frac{2}{3}
\right]\beta_{\rm F}^2
+
\mathcal{O}\!\left(\beta_{\rm F}^4\right),
\end{align}
where the terms explicitly retained include the leading
relativistic correction. The electron number density, Fermi
momentum, total Fermi energy, and velocity at the Fermi surface
are
\begin{align}
n_e
&=
\frac{\rho Y_e}{m_u}
=
\frac{\rho}{\mu_em_u}
=
\frac{p_{\rm F}^3}{3\pi^2\hbar^3},
&
p_{\rm F}
&=
\hbar(3\pi^2n_e)^{1/3},
\\
E_{\rm F}
&=
\sqrt{p_{\rm F}^2c^2+m_e^2c^4},
&
\beta_{\rm F}
&=
\frac{p_{\rm F}c}{E_{\rm F}},
\end{align}
where
\begin{equation}
Y_e
\equiv
\sum_j\frac{X_jZ_j}{A_j}
=
\frac{1}{\mu_e}.
\end{equation}

The dimensionless screening parameter is
\begin{equation}
\kappa^2
=
\frac{\hbar^2k_{{\rm D},i}^2}{2p_{\rm F}^2},
\qquad
k_{{\rm D},i}^2
=
\frac{4\pi\alpha_{\rm em}\hbar c}{k_{\rm B}T}
\frac{\rho}{m_u}
\sum_j\frac{X_jZ_j^2}{A_j},
\end{equation}
where $k_{{\rm D},i}$ is the ionic Debye wavenumber,
$\alpha_{\rm em}$ is the electromagnetic fine-structure constant,
and $m_u$ is the atomic mass unit. Thus, $\kappa$ describes ionic
screening, whereas $\beta_{\rm F}$ accounts for the relativistic
motion of electrons at the Fermi surface.

For the intermediate regime between degenerate and
non-degenerate conditions, we use the interpolation proposed by
Raffelt and Weiss \cite{Raffelt:1994ry},
\begin{equation}
\epsilon_{\rm B}
=
\left(
\epsilon_{\rm B,D}^{-1}
+
\epsilon_{\rm B,ND}^{-1}
\right)^{-1}.
\end{equation}

Another important axion-production mechanism in stars is the
Compton process. Its energy-loss rate in the  non-degenerate  limit
can be written as
\begin{equation}
\epsilon_{\rm C,ND}
=
2.7\times10^{-22}\,
g_{ae}^{\,2}
\frac{T^6}{\mu_e}
\ {\rm erg\,g^{-1}\,s^{-1}}.
\end{equation}

The Compton process is suppressed in degenerate matter by Pauli
blocking. States well below the Fermi energy are occupied, so
only electrons within an energy interval of order $k_{\rm B}T$
around the Fermi surface can efficiently scatter into available
final states. In the strongly degenerate limit, the corresponding
suppression factor is approximately \cite{Raffelt:1994ry}
\begin{equation}
F_{\rm deg}
\simeq
\frac{3E_{\rm F}k_{\rm B}T}
     {p_{\rm F}^2c^2}.
\end{equation}
For intermediate degeneracy, the Compton rate is calculated using
\begin{equation}
P
=
\left(1+F_{\rm deg}^{-2}\right)^{-1/2},
\qquad
\epsilon_{\rm C}
=
\epsilon_{\rm C,ND}P,
\end{equation}

\begin{figure}[tbp]
\centering
\includegraphics[width=0.75\linewidth, clip]
{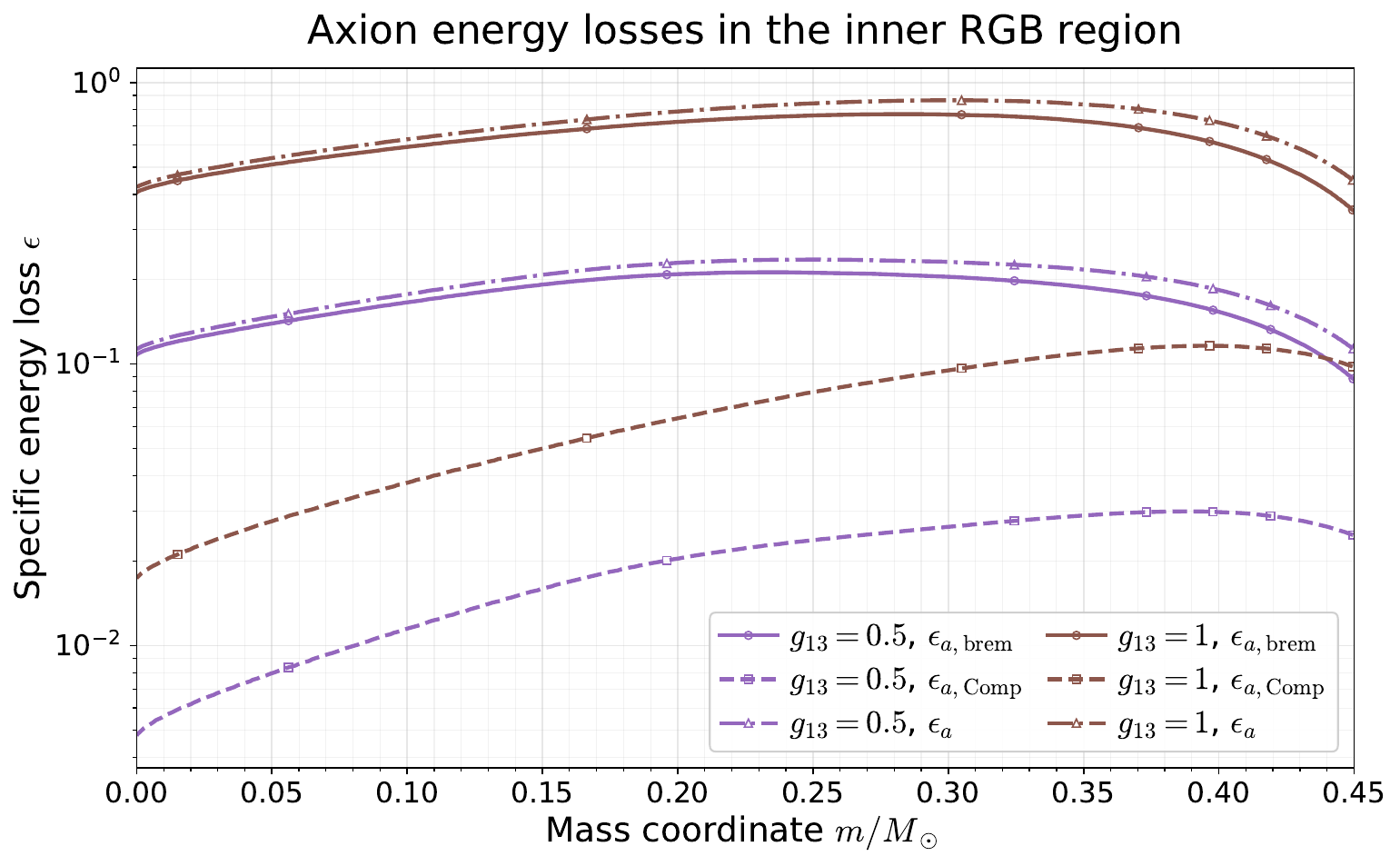}
\caption{%
Specific axion energy-loss rates along the inner region of an
RGB model shortly before helium ignition. The bremsstrahlung,
Compton, and total axion--electron contributions are shown for
$g_{13}=0.5$ and $g_{13}=1$.  The two couplings are shown both to compare the production channels and to verify the expected perturbative scaling: doubling $g_{13}$ raises every axion emissivity by a factor of four.  Over the displayed mass range,
bremsstrahlung exceeds the Pauli-suppressed Compton contribution
and therefore determines most of the total axion energy loss.}
\label{fig:axion-loss-components}
\end{figure}

Consequently, Compton production is strongly suppressed in the
degenerate helium core, and bremsstrahlung provides the dominant
contribution to the axion energy loss, as shown in Fig.~\ref{fig:axion-loss-components}.
 Figure~3 of Ref.~\cite{Straniero2020} provides a complementary absolute comparison of axion, standard-neutrino, and photon luminosities near the TRGB.  We keep our Fig.~\ref{fig:axion-loss-components} focused on the internal axion-rate and coupling-scaling checks.

A more recent calculation by Carenza and Lucente provides the electron--ion bremsstrah\-lung emissivity for a non-relativistic plasma with an arbitrary degree of electron degeneracy and for arbitrary axion masses \cite{Carenza:2021qti}. In the light-axion limit relevant to the present analysis, the authors found that, under representative RGB-core conditions, the Raffelt--Weiss interpolation used in our calculations underestimates the axion emissivity by less than $10\%$ thereby leading to a conservative upper limit on $g_{ae}$. Since the emissivity scales as $\epsilon_a\propto g_{ae}^{2}$, a $10\%$ increase in the emission rate would strengthen the corresponding upper limit on $g_{ae}$ by only about $5 \%$. We retain the Raffelt-Weiss interpolation and quote the conservative limit as our main result.

\subsubsection{Additional energy losses in canonical clusters}
\label{sec:canonical-energy-loss}

Additional cooling has a well-understood effect on the
canonical evolution of a low-mass RGB star
\cite{DiLuzio:2020wdo,Straniero2020}. Because the pressure of the
degenerate helium core depends only weakly on temperature,
axion emission is not efficiently compensated by the usual
expansion and cooling response of a non-degenerate gas. At a
given core mass, the axion-cooled core is consequently colder
than in the standard case, and helium ignition is postponed.

During this delay, the hydrogen-burning shell remains active
and continues to deposit helium ash onto the inert core.
The critical conditions for the off-centre helium flash are
therefore reached at a larger helium-core mass.

Shell-source homology relations show that, at fixed composition
and input physics, the luminosity of an RGB star with a
degenerate helium core is a steeply increasing function of its
core mass \cite{Kippenhahn:2012}.  Consequently, the increase in the core mass at helium ignition
caused by axion cooling produces a brighter TRGB. Our numerical
results fully reproduce this causal sequence: both the
helium-core mass and the maximum RGB luminosity increase
monotonically with $g_{13}$. Since the implemented
emissivities are proportional to $g_{13}^2$, the stellar response is also approximately quadratic at sufficiently small
coupling.

\red
For the representative canonical NGC~6093 model, it is useful to write the response directly in terms of $q=g_{13}^2$ and $\Delta M_{\rm TRGB}(q)\equiv M_{\rm TRGB}(q)-M_{\rm TRGB}(0)$.  A fit constrained to pass through the standard-cooling point, using all calculated MESA points over $0\le g_{13}\le1$, gives
\begin{equation*}
\Delta M_{\rm TRGB}(q)=-0.05817\,q\ {\rm mag}.
\end{equation*}
The maximum absolute residual of the calculated points from this linear relation is $6.3\times10^{-4}$~mag over the full grid.  Thus the squared-coupling variable gives an essentially linear description of the representative response throughout the MESA-calculated range relevant here.  

\red
As a complementary response diagnostic, we repeated the same NGC~6093 calculation at $g_{13}=0$ while multiplying the standard thermal-neutrino cooling rate by a common factor $r_\nu$ over $0.5\le r_\nu\le1.2$.  Defining $\delta r_\nu=r_\nu-1$ and $\Delta M_{\rm TRGB}(r_\nu)\equiv M_{\rm TRGB}(r_\nu)-M_{\rm TRGB}(1)$, the eight calculated points are described over the full interval by
\begin{equation}
\Delta M_{\rm TRGB}(r_\nu)
\simeq -0.2997\,\delta r_\nu
      +0.1897\,\delta r_\nu^2\ {\rm mag},
\label{eq:rnu-response}
\end{equation}
with a maximum absolute residual of $2.5\times10^{-3}$~mag.  The response is therefore monotonic but appreciably nonlinear once the standard cooling is reduced by several tens of percent.  For example, the direct MESA shifts are $\Delta M_{\rm TRGB}=+0.1048$~mag at $r_\nu=0.7$ and $-0.02835$~mag at $r_\nu=1.1$.  A symmetric finite difference about the standard point gives a local slope $d\Delta M_{\rm TRGB}/dr_\nu\simeq-0.299$~mag.  This representative calculation is intended as a response check rather than a cluster-independent calibration.
\black

\begin{figure}[tbp]
\centering
\includegraphics[width=0.78\linewidth]
{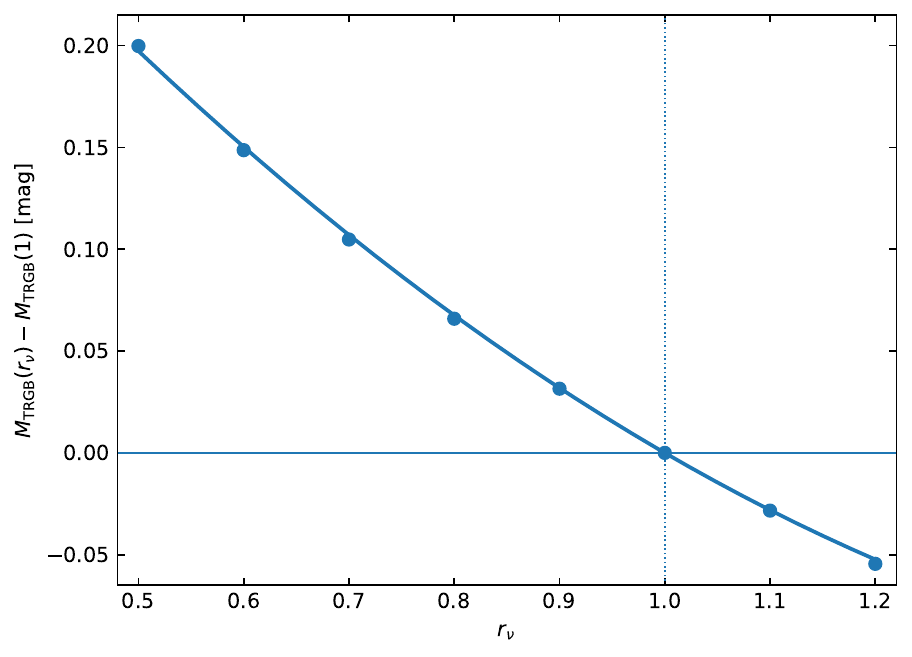}
\caption{\red
TRGB magnitude response of the representative canonical NGC~6093 model to a rescaling $r_\nu$ of the standard thermal-neutrino cooling rate, with $g_{13}=0$.  The points are the eight calculated MESA models for $0.5\le r_\nu\le1.2$, and the solid curve is the constrained quadratic approximation of Eq.~\eqref{eq:rnu-response}.  Reduced neutrino cooling produces a fainter TRGB, while enhanced cooling produces a brighter TRGB; the curvature becomes visible for large cooling deficits.\black}
\label{fig:canonical-rnu-response}
\end{figure}

\subsubsection{Additional energy losses in hot-flasher clusters}
\label{sec:hf-energy-loss}

Additional cooling has a qualitatively different effect on the HF clusters discussed in Sec.~\ref{sec:hot_flasher_clusters}.
It increases the maximum RGB luminosity, which enhances stellar mass loss through the Reimers prescription
\cite{Reimers:1975},
\begin{equation}
\left|\dot{M}_{\rm R}\right|
=
4\times10^{-13}\,\eta_{\rm R}
\left(\frac{L}{L_\odot}\right)
\left(\frac{R}{R_\odot}\right)
\left(\frac{M}{M_\odot}\right)^{-1}
M_\odot\,{\rm yr}^{-1}.
\end{equation}

Because an HF model already possesses an extremely small residual hydrogen-rich envelope, even a modest increase in the integrated RGB mass loss can appreciably weaken the hydrogen-burning shell and cause the star to leave the RGB earlier.
This gives rise to two competing effects. On the one hand,
additional axion cooling tends to delay helium ignition,
allowing the helium core to grow and increasing the stellar
luminosity. On the other hand, the increased luminosity
strengthens the Reimers wind, which removes the residual
hydrogen-rich envelope and limits the subsequent growth of
the helium core. As a result, the maximum RGB luminosity and
the helium-core mass at the TRGB remain nearly constant over
the considered range of axion--electron coupling strengths.

At the same time, the evolutionary sequence changes from the early HF to the late HF regime (see about classification Sec.~\ref{sec:hot_flasher_clusters}) between the different $g_{13}$. Thus, varying the axion--electron coupling can change not only the quantitative TRGB properties but also the HF classification of the model. An example of this behaviour for NGC~6352
is presented in Table~\ref{tab:ngc6352-hf-grid}.

\begin{table}[t]
\centering
\caption{%
TRGB luminosity, helium-core mass at the TRGB, residual
envelope mass at the onset of the helium flash, and HF
classification of the representative NGC~6352 model as a
function of $g_{13}$. Here, $L_{\rm TRGB}$ is the maximum
surface luminosity attained before the model leaves the RGB,
$M_{\rm c,He}^{\rm TRGB}$ is the helium-core mass evaluated
at the same evolutionary point, and
$M_{\rm env}^{\rm He\,onset}$ is the mass of the residual
envelope at helium ignition. EHF and LHF denote early HF and
late HF, respectively.}
\label{tab:ngc6352-hf-grid}
\begin{tabular}{ccccc}
\toprule
$g_{13}$ &
$\log_{10}(L_{\rm TRGB}/L_\odot)$ &
$M_{\rm c,He}^{\rm TRGB}/M_\odot$ &
$M_{\rm env}^{\rm He\,onset}/M_\odot$ &
Classification \\
\midrule
0.0 & 3.4001540 & 0.482797564 & 0.000994957 & EHF \\
0.1 & 3.4001780 & 0.482800335 & 0.000948258 & EHF \\
0.2 & 3.4003534 & 0.482777537 & 0.000889046 & EHF \\
0.3 & 3.4005846 & 0.482759351 & 0.000787450 & EHF \\
0.4 & 3.4009062 & 0.482720312 & 0.000772392 & LHF \\
0.5 & 3.4012761 & 0.482676996 & 0.000766955 & LHF \\
1.0 & 3.4034996 & 0.482319296 & 0.000762037 & LHF \\
\bottomrule
\end{tabular}
\end{table}

\subsection{Estimate of theoretical uncertainties}
\label{sec:theory_uncertainties}
Theoretical uncertainties in the predicted luminosity of the TRGB are important not only for the present analysis, but also, potentially, for other studies, since the TRGB is widely used as a standard candle for old stellar populations. These uncertainties have been extensively discussed in some previous studies. Here we perform an independent updated estimate tailored to our analysis.

Previous studies have shown that the dominant sources of theoretical uncertainty remain broadly similar over the range of stellar parameters relevant to old globular clusters, although the magnitude of the total uncertainty varies somewhat with initial mass and metallicity \citep[e.g.][]{Saltas:2022}. So we adopt NGC~288 as a representative cluster and use it to estimate the theoretical uncertainty budget, unless stated otherwise.

In this section, we consider uncertainties arising from theoretical modeling and, separately, the uncertainty associated with empirical estimates of the RGB mass loss in the clusters. All remaining sources of uncertainty are already included in the observational error budget.

\subsubsection{Thermal neutrinos}
\label{sec:sim:uncert:neutrinos}

The evolution of RGB stars is sensitive to energy losses through both standard neutrino emission and the production of hypothetical weakly interacting particles. One of the clearest manifestations of standard neutrino cooling is the off-center ignition of helium. In low-mass RGB stars, the helium flash occurs at an enclosed mass coordinate of approximately
\(m \simeq 0.15\text{--}0.20\,M_\odot\), rather than at the stellar center \citep{Kippenhahn:2012}. This is a consequence of efficient neutrino cooling in the densest central layers, which produces a central temperature inversion and shifts the temperature maximum away from the center.

Under the thermodynamic conditions characteristic of degenerate helium cores on the RGB, plasmon decay,
\(\gamma^{*}\rightarrow\nu\bar{\nu}\), is the dominant thermal-neutrino emission process. Its contribution becomes increasingly important as the core grows denser and hotter, reaching its largest values near the TRGB, shortly before the helium flash. We therefore include the uncertainty in the plasmon-neutrino emission rate in our theoretical uncertainty budget.

As shown in Table~\ref{tab:baseline_input_physics}, MESA calculates thermal-neutrino energy losses using the analytic fitting formulae of Itoh et al.~\cite{Itoh:1996}. To estimate the uncertainty associated with the treatment of plasmon decay, we compared the MESA plasmon-neutrino emissivity with the numerical tables of Kantor and Gusakov~\cite{Kantor:2007}. These tables were calculated over a wide range of temperatures and densities without imposing limiting assumptions on the degree of degeneracy or relativity of the electron gas.

\begin{figure}[t]
    \centering
    \includegraphics[
        width=\linewidth,
        height=0.42\textheight,
        keepaspectratio
    ]{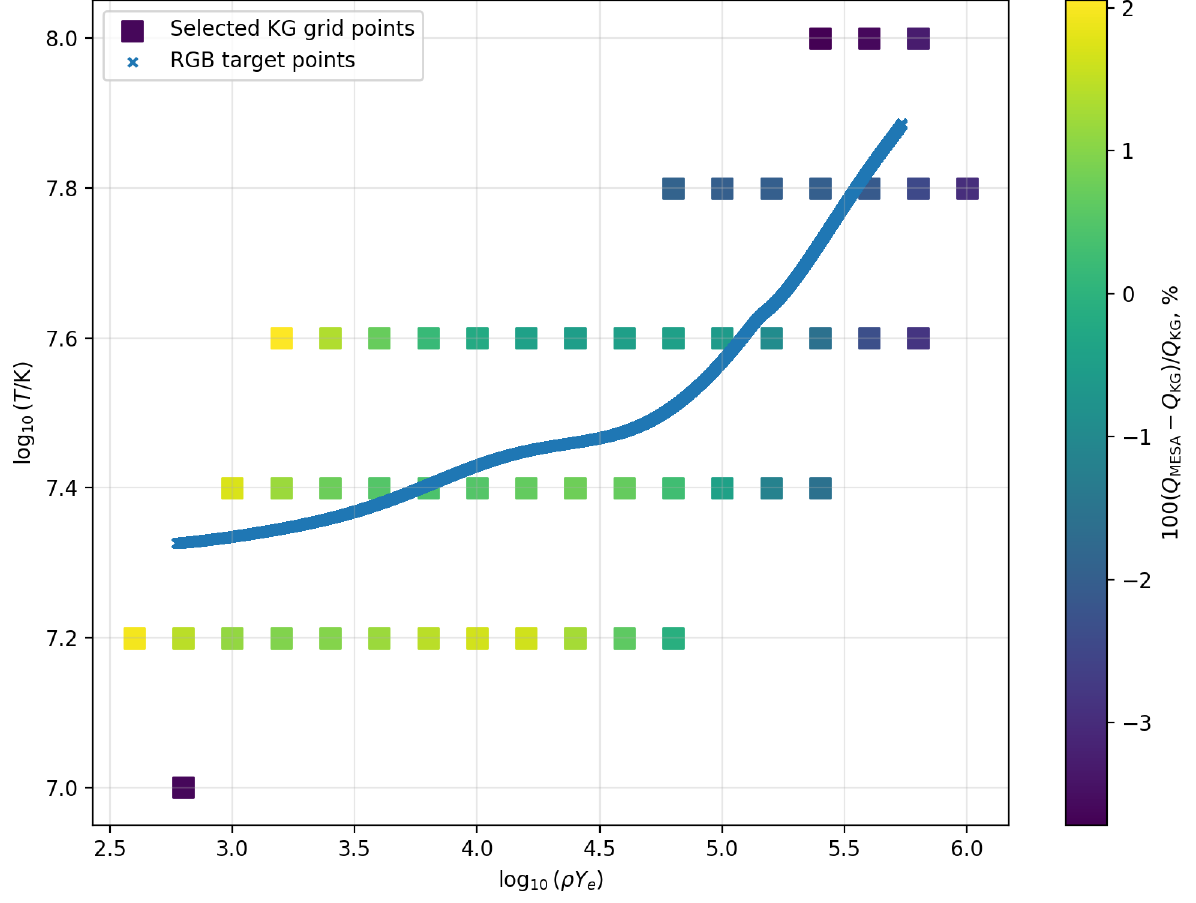}
    \caption{Representative RGB-core evolutionary track and the surrounding points from the tables of Kantor and Gusakov~\cite{Kantor:2007} used to compare their numerical plasmon-neutrino emissivities with the MESA approximation based on the fitting formulae of Itoh et al.~\cite{Itoh:1996}}
    \label{fig:plasmon_neutrino_rgb}
\end{figure}

Fig.~\ref{fig:plasmon_neutrino_rgb} shows a representative RGB-core evolutionary track in the \(\log T\)--\(\log(\rho Y_e)\) plane, which corresponds to the thermodynamic variables used to calculate the plasmon-neutrino emissivity. We selected the tabulated grid points closest to the track, together with the adjacent grid points, and compared the corresponding Kantor--Gusakov emissivities with the Itoh fitting formula used by MESA. We emphasize that, as can be seen in Fig.~\ref{fig:plasmon_neutrino_rgb}, both the largest negative and the largest positive deviations occur at tabulated points located some distance away from the representative RGB-core evolutionary track. The adopted range is therefore broader than the range sampled directly along the track and can be regarded as a conservative estimate of the uncertainty in the plasmon-neutrino energy-loss rate.

Defining the relative difference between the two calculations as
\(\delta_\nu\equiv Q_{\rm MESA}/Q_{\rm KG}-1\),
we find \(-3.72\%\leq\delta_\nu\leq+2.05\%\) over the selected
RGB-core region.

To propagate this difference through the stellar-evolution calculations, we rescaled the standard MESA plasmon-neutrino energy-loss rate according to
\begin{equation}\epsilon_\nu
=
r_\nu\,\epsilon_{\nu,\mathrm{MESA}},
\qquad
r_\nu=\frac{1}{1+\delta_\nu}.
\end{equation}The limiting relative differences correspond to
\(r_\nu=1.0386\simeq1.04\) and \(r_\nu=0.9799\simeq0.98\),
respectively.
For the representative NGC~288 model, these variations produce the following shifts in the predicted TRGB luminosity:
\begin{equation}\Delta\log_{10}L_{\mathrm{TRGB}}
=
\begin{cases}
-0.0018~\mathrm{dex}, & r_\nu=0.98,\\[2mm]
 +0.0037~\mathrm{dex}, & r_\nu=1.04.
\end{cases}
\end{equation}Thus, enhanced neutrino cooling delays helium ignition and produces a brighter TRGB, whereas reduced neutrino cooling has the opposite effect.
 For comparison with the axion sensitivity, at the 95\% power-constrained bound obtained below, $g_{13}\approx 0.64$, the NGC~288 MESA sequence brightens by $\Delta\log_{10}L_{\rm TRGB}\simeq0.0088$~dex relative to $g_{13}=0$; the median brightening of the 24 clusters in the primary sample is $\simeq0.0107$~dex.  These shifts are larger than the $+0.0037$~dex upper thermal-neutrino uncertainty quoted above. The latter is nevertheless propagated as a single common theoretical nuisance in the joint likelihood.

\subsubsection{Electron conduction}
\label{sec:sim:uncert:conduction}

Electron conduction is one of the dominant energy-transport mechanisms in the degenerate helium cores of RGB stars. The conductive opacity determines the thermal stratification of the core and affects the helium-core mass at the onset of helium burning, thereby influencing the predicted TRGB luminosity.

As shown in Table~\ref{tab:baseline_input_physics}, in the setup adopted here, MESA combines the conductive opacities of Cassisi et al.~\cite{Cassisi:2007} with the correction proposed by Blouin et al.~\cite{Blouin:2020} for moderately degenerate electrons. The degree of electron degeneracy can be characterized by
\(\theta\equiv T/T_{\mathrm{F}}\), where \(T_{\mathrm{F}}\) is the
electron Fermi temperature. The Blouin et al. correction is relevant
mainly in the moderately degenerate regime, \(\theta\sim0.1\text{--}1\),
which is encountered primarily in the outer part of the degenerate helium core. Their calculation is based on a quantum Landau--Fokker--Planck approach and provides an improved treatment of electron--electron scattering in this regime.

The Blouin et al.\ prescription is not expected to remain valid in the limit of strong degeneracy, \(\theta\ll1\), where the traditional results recover the appropriate asymptotic behaviour. Ref.~\cite{Cassisi:2021} therefore considered different prescriptions for bridging the moderately and strongly degenerate regimes. These prescriptions smoothly suppress the Blouin et al.\ correction as \(\theta\) decreases, allowing the conductive opacity to approach the strongly degenerate limit.

 The baseline calculation uses the undamped Blouin et al.\ prescription.  To define a one-sided uncertainty relative to this baseline, we also calculate the weak-damping prescription introduced in Ref.~\cite{Cassisi:2021}.  The general damping factor is
\begin{equation}D(\theta)
=
\left(1+a\theta^{-b}\right)^{-1},
\end{equation}with \(a=0.01\) and \(b=2\) for weak damping. 

The original Blouin et al.\ enhancement factor
$
\left(
\kappa_{\mathrm{c}}^{\text{\cite{Blouin:2020}}}
=
\kappa_{\mathrm{c}}^{\text{\cite{Cassisi:2007}}}/
     {F_{\mathrm{B}}},
~~
\lambda_{\mathrm{e}}^{\text{\cite{Blouin:2020}}}
=
F_{\mathrm{B}}
\lambda_{\mathrm{e}}^{\text{\cite{Cassisi:2007}} }
\right)
$
is written as
\begin{equation}F_{\mathrm{B}}
=
1+g(x,y)H[g(x,y)],
\end{equation}where \(\kappa_{\mathrm{c}}\) denotes the conductive opacity and \(\lambda_{\mathrm{e}}\) the electron thermal conductivity. Here
\(x=\log_{10}(\rho/\rho_0)\) and
\(y=\log_{10}(T/T_0)\), with
\((\rho_0,T_0)=(10^{5.45}\ {\rm g\,cm^{-3}},10^{8.40}\ {\rm K})\)
for hydrogen and
\((\rho_0,T_0)=(10^{6.50}\ {\rm g\,cm^{-3}},10^{8.57}\ {\rm K})\)
for helium. For the explicit expressions of $g(x,y)$ and $H(x,y)$, and for the corresponding coefficients, see Ref.~\cite{Blouin:2020}. 

The damping factor modifies this expression,
\begin{equation}F
=
1+
D(\theta)\,
g(x,y)H[g(x,y)].
\end{equation}This prescription preserves the correction \cite{Blouin:2020} in the moderately degenerate regime while gradually restoring the traditional strongly degenerate limit as \(\theta\) decreases.

To obtain a conservative estimate of this uncertainty, we use NGC~1261 rather than NGC~288, which serves as the representative cluster for most of the other uncertainty estimates. The effect of the conductive-opacity correction is expected to be largest in a stellar model with the least degenerate helium core. Within the parameter range covered by our cluster sample, weaker degeneracy is favoured by a combination of a higher stellar mass and a relatively high metallicity. Based on these two properties, we selected NGC~1261 as the cluster expected to be most sensitive to the treatment of electron conduction.

For the NGC~1261 model, replacing the  baseline undamped   Ref.~\cite{Blouin:2020} prescription with the weak-damping prescription results in
\begin{equation}\Delta\log_{10}L_{\mathrm{TRGB}}
\equiv
\log_{10}L_{\mathrm{TRGB}}^{\mathrm{weak}}
-
\log_{10}L_{\mathrm{TRGB}}^{\mathrm{B20}}
=
-0.013~\mathrm{dex}.
\end{equation}We adopt this shift as a conservative estimate of the uncertainty associated with the treatment of electron conduction in the transition between moderate and strong degeneracy.
This uncertainty is consistent with that reported by Cassisi et al.~\cite{Cassisi:2021}, who found a shift of $-0.012$ dex for a $0.8\,M_\odot$ metal-poor stellar model.

\subsubsection{Radiative opacity}
This uncertainty is perhaps the most ambiguous one. Several studies have shown that uncertainties in radiative opacity are among the most important sources of theoretical uncertainty \cite{Valle:2013b,Saltas:2022}. In other studies, however, the uncertainties associated with radiative opacity were found to be negligible \cite{Serenelli:2017,Straniero2020}.

Radiative opacity is expected to affect the TRGB luminosity primarily within a thin layer between the hydrogen-burning shell and the helium core, where electron conduction does not yet dominate the energy transport. The low-temperature opacity of the outer envelope affects only the effective temperature, $T_{\mathrm{eff}}$, as discussed in Ref.~\cite{Salaris:1993}.

In most studies, this uncertainty is estimated by comparing different opacity tables. For example, Ref.~\cite{Saltas:2022} found that the OP and OPAL opacities differ by no more than approximately $5\%$ under the conditions characteristic of this thin layer.

In the present work, we estimate this uncertainty by computing models with three different sets of high-temperature opacity tables: OP \cite{Badnell:2005}, OPAL \cite{Iglesias:1996}, and OPLIB \cite{Colgan:2016}. Our baseline models use $\alpha$-enhanced OPAL tables. However, because $\alpha$-enhanced OPLIB tables were not available to us, the comparison between the three opacity datasets was performed using non-$\alpha$-enhanced tables. The low-temperature opacity tables were kept unchanged in all three calculations.

We use the non-$\alpha$-enhanced OPAL calculation as the reference and define
\begin{equation}\Delta_{\kappa}\log_{10}L_{\mathrm{TRGB}}
\equiv
\log_{10}L_{\mathrm{TRGB}}^{\kappa}
-
\log_{10}L_{\mathrm{TRGB}}^{\mathrm{OPAL}}.
\end{equation}The resulting differences are
\begin{equation}\Delta_{\kappa}\log_{10}L_{\mathrm{TRGB}}
=
\begin{cases}
-0.00805~\mathrm{dex}, & \mathrm{OP},\\
+0.007288~\mathrm{dex}, & \mathrm{OPLIB},\\
0, & \mathrm{OPAL}.
\end{cases}
\end{equation}Thus, relative to the non-$\alpha$-enhanced OPAL calculation, the OP tables produce a lower TRGB luminosity, whereas the OPLIB tables produce a higher TRGB luminosity. Our final result is close to the result of Ref.~\cite{Saltas:2022}.

\subsubsection{Nuclear reactions}
\medskip
\noindent
\textbf{$^{14}\mathrm{N}(p,\gamma)^{15}\mathrm{O}$ reaction.}
\par\smallskip
The \(^{14}\mathrm{N}(p,\gamma)^{15}\mathrm{O}\) reaction is the
bottleneck of the CNO cycle of hydrogen burning. In low-mass
globular-cluster stars at evolutionary stages close to the TRGB,
hydrogen burning takes place in a thin shell surrounding the helium
core. Under these conditions, the CNO cycle provides the dominant
contribution to nuclear energy generation in the shell.

At the energies relevant to stellar hydrogen burning, the uncertainty
in this reaction rate is commonly characterized in terms of the
zero-energy astrophysical $S$-factor, $S_{1\,14}(0)$. The rate used in
the baseline MESA calculations is based on the LUNA result,
\(S_{1\,14}(0)=1.61\pm0.08\ \mathrm{keV\,b}\)
\cite{Imbriani2005}. To estimate the corresponding nuclear-physics
uncertainty, we adopted the Solar Fusion III recommendation,
\(S_{1\,14}(0)=1.68\pm0.14\ \mathrm{keV\,b}\)
\citep{Acharya2025}.

To estimate the resulting shift and uncertainty, we rescaled the
MESA ${}^{14}\mathrm{N}(p,\gamma){}^{15}\mathrm{O}$ reaction rate
by the factors $r=1.13043,\ 1.0435,\ 0.9565$, corresponding to the
upper, central, and lower values of the updated $S$-factor, respectively. The resulting shift in the TRGB luminosity is  (rounded to three decimal places):
\begin{equation}
    \Delta \log_{10}
    \left(
        \frac{L_{\mathrm{TRGB}}}{L_{\odot}}
    \right)
    =
    0.001^{+0.002}_{-0.002}\ \mathrm{dex}.
\end{equation}

\noindent\textbf{Triple-\(\alpha\) reaction.} 

Although helium burning makes a negligible contribution to the stellar luminosity on the red-giant branch, uncertainties in the triple-\(\alpha\) reaction rate are important for the predicted TRGB luminosity: a lower triple-\(\alpha\) reaction rate delays the helium flash. Consequently, hydrogen-shell burning continues for longer, the mass of the helium core increases, and the total stellar luminosity at the tip of the red-giant branch becomes higher.

For the present analysis, this reaction is of greatest interest at temperatures close to helium ignition, \(T\simeq10^{8}\,\mathrm{K}\). At these temperatures, the reaction proceeds through the narrow Hoyle-state resonance in \(^{12}\mathrm{C}\). The resonant contribution to the reaction rate is
\begin{equation}
    r_{3\alpha}^{(\mathrm{H})}
    \propto
    T^{-3}
    \exp\left(-\frac{E_{\mathrm{H}}}{k_{\mathrm{B}}T}\right)
    \frac{\Gamma_{\alpha}\Gamma_{\mathrm{rad}}}{\Gamma},
    \qquad
    \frac{\Gamma_{\alpha}}{\Gamma}\simeq 1,
\end{equation}
where \(E_{\mathrm{H}}\) is the resonance energy above the three-\(\alpha\) threshold, \(\Gamma_{\alpha}\) and \(\Gamma_{\mathrm{rad}}\) are the \(\alpha\)-decay and radiative widths, respectively, and \(\Gamma=\Gamma_{\alpha}+\Gamma_{\mathrm{rad}}\) is the total width. Thus, at fixed \(E_{\mathrm{H}}\), the rate is approximately proportional to \(\Gamma_{\mathrm{rad}}\).

The greatest uncertainty in the reaction rate comes from the radiative width of this resonance. It cannot be measured directly and is instead reconstructed from three independently measured quantities,
\begin{equation}
    \Gamma_{\mathrm{rad}}
    =
    \left(\frac{\Gamma_{\mathrm{rad}}}{\Gamma}\right)
    \left(\frac{\Gamma}{\Gamma_{\pi}(E0)}\right)
    \Gamma_{\pi}(E0)
    =
    \frac{\Gamma_{\mathrm{rad}}/\Gamma}
         {\Gamma_{\pi}(E0)/\Gamma}
    \Gamma_{\pi}(E0).
    \label{eq:gamma_rad_reconstruction_close}
\end{equation}

For the pair-decay quantities, we adopted
\begin{align}
    \frac{\Gamma_{\pi}(E0)}{\Gamma}
    &=
    (7.6\pm0.4)\times10^{-6},
    &
    \frac{\sigma_{\Gamma_{\pi}/\Gamma}}
         {\Gamma_{\pi}/\Gamma}
    &=5.3\%
    \quad\text{\cite{Eriksen:2020}},
    \\
    \Gamma_{\pi}(E0)
    &=
    (62.3\pm2.0)\,\mu\mathrm{eV},
    &
    \frac{\sigma_{\Gamma_{\pi}}}{\Gamma_{\pi}}
    &=3.2\%
    \quad\text{\cite{Chernykh:2010}}.
\end{align}

The measurements used to determine \(\Gamma_{\mathrm{rad}}/\Gamma\) are summarised in Table~\ref{tab:hoyle_branching_close}. 
\begin{table*}
    \centering
    \footnotesize
    \caption{Measurements of the radiative branching ratio of the Hoyle state (in units of $10^{-4}$). The original result of Kib\'edi et al. is shown for completeness but was not included in the averaging procedure because it has been superseded by the corrected analysis of Paulsen et al.}
    \label{tab:hoyle_branching_close}
    \begin{tabularx}{\textwidth}{
        @{}
        >{\raggedright\arraybackslash}X
        c
        >{\raggedright\arraybackslash}X
        c
        @{}
    }
        \hline
        Earlier data &
        \(\Gamma_{\mathrm{rad}}/\Gamma\) &
        Recent data &
        \(\Gamma_{\mathrm{rad}}/\Gamma\) \\
        \hline
        Alburger (1961) \cite{Alburger:1961}
            & \(3.3\pm0.9\)
            & Tsumura et al. (2021) \cite{Tsumura:2021}
            & \(4.3\pm0.8\) \\
        Hall and Tanner (1964) \cite{HallTanner:1964}
            & \(3.3\pm1.2\)
            & Luo et al. (2024) \cite{Luo:2024}
            & \(4.0\pm0.3_{\mathrm{stat}}\pm0.16_{\mathrm{syst}}\) \\
        Chamberlin et al. (1974) \cite{Chamberlin:1974}
            & \(4.20\pm0.22\)
            & Dell'Aquila et al. (2024) \cite{DellAquila:2024}
            & \(4.2\pm0.6\) \\
        Davids et al. (1975) \cite{Davids:1975}
            & \(4.14\pm0.21\)
            & Rana et al. (2024) \cite{Rana:2024}
            & \(4.04\pm0.30\) \\
        Mak et al. (1975) \cite{Mak:1975}
            & \(4.15\pm0.34\)
            & Paulsen et al. (2025) \cite{Paulsen:2025}
            & \(4.1\pm0.4\) \\
        Obst and Braithwaite (1976) \cite{ObstBraithwaite:1976}
            & \(4.09\pm0.27\)
            & Sakanashi et al. (2025) \cite{Sakanashi:2025}
            & \(4.07\pm0.28\) \\
        Markham et al. (1976) \cite{Markham:1976}
            & \(3.87\pm0.25\)
            & Kib\'edi et al. (2020) \cite{Kibedi:2020}
            & \(6.2\pm0.6\) (superseded) \\
        Seeger and Kavanagh (1963) \cite{SeegerKavanagh:1963}
            & \(2.82\pm0.29\)
            & & \\
        \hline
    \end{tabularx}
\end{table*}

To determine \(\Gamma_{\mathrm{rad}}/\Gamma\), we followed the official recommendations of the Evaluated Nuclear Structure Data File (ENSDF) \cite{ENSDFGuidelines:2021}. We excluded the original value reported by Kib\'edi et al. \cite{Kibedi:2020}, because the reanalysis by Paulsen et al. \cite{Paulsen:2025} identified several corrections to the original analysis. 

For a conservative treatment of discrepant data, we used the limitation of relative statistical weight (LRSW, also known as LWM), normalised-residual (NR), and Rajeval-technique (RT) methods recommended in the ENSDF guidelines and implemented in V.AveLib \cite{RajputMacMahon:1992,James:1992,VAveLib:2015}. The LRSW method gave the largest uncertainty and was therefore adopted as conservative,
\begin{equation}
    \frac{\Gamma_{\mathrm{rad}}}{\Gamma}
    =
    (4.01\pm0.19)\times10^{-4},
    \qquad
    \frac{\sigma_{\Gamma_{\mathrm{rad}}/\Gamma}}
         {\Gamma_{\mathrm{rad}}/\Gamma}
    =
    4.7\%.
\end{equation}

Assuming that the three inputs in Eq.~\eqref{eq:gamma_rad_reconstruction_close} are independent, their uncertainties sum up in quadrature to
\(\sigma_{\Gamma_{\mathrm{rad}}}/\Gamma_{\mathrm{rad}}=7.8\%\), and hence
\(\Gamma_{\mathrm{rad}}=(3.287\pm0.256)\,\mathrm{meV}\).

The baseline triple-\(\alpha\) rate in our MESA calculations is based on Fynbo et al. \cite{Fynbo:2005} and uses the reference Hoyle-state radiative width \(\Gamma_{\mathrm{rad,ref}}=3.675\,\mathrm{meV}\) adopted in the NACRE-based normalisation \cite{Angulo:1999}.
To estimate the uncertainty, we introduced multiplicative factors for the MESA triple-\(\alpha\) rate corresponding to the updated central value and its uncertainty:
\begin{equation}
    f_{3\alpha}^{+}=0.9575,
    \qquad
    f_{3\alpha}^{\mathrm{central}}=0.8884,
    \qquad
    f_{3\alpha}^{-}=0.8193.
\end{equation}

Adopting the updated rate instead of the baseline value, and propagating its uncertainty, gives (rounded to three decimal places):
\begin{equation}
    \Delta\log_{10}\left(
        \frac{L_{\mathrm{TRGB}}}{L_{\odot}}
    \right)
    =
    +0.003^{+0.002}_{-0.002}\,\mathrm{dex}.
\end{equation}

\subsubsection{Convection}
\label{sec:theory_convection}
To estimate the uncertainty associated with the treatment of convection, we computed a grid of models with different values of the convective parameters. The adopted parameter grid is summarised in Table~\ref{tab:convection_grid}.
\begin{table}[t]
\centering
\caption{Convective parameters varied in the model grid.}
\label{tab:convection_grid}
\begin{tabular}{lc}
\hline
Parameter & Adopted values \\
\hline
$\alpha_{\mathrm{MLT}}$ & $1.60,\ 1.82,\ 2.00$ \\
$\alpha_{\mathrm{sc}}$  & $0,\ 0.1$ \\
$f_{\mathrm{ov}}$       & $0,\ 0.016,\ 0.125$ \\
\hline
\end{tabular}
\end{table}

Convective energy transport is described using the mixing-length formalism of Cox~\cite{Cox:1968}. We recall that
$\alpha_{\mathrm{MLT}}=\ell_{\mathrm{MLT}}/H_P$, where
$\ell_{\mathrm{MLT}}$ is the mixing length and $H_P$ is the pressure scale height. The parameter $\alpha_{\mathrm{sc}}$ is a dimensionless coefficient that controls the efficiency of semiconvective mixing treated as a diffusive process; MESA implements the prescription of Langer et al.~\cite{Langer:1983}. Finally, $f_{\mathrm{ov}}$ is the dimensionless overshooting parameter in the exponential prescription and determines the exponential decay length of the overshooting diffusion coefficient rather than a sharply defined geometrical extent of the overshooting region. Detailed descriptions of these implementations and a discussion of possible values of the convective parameters are given by Paxton et al.~\cite{Paxton:2013}. We note that commonly adopted values are $\alpha_{\mathrm{sc}}=0.01$ and $f_{\mathrm{ov}}=0.016$, although both parameters depend on the stellar regime and model calibration. The upper value $f_{\mathrm{ov}}=0.125$ was deliberately chosen as an extreme limiting case.

We find that increasing the semiconvection parameter has a negligible effect on the TRGB luminosity. For example, increasing $\alpha_{\mathrm{sc}}$ from $0.01$ to $0.1$ increases the predicted luminosity by only
\(\Delta\log_{10}L_{\mathrm{TRGB}}\simeq2\times10^{-6}\ {\rm dex}\).

To estimate the corresponding uncertainty, we selected two limiting combinations of the convective parameters:
\begin{equation}\begin{aligned}
\mathcal{C}_{1}:&
\quad
(\alpha_{\mathrm{MLT}},\alpha_{\mathrm{sc}},f_{\mathrm{ov}})
=
(1.60,0,0),
\\
\mathcal{C}_{2}:&
\quad
(\alpha_{\mathrm{MLT}},\alpha_{\mathrm{sc}},f_{\mathrm{ov}})
=
(2.00,0.1,0.125).
\end{aligned}
\end{equation}We used the luminosities predicted by these limiting models to define the lower and upper bounds of a uniform distribution in the Monte Carlo simulations. The resulting uncertainty interval is
\begin{equation}\Delta_{\mathrm{conv}}
\log_{10}L_{\mathrm{TRGB}}
=
{}^{+0.006}_{-0.011}\ {\rm dex}.
\end{equation}\subsubsection{Other uncertainties} 
\label{sec:theory_other_uncertainties}
We finally consider the remaining sources of theoretical uncertainty. 

\noindent\textbf{Stellar wind.} For clusters included in the analysis of Tailo et al.~\cite{Tailo:2020}, we adopt their cluster-specific constraints on RGB mass loss. For clusters without an available measurement, we use the Reimers prescription \cite{Reimers:1975} with a reference scaling factor $\eta_{\mathrm{R}}=0.3$. For most canonical clusters, the resulting uncertainty in the predicted TRGB luminosity is small. We therefore use the uncertainty obtained for NGC~288 as representative of this group. The dependence on the adopted wind prescription becomes substantially stronger for models undergoing hot-flasher evolution. For these clusters, the mass-loss uncertainty can become one of the dominant contributions to the total error budget. We therefore propagate the cluster-specific mass-loss uncertainty separately for every hot-flasher cluster. The result for NGC~6366 is shown in Table~\ref{tab:theory_uncertainties} as a representative example.

\begin{table}
\centering
\caption{Uncertainty components included in the Monte Carlo
propagation. For each component,
$\delta\equiv
\Delta\log_{10}(L_{\rm TRGB}/L_{\odot})$. Here
$d_M\equiv|\Delta M_{\rm init}|/(0.01\,M_{\odot})$
is the initial-mass difference expressed in units of the adopted
$0.01\,M_{\odot}$ mass step. All numerical values are given in dex.  In the joint likelihood, the stellar-physics components are shared among clusters, whereas numerical resolution, RGB wind, and the residual initial-mass contributions are treated as cluster-specific.}
\label{tab:theory_uncertainties}

\small
\renewcommand{\arraystretch}{1.12}
\setlength{\tabcolsep}{3.5pt}

\begin{tabularx}{\textwidth}{
    @{}
    >{\raggedright\arraybackslash}X
    >{\raggedright\arraybackslash}p{2.30cm}
    >{\centering\arraybackslash}p{1.50cm}
    >{\centering\arraybackslash}p{1.30cm}
    >{\centering\arraybackslash}p{1.30cm}
    @{}
}
\toprule
Source
& Distribution
& $\delta_{0}$
& $\delta_{-}$
& $\delta_{+}$ \\
\midrule

\multicolumn{5}{@{}l}{\textit{Stellar-physics and numerical components}}\\
Numerical resolution
& Uniform
& $0$
& $0.0010$
& $0.0010$ \\

Convection
& Uniform
& $0$
& $0.011$
& $0.0056$ \\

Electron conduction
& Uniform
& $0$
& $0.013$
& $0$ \\

Radiative-opacity tables
& Uniform
& $0$
& $0.0080$
& $0.0073$ \\

${}^{14}\mathrm{N}(p,\gamma){}^{15}\mathrm{O}$ rate
& Shifted split normal
& $+0.0011$
& $0.0023$
& $0.0021$ \\

Triple-$\alpha$ rate
& Shifted split normal
& $+0.0033$
& $0.0022$
& $0.0022$ \\

Atmospheric boundary condition
& Uniform
& $0$
& $0$
& $0.0002$ \\

Equation of state
& Uniform
& $0$
& $0$
& $0.0028$ \\

Electron screening:
\texttt{extended} versus \texttt{chugunov}
& Uniform
& $0$
& $0.0114$
& $0$ \\

Thermal-neutrino rates
& Uniform
& $0$
& $0.0018$
& $0.0037$ \\

\midrule
\multicolumn{5}{@{}l}{\textit{Cluster-dependent RGB-wind components}}\\

RGB wind: NGC~288
(representative canonical cluster)
& Split normal
& $0$
& $0.0023$
& $0.0015$ \\

RGB wind: NGC~6366
& Split normal
& $0$
& $0.0304$
& $0.0092$ \\

RGB wind: NGC~6752
& Split normal
& $0$
& $0.0166$
& $0.0035$ \\

RGB wind: NGC~6352
& Split normal
& $0$
& $0.0309$
& $0.0111$ \\

\midrule
\multicolumn{5}{@{}l}{\textit{Initial stellar-mass components}}\\

$d_M<1$
& Neglected
& $0$
& $0$
& $0$ \\

$1\leq d_M\leq2$
(NGC~5024 representative case)
& Uniform
& $0$
& $0$
& $0.0008$ \\

$2<d_M\leq3$
(NGC~6809 representative case)
& Uniform
& $0$
& $0$
& $0.0017$ \\

$d_M>3$
(NGC~6341 and NGC~6541)
& Uniform
& $0$
& $0$
& $0.0026$ \\

Hot-flasher clusters
(NGC~6352 representative mass test)
& Uniform
& $0$
& $0.0050$
& $0.0050$ \\

\bottomrule
\end{tabularx}
\end{table}

\medskip \noindent\textbf{Electron screening.} To obtain a conservative estimate of the uncertainty associated with plasma screening, we compare two alternative MESA implementations that cover both weak- and strong-screening regimes. Our reference calculation uses the \texttt{chugunov} prescription, based on Chugunov et al.~\cite{Chugunov:2007}. The alternative \texttt{extended} prescription combines the intermediate-screening treatment of Graboske et al.~\cite{Graboske:1973} with the strong-screening results of Alastuey and Jancovici~\cite{Alastuey:1978} and the plasma parameters of Itoh et al.~\cite{Itoh:1979}. The latter prescription was the default in MESA releases up to revision 11435. Relative to the \texttt{chugunov} calculation, the \texttt{extended} prescription changes the predicted TRGB luminosity by \(\Delta_{\mathrm{screen}}\log_{10}L_{\mathrm{TRGB}}=-0.0114\ {\rm dex}\). We do not use the MESA \texttt{salpeter} option \cite{Salpeter:1954} as an alternative endpoint because it is restricted to the weak-screening approximation and therefore does not provide a like-for-like comparison across all relevant screening regimes.

\medskip 
\noindent \textbf{Numerical uncertainty.} We assessed the numerical uncertainty by performing convergence tests
with progressively tighter resolution controls. Further improvements in
the numerical resolution changed the predicted TRGB luminosity by less
than $0.001\ \mathrm{dex}$. Moreover, throughout the stellar evolution
calculations, we imposed the hard limit
$
\left|
\Delta\log_{10}\left(\frac{L}{L_{\odot}}\right)
\right|
\leq 0.001
$
although other convergence settings in MESA (e.g. energy error limits) result in much higher luminosity resolution than 0.001 dex. We therefore conservatively assign a numerical
uncertainty of $0.001\ \mathrm{dex}$ to the predicted TRGB luminosity
(Table~\ref{tab:theory_uncertainties}).

\medskip
\noindent
\textbf{Initial stellar-mass uncertainty.} The initial stellar mass assigned to each globular cluster was inferred
from its age using the $\alpha$-enhanced MIST evolutionary models
\cite{Dotter2026}. For most clusters, the corresponding initial-mass
interval was no wider than $0.01\,M_{\odot}$. We deliberately adopted
a conservative conversion between the age and mass intervals: over
the relevant region of the MIST grid, a difference of
$0.01\,M_{\odot}$ corresponds approximately to
$4\times10^{8}\ {\rm yr}$.

For the compositions and ages considered here, the initial masses
inferred from MIST were systematically higher than the values required
by our baseline MESA calculations. We therefore treated the resulting
TRGB-luminosity uncertainty as one-sided. Using representative clusters,
we assigned the following uniformly distributed luminosity shifts:
\begin{equation}
\delta_{\rm mass}
\equiv
\Delta\log_{10}
\left(
    \frac{L_{\mathrm{TRGB}}}{L_{\odot}}
\right)
\in
\begin{cases}
\{0\},
& \Delta M_{\rm ini}<0.01\,M_{\odot}, \\[2mm]
[0,\,+0.0008]\ {\rm dex},
& 0.01\leq\Delta M_{\rm ini}/M_{\odot}<0.02,
\quad \text{NGC~5024}, \\[2mm]
[0,\,+0.0017]\ {\rm dex},
& 0.02\leq\Delta M_{\rm ini}/M_{\odot}<0.03,
\quad \text{NGC~6809}, \\[2mm]
[0,\,+0.0026]\ {\rm dex},
& 0.03\leq\Delta M_{\rm ini}/M_{\odot}\leq0.033,
\quad \text{NGC~6541}.
\end{cases}
\end{equation}

The hot-flasher clusters were treated separately because their TRGB
luminosities are considerably more sensitive to the adopted initial
mass. For each of these clusters, we verified that the appropriate
mass was bracketed by models spanning no more than
$0.01\,M_{\odot}$. For the representative cluster NGC~6352, two
models separated by $0.01\,M_{\odot}$ differed in TRGB luminosity
by approximately $0.01\ {\rm dex}$. Interpreting this difference as
the full width of the allowed luminosity interval, we adopt
\begin{equation}
-0.005
\leq
\Delta \log_{10}L_{\rm mass}^{\rm HF}
\leq
+0.005\ {\rm dex},
\end{equation}
with a uniform distribution within this interval
(Table~\ref{tab:theory_uncertainties}).

\medskip
\noindent
\textbf{Equation of state.}
MESA uses the following hierarchy of equation-of-state
sources\footnote{\url{https://docs.mesastar.org/en/latest/eos/overview.html}}:
\begin{equation}
    \texttt{Skye}
    >
    \texttt{PC}
    >
    \texttt{FreeEOS}
    >
    \texttt{OPAL/SCVH}
    >
    \texttt{HELM},
\end{equation}
where the individual sources are described in
Refs.~\cite{Jermyn:2021,Potekhin:2010,Irwin:2004,Rogers:2002,Saumon:1995,Timmes2000}.

In the default MESA configuration, the trajectory of the central
thermodynamic conditions of our RGB model lies in the regions covered
by \texttt{FreeEOS} and \texttt{Skye}. A smooth blend is applied between
the two EOS sources.

We performed two tests:
\begin{enumerate}
    \item We disabled both \texttt{Skye} and \texttt{PC}, thereby leaving
    \texttt{FreeEOS} as the only EOS source contributing to our model.
    No blending with \texttt{OPAL/SCVH} occurred anywhere in the model.
    In this calculation, the TRGB luminosity was higher than in the
    baseline model by $0.0027\ {\rm dex}$.

    \item We disabled only \texttt{Skye}. In this case, the TRGB
    luminosity increased by $0.0028\ {\rm dex}$, and blending occurred
    between \texttt{PC} and \texttt{FreeEOS}. As follows from the
    difference between the two tests, the effect associated with this
    blend is smaller than $0.0001\ {\rm dex}$.
\end{enumerate}

For the final EOS uncertainty, we adopt the largest of these shifts,
corresponding to the most conservative case
(Table~\ref{tab:theory_uncertainties}).

\medskip
\noindent
\textbf{Atmospheres and diffusion.}
To estimate the uncertainty associated with the atmospheric boundary
condition, we replaced the grey Eddington atmosphere with a tabulated
atmosphere. The resulting change in the TRGB luminosity is reported in
Table~\ref{tab:theory_uncertainties}. The treatment of the atmosphere is a subdominant contribution to the TRGB-luminosity uncertainty.

The impact of uncertainties in microscopic diffusion was studied in Ref.~\cite{Saltas:2022}. By varying the diffusion velocities by $\pm15\%$, they obtained a maximum change in the predicted TRGB luminosity of approximately $3$--$5\times10^{-4}\ {\rm dex}$. They consequently found that microscopic diffusion is not a major source of uncertainty in the TRGB luminosity. Given the small magnitude of this effect, we do not include a separate diffusion-velocity contribution in our uncertainty budget.


\subsubsection{Monte Carlo propagation of theoretical uncertainties}
\label{sec:theory_monte_carlo}

Table~\ref{tab:theory_uncertainties} summarises the individual
TRGB-luminosity uncertainty components and the probability
distributions adopted for their propagation. Uniform distributions
were used when the quoted limits represent the range spanned by
alternative physical prescriptions or model choices. Split-normal
distributions were used when the lower and upper values represent
asymmetric statistical uncertainties. For the two nuclear-reaction
rates, the distributions were shifted because the recommended
central rates differ from those used in the baseline MESA
calculation.

 To determine the one-cluster marginal distributions reported below, for each cluster category we generated
$N_{\rm MC}=10^{7}$ Monte Carlo realisations.  These one-dimensional marginals are unchanged by the subsequent treatment of cross-cluster correlations. 
In all simulations, we found a characteristic downward shift in
the predicted theoretical TRGB luminosity. When expressed relative
to their shifted median, the resulting distributions were
approximately Gaussian.

The asymmetric and one-sided input distributions were used without
recentering. Consequently, the distribution
has a non-zero median, which represents a correction to the
baseline theoretical prediction rather than an additional
symmetric uncertainty.

For canonical clusters, the common uncertainty components were
combined with the representative NGC~288 wind distribution and the
initial-mass contribution corresponding to the appropriate value
of $d_M$. The mass contribution is one-sided because the
MIST-based initial masses were systematically higher than the
masses adopted in the baseline MESA calculations. Hot-flasher
clusters were treated individually: for each cluster, we used its
cluster-specific RGB-wind distribution together with a uniform
initial-mass contribution (see more details in the relevant sections).

 For the joint likelihood, we further distinguish physical inputs that are common to all stellar calculations from cluster-specific contributions.  Convection, electron conduction, radiative opacity, the ${}^{14}{\rm N}(p,\gamma){}^{15}{\rm O}$ and triple-$\alpha$ rates, the atmospheric boundary prescription, the equation of state, electron screening, and thermal-neutrino rates are represented by common nuisance draws: in each joint realization the same draw of a given physical input is applied to all clusters.  Numerical-resolution uncertainty, RGB-wind uncertainty, and the residual initial-mass contribution are kept cluster-specific.  Synchronizing the common draws leaves the one-cluster distributions in Table~\ref{tab:mc_theory_results} unchanged, but prevents common stellar-physics uncertainties from averaging down as independent errors.  Since the median corrections in Table~\ref{tab:mc_theory_results} are already included in the central theoretical predictions, the common theoretical nuisance used in the likelihood is centered on its own median.

Let $Q_p$ denote the $p$th percentile of the  distribution of 
$ \Delta_{\rm MC}= \log_{10} ( L_{\rm TRGB}^{\rm MC}/L_{\odot} ) - \log_{10} ( L_{\rm TRGB}^{\rm base}/L_{\odot} )$. We report the median shift and its
central 68\% interval as
\begin{equation}
    \Delta_{\rm MC}
    =
    Q_{50}\phantom{.}_{-\left(Q_{50}-Q_{16}\right)}
             ^{+\left(Q_{84}-Q_{50}\right)},
\end{equation}
with the lower and upper theoretical uncertainties determined by
$Q_{50}-Q_{16}$ and $Q_{84}-Q_{50}$, respectively. The resulting
corrections and uncertainties for the different cluster categories
are presented in Table~\ref{tab:mc_theory_results}.
\begin{table}[t]
\centering
\caption{Median Monte-Carlo shifts in the predicted TRGB luminosity
and their central 68\% intervals. 
$d_M\equiv|\Delta M_{\rm ini}|/(0.01\,M_{\odot})$.
All values are given in dex and rounded to three decimal places.}
\label{tab:mc_theory_results}

\small
\renewcommand{\arraystretch}{1.15}
\setlength{\tabcolsep}{6pt}

\begin{tabularx}{0.86\textwidth}{
    @{}
    >{\raggedright\arraybackslash}X
    >{\centering\arraybackslash}p{5.0cm}
    @{}
}
\toprule
Cluster class or mass category
& $\Delta_{\rm MC}$ (dex) \\
\midrule

\multicolumn{2}{@{}l}{\textit{Canonical clusters}}\\

$d_M \le 3$
& $-0.009_{-0.009}^{+0.009}$ \\

%

$d_M>3$
& $-0.008_{-0.009}^{+0.009}$ \\

\midrule
\multicolumn{2}{@{}l}{\textit{Hot-flasher clusters}}\\

NGC~6366
& $-0.023_{-0.026}^{+0.020}$ \\

NGC~6752
& $-0.018_{-0.015}^{+0.013}$ \\

NGC~6352
& $-0.022_{-0.027}^{+0.021}$ \\

\bottomrule
\end{tabularx}
\end{table}

\section{Comparison of theoretical and observed TRGB luminosities}
\label{sec:comparison}
Before turning to constraints on additional energy losses, we compare the observed TRGB magnitudes with the standard-cooling MESA predictions and examine the main trends in the data.

\red
For the adopted $[\alpha/{\rm Fe}]=+0.4$, we convert the iron abundance to global metallicity using \cite{Salaris:1993}
\begin{equation}
[\mathrm M/\mathrm H]=[\mathrm{Fe}/\mathrm H]+\log_{10}\!\left(0.638\,10^{[\alpha/{\rm Fe}]}+0.362\right)
=[\mathrm{Fe}/\mathrm H]+0.2933 .
\label{eq:mh-conversion}
\end{equation}
Table~\ref{tab:trgb-master} collects the corresponding metallicities, the central standard-cooling predictions, and the primary variable-inclusive observational estimates.  The theoretical central values include the median corrections from the theoretical Monte-Carlo uncertainty distributions discussed in Sec.~\ref{sec:theory_uncertainties}, together with the cluster-specific central helium adjustment where applicable.

\red\begin{table*}[t]
\centering
\small\red
\setlength{\tabcolsep}{7pt}
\caption{Summary of the direct theory--observation comparison for all 27 clusters.  The theoretical column gives the central $g_{13}=0$ standard-cooling MESA prediction after the median theoretical correction; its asymmetric uncertainties are the 68\% one-cluster theoretical intervals.  The last column gives the variable-inclusive observational TRGB estimator and its asymmetric 68\% interval.  Shared systematic components are correlated between clusters and are not treated as independent errors in the joint likelihood.  The three hot-flasher clusters, marked by $\dagger$, are excluded from the primary 24-cluster particle-physics analysis.}
\label{tab:trgb-master}
\begin{tabular}{lccc}
\toprule
Cluster & $[\mathrm M/\mathrm H]$ & $M_{\rm TRGB}^{\rm th}$ & $M_{\rm TRGB}^{\rm var}$ \\
        &                         & (mag)                    & (mag) \\
\midrule
NGC~288 & $-0.957\pm 0.050$ & $-3.669_{-0.023}^{+0.023}$ & $-3.703_{-0.270}^{+0.127}$ \\
NGC~362 & $-0.957\pm 0.050$ & $-3.684_{-0.023}^{+0.023}$ & $-3.623_{-0.170}^{+0.137}$ \\
NGC~1261 & $-1.057\pm 0.150$ & $-3.665_{-0.023}^{+0.023}$ & $-3.590_{-0.216}^{+0.142}$ \\
NGC~5024 & $-1.607\pm 0.050$ & $-3.563_{-0.023}^{+0.023}$ & $-3.476_{-0.069}^{+0.064}$ \\
NGC~5053 & $-1.927\pm 0.150$ & $-3.493_{-0.023}^{+0.023}$ & $-2.585_{-0.354}^{+0.174}$ \\
NGC~5272 & $-1.277\pm 0.050$ & $-3.615_{-0.023}^{+0.023}$ & $-3.540_{-0.131}^{+0.106}$ \\
NGC~5466 & $-1.707\pm 0.100$ & $-3.540_{-0.023}^{+0.023}$ & $-3.370_{-0.271}^{+0.129}$ \\
NGC~5897 & $-1.627\pm 0.050$ & $-3.559_{-0.023}^{+0.023}$ & $-4.065_{-0.548}^{+0.319}$ \\
NGC~5904 & $-1.007\pm 0.100$ & $-3.669_{-0.023}^{+0.023}$ & $-3.590_{-0.092}^{+0.089}$ \\
NGC~6093 & $-1.407\pm 0.050$ & $-3.598_{-0.023}^{+0.024}$ & $-3.652_{-0.197}^{+0.155}$ \\
NGC~6101 & $-1.687\pm 0.100$ & $-3.546_{-0.023}^{+0.023}$ & $-3.701_{-0.399}^{+0.221}$ \\
NGC~6171 & $-0.727\pm 0.100$ & $-3.711_{-0.023}^{+0.023}$ & $-3.463_{-0.308}^{+0.195}$ \\
NGC~6205 & $-1.307\pm 0.100$ & $-3.598_{-0.023}^{+0.023}$ & $-3.360_{-0.096}^{+0.088}$ \\
NGC~6218 & $-0.957\pm 0.050$ & $-3.669_{-0.023}^{+0.024}$ & $-3.546_{-0.155}^{+0.088}$ \\
NGC~6254 & $-1.307\pm 0.100$ & $-3.609_{-0.023}^{+0.023}$ & $-3.625_{-0.159}^{+0.110}$ \\
NGC~6341 & $-1.957\pm 0.050$ & $-3.488_{-0.024}^{+0.024}$ & $-3.324_{-0.097}^{+0.081}$ \\
NGC~6352$^{\dagger}$ & $-0.387\pm 0.150$ & $-3.705_{-0.053}^{+0.066}$ & $-3.234_{-0.245}^{+0.162}$ \\
NGC~6362 & $-0.737\pm 0.150$ & $-3.720_{-0.024}^{+0.025}$ & $-3.418_{-0.158}^{+0.103}$ \\
NGC~6366$^{\dagger}$ & $-0.337\pm 0.100$ & $-3.714_{-0.050}^{+0.064}$ & $-3.125_{-0.337}^{+0.195}$ \\
NGC~6397 & $-1.627\pm 0.100$ & $-3.559_{-0.023}^{+0.023}$ & $-3.492_{-0.235}^{+0.145}$ \\
NGC~6541 & $-1.507\pm 0.100$ & $-3.570_{-0.024}^{+0.024}$ & $-3.481_{-0.111}^{+0.094}$ \\
NGC~6723 & $-0.777\pm 0.150$ & $-3.718_{-0.023}^{+0.024}$ & $-3.205_{-0.210}^{+0.178}$ \\
NGC~6752$^{\dagger}$ & $-1.187\pm 0.100$ & $-3.593_{-0.033}^{+0.038}$ & $-3.577_{-0.168}^{+0.136}$ \\
NGC~6779 & $-1.707\pm 0.100$ & $-3.540_{-0.023}^{+0.023}$ & $-3.613_{-0.226}^{+0.131}$ \\
NGC~6809 & $-1.537\pm 0.050$ & $-3.570_{-0.023}^{+0.024}$ & $-3.487_{-0.122}^{+0.081}$ \\
NGC~6838 & $-0.437\pm 0.150$ & $-3.773_{-0.023}^{+0.023}$ & $-4.016_{-0.448}^{+0.228}$ \\
NGC~7099 & $-1.807\pm 0.150$ & $-3.523_{-0.023}^{+0.023}$ & $-3.695_{-0.373}^{+0.181}$ \\
\bottomrule
\end{tabular}
\end{table*}\black

For the primary 24-cluster sample, the central standard-cooling predictions are very accurately described by the anchored relation
\begin{equation}
M_{\rm TRGB}^{\rm th}= -3.5767-0.18376\,\bigl([\mathrm M/\mathrm H]+1.5\bigr)\quad {\rm mag},
\label{eq:trgb-metallicity-relation}
\end{equation}
with an rms cluster-to-cluster residual of 0.0057~mag about the straight line.  Propagating the same theoretical nuisance components as in the main likelihood gives a 68\% uncertainty of approximately $\pm0.023$~mag on the normalization at $[\mathrm M/\mathrm H]=-1.5$ and $\pm0.0030$~mag~dex$^{-1}$ on the slope.  Thus the standard-cooling prediction at the anchor metallicity is $M_{\rm TRGB}^{\rm th}=-3.577\pm0.023$~mag.

For orientation, this simple relation is numerically very close to earlier bolometric TRGB calibrations.  Valenti, Ferraro, and Origlia~\cite{Valenti2004TRGB} obtained the empirical globular-cluster relation $M_{\rm bol}^{\rm tip}=-3.85-0.19[\mathrm M/\mathrm H]$ from a homogeneous sample of 24 clusters; it gives $-3.565$~mag at $[\mathrm M/\mathrm H]=-1.5$ and differs from Eq.~\eqref{eq:trgb-metallicity-relation} by less than 0.015~mag over the metallicity range of our primary sample.  The FuNS standard-cooling parametrization reported by Straniero et al.~\cite{Straniero2018Patras}, $M_{\rm bol}^{\rm tip}=0.0161[\mathrm M/\mathrm H]^2-0.1716[\mathrm M/\mathrm H]-3.87$, gives $-3.576$~mag at the same anchor and differs from our relation by at most 0.021~mag over this range.  These are intended as simple scale comparisons: the former is an empirical brightest-star calibration, whereas the latter is a theoretical stellar-model relation.

As a direct one-parameter consistency test, we allow a common vertical displacement $\Delta M$ of the cluster-specific standard-cooling predictions, $M_{{\rm TRGB},i}^{\rm th}\to M_{{\rm TRGB},i}^{\rm th}+\Delta M$, while retaining the exact non-Gaussian finite-sampling distributions and profiling the same nuisance parameters as in the primary likelihood.  We find
\begin{equation}
\Delta M=+0.078^{+0.036}_{-0.037}\ {\rm mag},
\label{eq:trgb-common-offset}
\end{equation}
where a positive value means that the observed TRGB is fainter than the standard-cooling prediction.  The no-offset point has $-2\Delta\ln\mathcal L=4.41$ relative to the best fit, corresponding to about $2.1\sigma$ in the one-parameter asymptotic interpretation.  This simple direct comparison therefore reproduces both the direction and the approximate $0.07$--$0.08$~mag size of the deficit found with the signed cooling-response diagnostic discussed below.

Using the representative NGC~6093 neutrino-cooling response of Fig.~\ref{fig:canonical-rnu-response} as a diagnostic translation, the best-fit common displacement of Eq.~\eqref{eq:trgb-common-offset} corresponds to $r_\nu\simeq0.77$; the endpoints of its quoted 68\% interval map approximately to $r_\nu\simeq0.68$ and $0.87$.  The signed axion-response best fit $q=-1.19$ implies a dimming of $0.069$~mag and maps to $r_\nu\simeq0.79$ in the same representative model.  Conversely, the brightening at the power-constrained axion limit $g_{13}=0.64$, $\Delta M_{\rm TRGB}\simeq-0.0238$~mag for NGC~6093, corresponds to $r_\nu\simeq1.08$.  
\black
\begin{figure*}
   \centering
   \includegraphics[width=0.82\textwidth, angle=0]{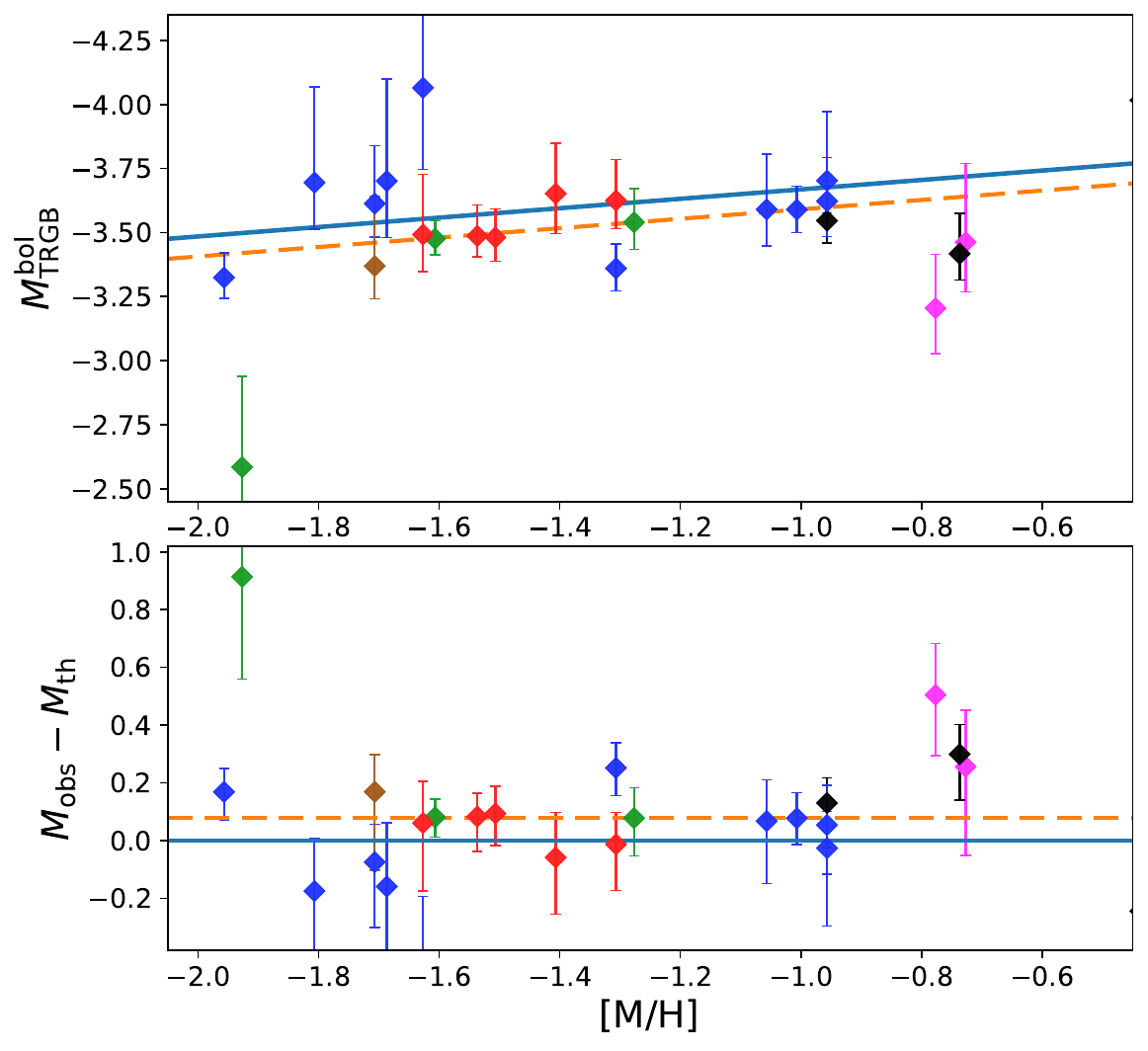}
   \caption{\red Direct comparison between the variable-inclusive observed TRGB bolometric magnitudes and the standard-cooling predictions for the primary 24-cluster sample as a function of metallicity [M/H]. The upper panel shows the observed cluster values together with the anchored theoretical relation of Eq.~\eqref{eq:trgb-metallicity-relation}; the dashed line shows the best-fit common vertical offset of Eq.~\eqref{eq:trgb-common-offset}. The lower panel shows the residuals $M_{\rm obs}-M_{\rm th}$ for the same clusters; the solid and dashed horizontal lines indicate zero offset and the best-fit common offset, respectively. The observational error bars are the updated asymmetric 68\% exact-discreteness intervals. The cluster associations are shown with the same color code as in Fig.~\ref{mbolvarnon}.\black }
\label{mbol_mh}
\end{figure*}
\begin{figure*}
   \centering
   \includegraphics[width=14cm, angle=0]{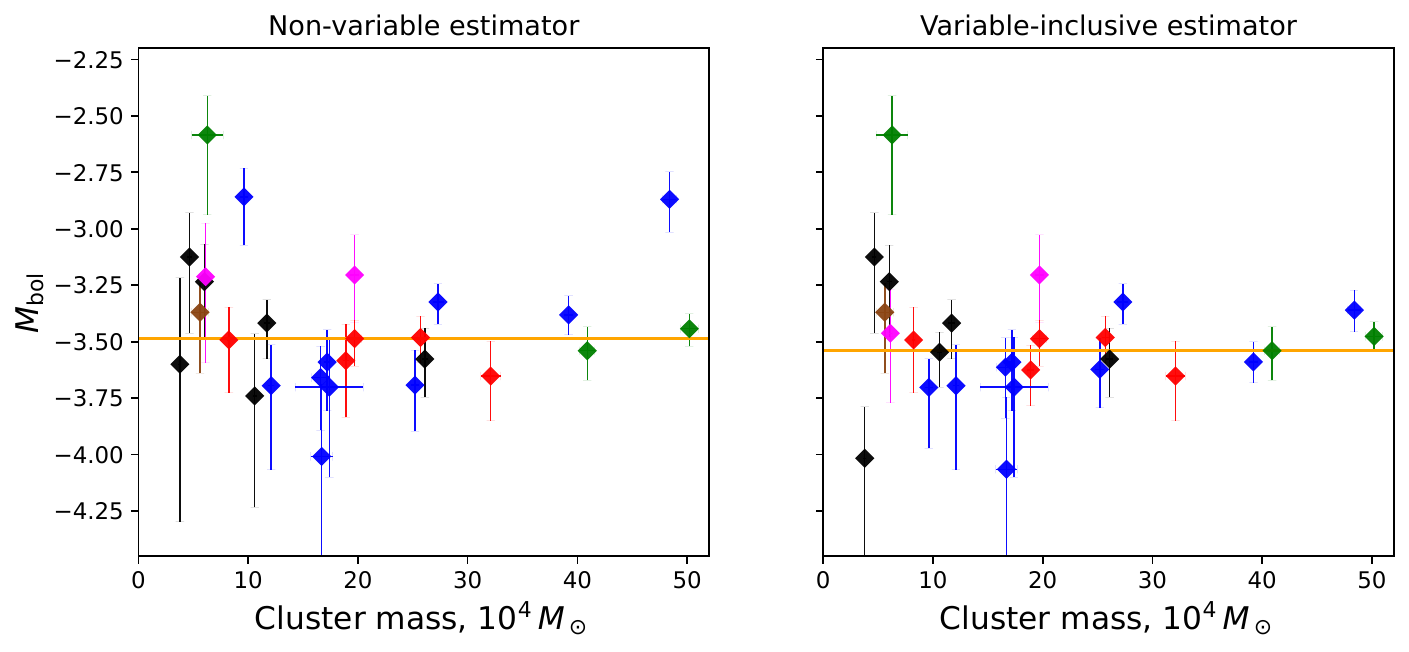}
   \caption{The observed corrected $M_{\rm bol}^{\rm non}$ (left) and $M_{\rm bol}^{\rm var}$ (right) as a function of cluster mass. The vertical error bars are the updated asymmetric 68\% exact-discreteness intervals; the horizontal error bars show the quoted cluster-mass uncertainties. Cluster associations are coloured as in Fig.~\ref{mbolvarnon}, and the orange lines show the sample medians. }
\label{mbol_mass}
\end{figure*}

\begin{figure*}
   \centering
   \includegraphics[width=0.48\textwidth, angle=0]{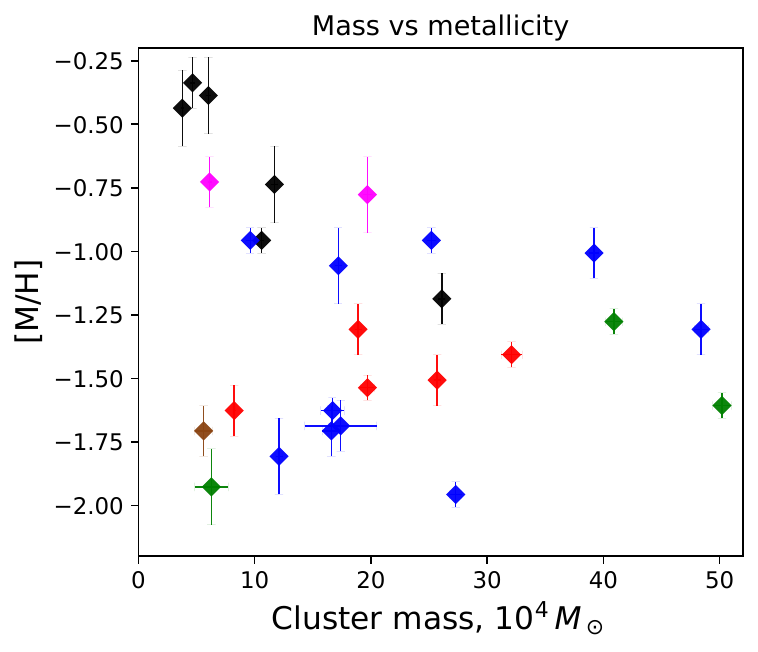}
   ~~~
   \includegraphics[width=0.48\textwidth, angle=0]{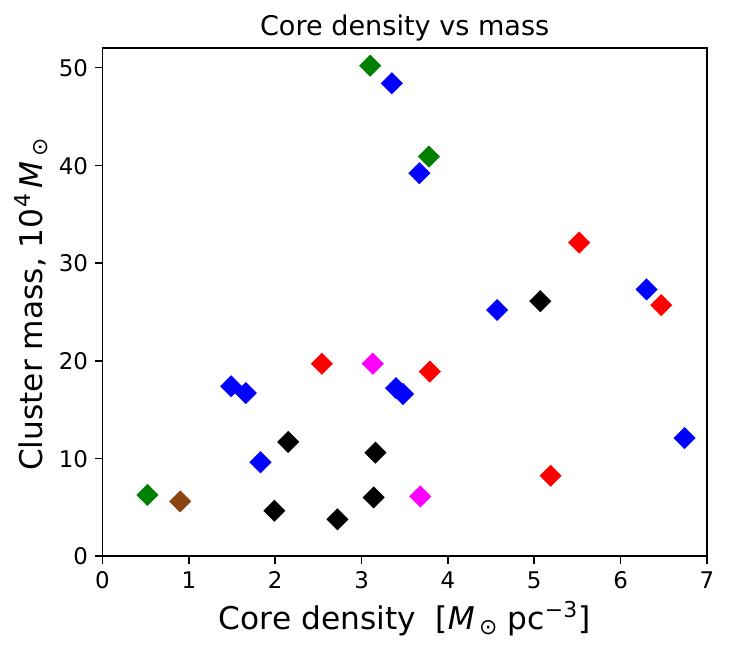}
\caption{The relations between cluster mass and metallicity [M/H] (left) and between cluster core density and mass (right).
   Different association of the clusters is coloured as in Fig.~\ref{mbolvarnon}.
   }
\label{mass_mh}
\end{figure*}
\begin{figure*}
   \centering
   \includegraphics[width=14cm, angle=0]{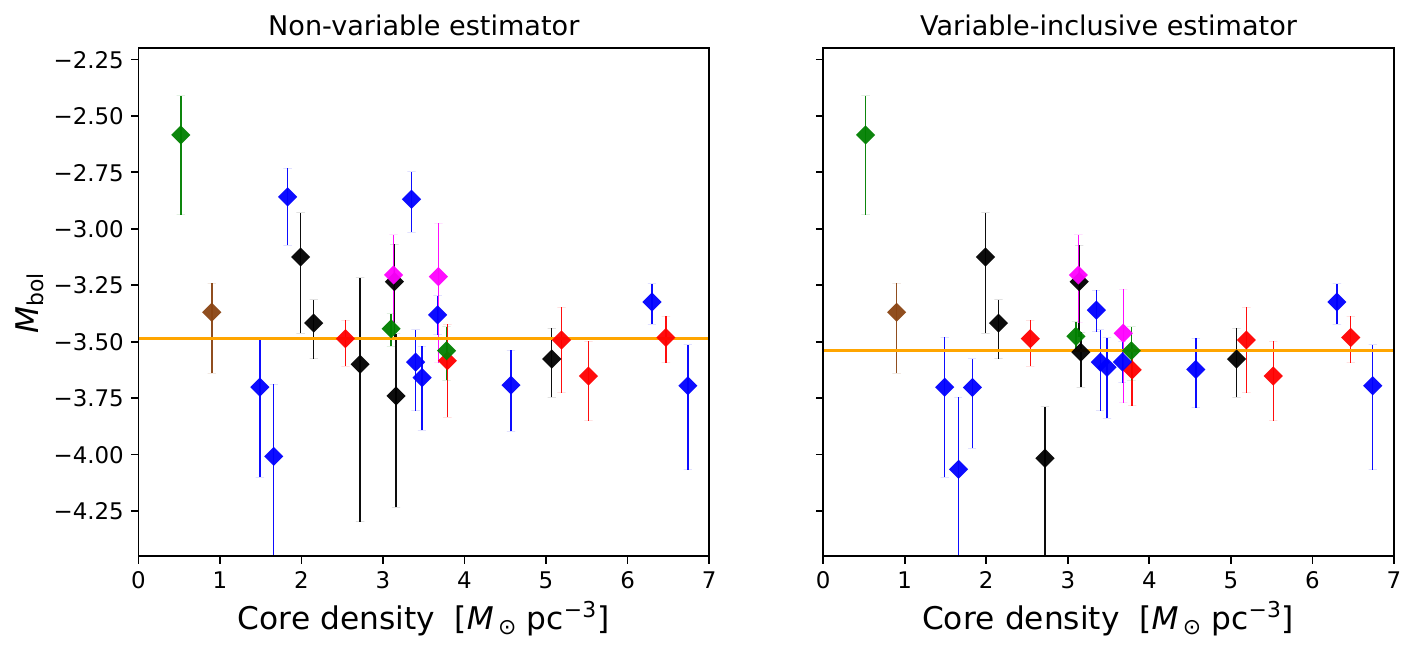}
   \caption{The observed corrected $M_{\rm bol}^{\rm non}$ (left) and $M_{\rm bol}^{\rm var}$ (right) as a function of cluster core density.  Cluster associations are coloured as in Fig.~\ref{mbolvarnon}, and the orange lines show the sample medians. }
\label{mbol_cd}
\end{figure*}
\begin{figure*}
   \centering
   \includegraphics[width=14cm, angle=0]{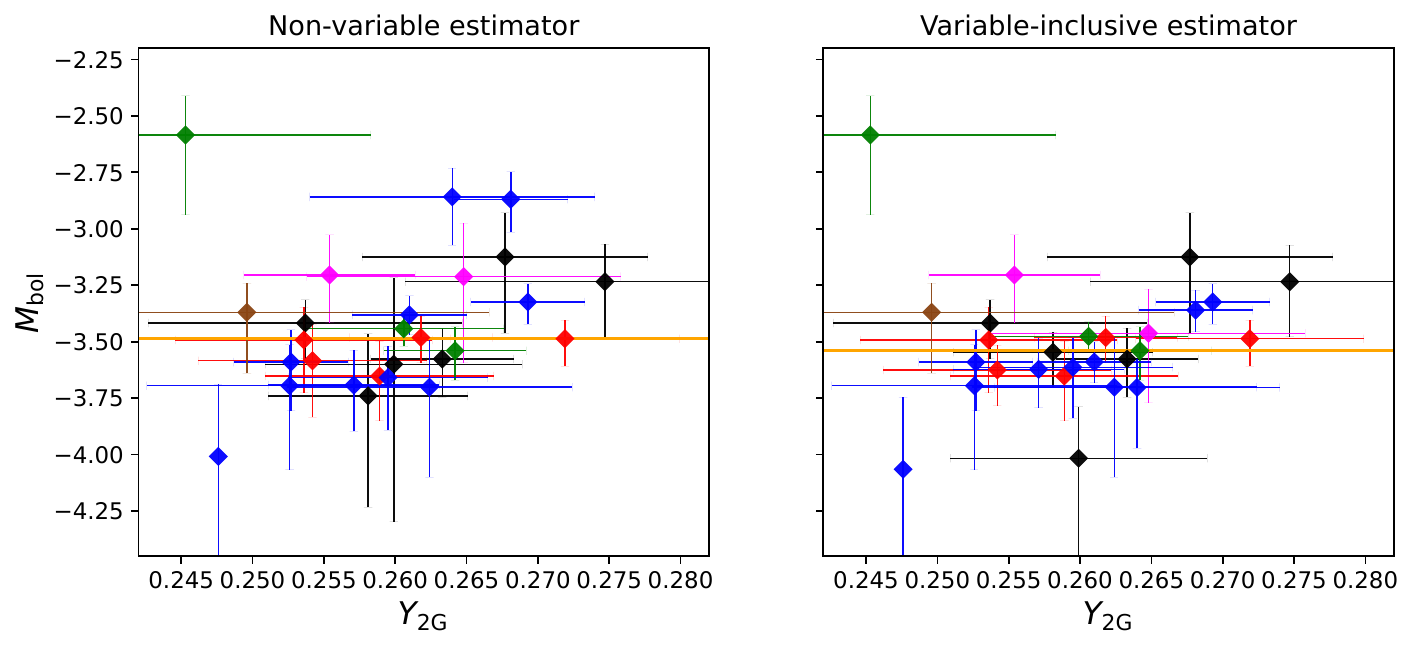}
   \caption{The observed corrected $M_{\rm bol}^{\rm non}$ (left) and $M_{\rm bol}^{\rm var}$ (right) as a function of the helium mass fraction $Y_{\rm 2G}$ of 2G stars. Cluster associations are coloured as in Fig.~\ref{mbolvarnon}, and the orange lines show the sample medians. }
\label{mbol_y2g}
\end{figure*}
\clearpage

\subsection{Bolometric TRGB magnitude versus cluster parameters}
\label{sec:comparison:parameters}
Table~\ref{tab:results} shows a noticeable scatter of $M_{\rm bol}^{\rm non}$ or $M_{\rm bol}^{\rm var}$ estimated from observations. Hence, relations between these estimates and orbital, structural, and other parameters of the clusters should be considered.
As in our previous paper \cite{gcP8}, we adopt $Y_\mathrm{1G}$, $\delta Y_\mathrm{2G,1G}$, and $Y_\mathrm{2G}$ from \cite{milone2018},  the fraction of 1G stars $N_\mathrm{1G}/N_\mathrm{TOT}$ from \cite{dondoglio2021} and \cite{jang2025}, mass loss estimates $\mu_\mathrm{1G}$ and $\mu_\mathrm{2G}$ from \cite{Tailo:2020}, cluster mass and core density from the database of \cite{BaumgardtVasiliev2021}, and estimates of the orbital parameters (apogalactic and perigalactic distance, eccentricity, inclination, period, and time after the last crossing of the Galactic disk) from \cite{bajkova2021}. \red The direct metallicity comparison with the standard-cooling prediction is shown separately in Fig.~\ref{mbol_mh}; additional relations involving cluster mass, metallicity, core density, helium abundance, and the observational TRGB estimators are shown in Figs.~\ref{mbol_mass}--\ref{mbol_y2g}.\black

\red In the panels comparing the TRGB estimators with cluster parameters, inclusion of well-defined variable stars generally reduces the scatter of the derived bolometric magnitudes.  The trends can be interpreted in terms of an increase of the magnitudes for clusters of the lowest mass, as seen in Fig.~\ref{mbol_mass}, in combination with the fact that only the most metal-poor and metal-rich clusters demonstrate the lowest mass, as seen in Fig.~\ref{mass_mh}. \black
Furthermore, the clusters of the lowest mass have rather low core density, as seen in the right panel of Fig.~\ref{mass_mh}. This leads to a  higher  magnitudes and higher scatter of the magnitudes derived for clusters with a low core density, as seen in Fig.~\ref{mbol_cd}.
Finally, Fig.~\ref{mbol_y2g} shows that the magnitudes tend to be  higher  for clusters with the highest helium enrichment of the 2G stars. This can be explained, if a helium enrichment makes the TRGB stars variable and force us to exclude many of them.

Altogether, these trends may indicate that the cluster mass is the key parameter: the TRGB of low-mass clusters (having also an extreme metallicity and rather low core density) is depopulated due to an intensive mass loss and loss of low-mass members in cluster evolution, hence, providing a biased statistics and a higher scatter for the TRGB magnitude estimates.
\red
Consistently with the visual comparison in Fig.~\ref{mbol_mh}, a separate descriptive fit confirms that the variable-inclusive estimator follows the expected metallicity dependence.  In the 24-cluster sample, the central standard-cooling prediction has the slope $dM_{\rm bol}/d[{\rm M/H}]=-0.184\pm0.003$~mag~dex$^{-1}$, consistently with Eq.~\eqref{eq:trgb-metallicity-relation}. For this descriptive trend test, we form a covariance from the second moments of the exact cluster kernels together with the shared nuisance modes. Generalized least-squares fits give $-0.132\pm0.077$~mag~dex$^{-1}$ for the variable-inclusive estimator and $+0.029\pm0.082$~mag~dex$^{-1}$ for the non-variable control estimator. The corresponding residual slopes relative to the cluster-specific MESA predictions are $+0.053\pm0.077$ and $+0.215\pm0.082$~mag~dex$^{-1}$, respectively. Thus the fiducial estimator is consistent with the theoretical metallicity trend. These auxiliary Gaussianized slope fits use only the second moments of the finite-sampling kernels; we use the exact-likelihood common-offset result of Eq.~\eqref{eq:trgb-common-offset}, rather than a Gaussian fixed-slope intercept fit, for the quantitative normalization comparison.
\black

The median corrected value $M_{\rm bol}^{\rm var}=-3.54\pm0.05$~mag for all 27 clusters remains a useful descriptive summary. The same value is obtained for the sample of 24 clusters without hot flashers. However, the bolometric TRGB is not a metallicity-independent standard candle.  We therefore do not use this median in the particle-physics likelihood or interpret it as a universal calibration; every observed cluster is compared with its own MESA prediction. 

\subsection{Comparison with previous TRGB studies}
\label{sec:comparison:otherTRGB}
The luminosity of the TRGB has long been used to constrain anomalous energy losses from degenerate stellar cores. An early detailed analysis of the globular cluster M5 obtained
$g_{ae}<4.3\times10^{-13}$ (95\%~{\rm CL}), with the uncertainty dominated by poorly known cluster distance \cite{Viaux:2013lha}. Subsequent improvements in geometric distance determinations and stellar modeling strengthened this limit by a factor of a few. Using updated geometric calibrations of the $I$-band TRGB, Ref.~\cite{CapozziRaffelt2020} derived $g_{ae}<1.3\times10^{-13}$ (95\%~{\rm CL}) from $\omega$~Cen and $g_{ae}<1.6\times10^{-13}$ (95\%~{\rm CL}) from NGC~4258. While the former value has become a widely used benchmark, it should be interpreted with caution because $\omega$~Cen is an exceptionally complex stellar system rather than a typical chemically homogeneous globular cluster. It contains a broad metallicity distribution and numerous stellar subpopulations \cite{Nitschai:2024qje}, with non-trivial age--metallicity structure \cite{Clontz:2024age} and substantial helium variations \cite{Milone:2017omegacen}. 

In Ref.~\cite{Straniero2020}, bolometric TRGB magnitudes were determined for 22 Galactic globular clusters\footnote{\red A later unpublished Gaia-era update was presented in conference slides; because no corresponding paper or preprint is available and the original presentation is no longer publicly accessible, we do not use it in the quantitative comparisons below.}. Their combined likelihood gave the upper limit $g_{ae}<1.48\times10^{-13}$ (95\%~{\rm CL}) with a one-sigma preference for a nonzero value. This analysis demonstrated the advantage of combining several clusters, while also showing the importance of distances, bolometric corrections, discrete sampling of the upper RGB, and theoretical stellar uncertainties.

\red
For clarity, Table~\ref{tab:straniero-methods} summarizes the main methodological differences between the published analysis of Ref.~\cite{Straniero2020} and the present work.  We emphasize that both analyses include correlated systematic uncertainties; the difference is primarily in how the cluster information, finite-sampling correction, stellar response, and correlated nuisance structure are constructed and propagated.

One notable difference concerns the finite-sampling correction.  In Ref.~\cite{Straniero2020}, the calibration from synthetic CMDs is defined for a specified number of stars within 2.5~mag below the true model TRGB, whereas the observed richness is counted within 2.5~mag below the brightest observed RGB star.  Because the latter window is displaced toward fainter magnitudes by the unknown brightest-star offset, it generally contains additional faint stars.  Using this larger observed count directly in a correction calibrated for the true TRGB reference window therefore tends to underestimate the discreteness correction.  Our conditional construction explicitly accounts for this window shift through the distinction between $N_{\rm tip}$ and $N_{\rm obs}$.
\black

\red
For the clusters in which our two estimators differ, the published Straniero et al.\ values are generally closer to the variable-inclusive result, most clearly for NGC~288 and NGC~6205.  The published 22-cluster analysis gave $g_{13}<1.48$ at 95\% confidence, compared with the power-constrained $g_{13}<0.64$ from our primary 24-cluster sample.  
\black
\begin{table}[t]
\centering

\caption{Bolometric TRGB magnitudes for the ten clusters common to our sample and Table~6 of the published analysis of Ref.~\cite{Straniero2020}. The second column uses their values based on the Baumgardt et al.\ kinematic distance scale. The last two columns give our non-variable and variable-inclusive estimators.}
\label{tab:straniero-comparison}

\begin{tabular}{lccc}
\toprule
Cluster & Straniero et al.\ (2020) & This work: non-var. & This work: var.-incl. \\
\midrule
NGC~288  & $-4.07\pm0.16$ & $-2.859^{+0.132}_{-0.217}$ & $-3.703^{+0.132}_{-0.272}$ \\
NGC~362  & $-3.66\pm0.14$ & $-3.692^{+0.159}_{-0.208}$ & $-3.623^{+0.142}_{-0.174}$ \\
NGC~5272 & $-3.45\pm0.17$ & $-3.540^{+0.112}_{-0.136}$ & $-3.540^{+0.112}_{-0.136}$ \\
NGC~5904 & $-3.61\pm0.16$ & $-3.381^{+0.091}_{-0.095}$ & $-3.590^{+0.096}_{-0.099}$ \\
NGC~6171 & $-3.47\pm0.28$ & $-3.212^{+0.240}_{-0.383}$ & $-3.463^{+0.198}_{-0.310}$ \\
NGC~6205 & $-3.28\pm0.20$ & $-2.870^{+0.127}_{-0.150}$ & $-3.360^{+0.095}_{-0.102}$ \\
NGC~6341 & $-3.51\pm0.16$ & $-3.324^{+0.089}_{-0.104}$ & $-3.324^{+0.089}_{-0.104}$ \\
NGC~6752 & $-3.61\pm0.16$ & $-3.577^{+0.141}_{-0.172}$ & $-3.577^{+0.141}_{-0.172}$ \\
NGC~6809 & $-3.44\pm0.18$ & $-3.487^{+0.089}_{-0.127}$ & $-3.487^{+0.089}_{-0.127}$ \\
NGC~7099 & $-3.41\pm0.31$ & $-3.695^{+0.185}_{-0.375}$ & $-3.695^{+0.185}_{-0.375}$ \\
\bottomrule
\end{tabular}

\normalsize
\end{table}

\begin{table}[t]
\centering
\small
\red
\setlength{\tabcolsep}{3.5pt}
\renewcommand{\arraystretch}{1.02}

\caption{Main methodological differences between the published analysis of Ref.~\cite{Straniero2020} and the present work.}
\label{tab:straniero-methods}

\begin{tabular}{@{}%
>{\raggedright\arraybackslash}p{0.18\textwidth}%
>{\raggedright\arraybackslash}p{0.385\textwidth}%
>{\raggedright\arraybackslash}p{0.385\textwidth}@{}}

\toprule
Aspect & Ref.~\cite{Straniero2020} & This work \\
\midrule

GC sample
&
22 Galactic GCs
&
27 GCs; 24 in the primary particle-physics likelihood after excluding three hot-flasher systems
\\
\addlinespace[2.5pt]

Observational data and membership
&
HST and ground-based optical/near-IR photometry; the main 22-cluster sample is based on near-IR data, with combined optical and near-IR analyses for three best-studied clusters
&
Stetson, HST, and Gaia DR3 photometry; cluster membership selected using HST or Gaia astrometry
\\
\addlinespace[2.5pt]

Cluster inputs
&
Reference stellar grid with age $\simeq13$~Gyr and $Y=0.25$; cluster-to-cluster metallicities and uncertainties in age, $Y$, and $[\mathrm{M/H}]$ propagated around this grid
&
Age, $[\mathrm{Fe/H}]$, distance, reddening, helium content, and RGB-star mass determined cluster by cluster from the homogeneous isochrone analysis
\\
\addlinespace[2.5pt]

Stellar models
&
FuNS models on a grid in $[\mathrm{M/H}]$ and $g_{13}$, with cluster-parameter variations treated as uncertainties
&
Dedicated MESA calculations for every cluster, using its inferred mass, helium abundance, metallicity, and $\alpha$ enhancement, repeated as a function of $g_{13}$
\\
\addlinespace[2.5pt]

Finite-sampling richness
&
Synthetic-CMD calibration uses the number $N$ of stars within 2.5~mag below the model TRGB; for observed clusters, $N$ is counted within 2.5~mag below the brightest observed RGB star
&
Distinguishes explicitly the latent richness $N_{\rm tip}$ in the true TRGB-referenced interval from the observable $N_{\rm obs}$ in the shifted interval below the brightest retained star
\\
\addlinespace[2.5pt]

Finite-sampling distribution
&
Median correction and mean deviation from the median derived from synthetic CMDs; a symmetric uncertainty is used in the likelihood
&
Full conditional order-statistics density $p(\delta_k\mid N_{\rm obs},k)$, including the shifted faint boundary, removal of brighter variables, and the non-Gaussian tail
\\
\addlinespace[2.5pt]

Variable stars
&
Near-IR photometry reduces the amplitudes of long-period variables; no separate variable/non-variable TRGB estimators
&
Separate variable-inclusive and non-variable estimators; well-sampled variables enter the primary estimator, while the non-variable result is retained as a population-selection control
\\
\addlinespace[2.5pt]

Correlated systematics
&
Covariance matrix includes correlations from the distance-scale zero point, bolometric corrections, and common theoretical uncertainties
&
Shared signed response modes and cluster-specific errors are represented explicitly, with common nuisance amplitudes jointly profiled
\\
\addlinespace[2.5pt]

Statistical inference
&
Cumulative likelihood; the upper bound is defined by integrating the normalized likelihood above the quoted limit
&
Frequentist profile likelihood with pseudoexperiment calibration, goodness-of-fit test, signed-response diagnostic, and a power-constrained upper limit
\\

\bottomrule
\end{tabular}

\black
\normalsize
\setlength{\tabcolsep}{6pt}
\renewcommand{\arraystretch}{1}
\end{table}

\section{New-physics constraints and discussion}
\label{sec:constraints}
\subsection{Statistical method}
\label{sec:constraints:likelihood}

Our primary inference is based on a frequentist profile-likelihood analysis. We combine the cluster measurements in a joint likelihood in which shared observational, method, and stellar-physics uncertainties are represented by common nuisance parameters rather than by independent cluster-by-cluster errors. The primary sample contains 24 clusters, after excluding the hot-flasher clusters NGC~6352, NGC~6366, and NGC~6752. The primary TRGB estimator is variable-inclusive: for clusters with an accepted well-sampled variable it uses that star, and otherwise it coincides with the non-variable estimator.

For cluster $i$, we define
\begin{equation}
 r_i(g_{ae},\boldsymbol\eta,t)
 =M_{{\rm bol},i}^{\rm th}(g_{ae})+t
 -M_{{\rm bol},i}^{\rm obs}
 -\sum_a s_{a,i}\eta_a ,
\label{eq:correlated-residual}
\end{equation}
where $s_{a,i}$ are the signed observational and method responses defined in Eq.~\eqref{eq:obs-common-nuisance}, $\eta_a$ are the corresponding Gaussian-constrained nuisance amplitudes, and $t$ is the common theoretical shift. The constraint density for $t$ is obtained by Monte Carlo from the shared stellar-physics components of Table~\ref{tab:theory_uncertainties}, converted through $\Delta M_{\rm bol}=-2.5\Delta\log_{10}L$ and centered on its median.

The cluster-specific density $p_i^0$ contains the one-star observational uncertainty and the cluster-specific part of the theoretical uncertainty, but not the finite-sampling correction. Its small residual asymmetry is represented by
\begin{equation}
 p_i^0(x)=\frac{\sqrt{2/\pi}}{\sigma_{i,-}^0+\sigma_{i,+}^0}
 \begin{cases}
 \exp[-x^2/(2(\sigma_{i,-}^0)^2)], & x<0,\\[1mm]
 \exp[-x^2/(2(\sigma_{i,+}^0)^2)], & x\ge0,
 \end{cases}
\label{eq:two-piece-core}
\end{equation}
where the widths contain only the remaining cluster-specific terms. The finite-sampling distribution is included exactly by convolution. If $\widetilde\delta_i=M_{\star,i}-M_{{\rm bol},i}^{\rm obs}$ denotes the median correction used only to center the tabulated $M_{{\rm bol},i}^{\rm obs}$, the cluster residual kernel is
\begin{equation}
 K_i(r)=\int_0^\infty d\delta\,
 p_i(\delta\mid N_{{\rm obs},i},k_i)\,
 p_i^0\!\left(r+\delta-\widetilde\delta_i\right).
\label{eq:exact-cluster-kernel}
\end{equation}

The likelihood used for the inference is
\begin{equation}
 {\cal L}(g_{ae},\boldsymbol\eta,t)
 =p_{\rm th}^{\rm com}(t)\prod_a\phi(\eta_a)
 \prod_{i=1}^{24}K_i\!\left[r_i(g_{ae},\boldsymbol\eta,t)\right],
\label{eq:correlated-likelihood}
\end{equation}
where $\phi$ is the unit-Gaussian constraint density and $p_{\rm th}^{\rm com}$ is the constraint density for the common theoretical shift.

We take
\[
\mu\equiv g_{13}^2\ge0
\]
as the physical signal-strength parameter and construct the upper limit from the one-sided profile-likelihood statistic \cite{Cowan:2010js}
\begin{equation}
 q_\mu=\begin{cases}
 -2\ln\lambda(\mu), & \widehat\mu\le\mu,\\
 0, & \widehat\mu>\mu,
 \end{cases}
 \qquad
 \lambda(\mu)=
 \frac{{\cal L}(\mu,\widehat{\widehat{\boldsymbol\nu}}_\mu)}
 {{\cal L}(\widehat\mu,\widehat{\boldsymbol\nu})},
\label{eq:qmu-profile}
\end{equation}
where $\boldsymbol\nu$ denotes all nuisance parameters and hats indicate the corresponding maximum-likelihood estimates.

Because a downward fluctuation of the observed TRGB luminosities can produce an anomalously strong upper limit on additional cooling, we impose the power constraint of Ref.~\cite{Cowan:2011PCL}. For each tested signal strength $\mu$, we determine from background-only pseudoexperiments the probability of excluding that value. We adopt the conventional PCL choice $M_{\min}=\Phi(-1)\simeq0.16$ advocated in Ref.~\cite{Cowan:2011PCL}. This corresponds to the lower $1\sigma$ quantile of the background-only expected-limit distribution: an observed limit stronger than this boundary is therefore replaced by the corresponding sensitivity floor. The test size is $\alpha=0.05$, and the critical distributions and power are calibrated with pseudoexperiments using conditional maximum-likelihood estimates of the nuisance parameters. All physical fits and limits are restricted to the calculated MESA range $0\le g_{13}\le1$.

The global goodness of fit is assessed independently with a generalized maximum-likelihood deviance. Its sampling distribution is obtained from a parametric bootstrap generated from the same exact non-Gaussian cluster kernels and with the same shared nuisance structure as in the data fit.

For secondary diagnostics only, we also construct nuisance-integrated likelihoods by importance-sampled scrambled Sobol quasi-Monte-Carlo integration. These are used for the individual-cluster leverage plots and for the Bayesian prior-sensitivity check discussed below; they do not enter the quoted frequentist confidence limit.

\subsection{Constraints on the axion-electron coupling}
\label{sec:constraints:result}

Applying the profile-likelihood construction of Sec.~\ref{sec:constraints:likelihood} to the primary 24-cluster variable-inclusive sample, we find a physical maximum at
\[
\widehat g_{13}=0.0377,
\]
statistically indistinguishable from zero. The ordinary observed 95\% profile-likelihood limit is \(g_{13}<0.607\), while the \(M_{\min}=0.16\) power constraint corresponds to a sensitivity floor at \(g_{13}=0.640\). Since the observed limit falls below this floor, the power-constrained limit determines our primary result,
\begin{equation}
 \boxed{g_{ae}<6.4\times10^{-14}}\qquad
 (95\%~{\rm CL,\ power\mbox{-}constrained}),
\label{eq:main-gae-limit}
\end{equation}
or equivalently $g_{13}<0.64$. The limit lies fully within the calculated MESA interval $0\le g_{13}\le1$ and does not depend on a prior for $g_{ae}$.

\red
The same background-only pseudoexperiments also provide a simple measure of the expected sensitivity. The ordinary observed limit, $g_{13}<0.607$, lies near the lower $2.6\%$ tail of the expected-limit distribution, corresponding to approximately a $2\sigma$ downward fluctuation in Gaussian terms. We therefore apply the conventional power-constrained prescription \cite{Cowan:2011PCL} with $M_{\min}=\Phi(-1)\simeq0.16$, which prevents the quoted limit from fluctuating below the lower edge of the central $68\%$ range of expected limits. This gives the sensitivity floor $g_{13}=0.640$. The median expected 95\% upper limit is approximately $g_{13}\simeq1$.
\black

\begin{figure*}[t]
   \centering
   \includegraphics[width=0.93\textwidth]{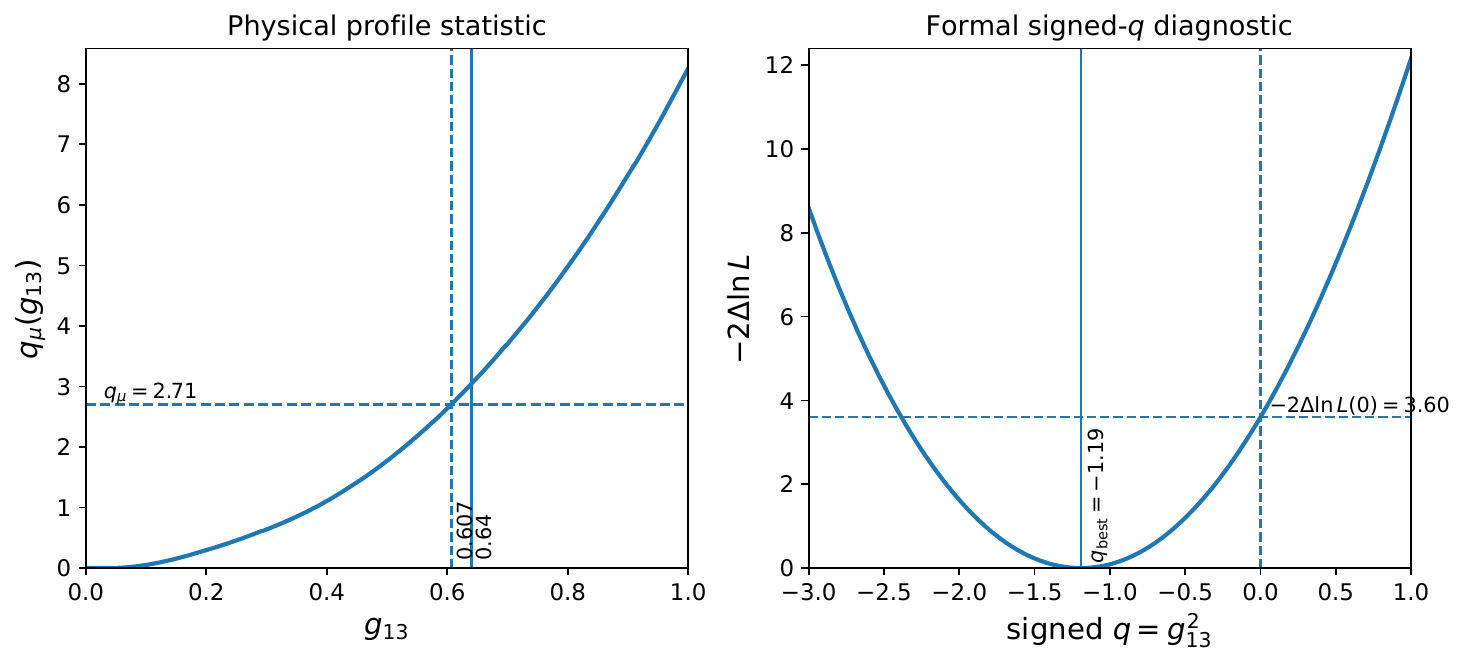}
\caption{ Left: the observed one-sided profile-likelihood statistic $q_\mu(g_{13})$ for the primary 24-cluster variable-inclusive sample over the MESA-calculated interval $0\le g_{13}\le1$. The dashed horizontal line marks the asymptotic 95\% threshold $q_\mu=2.71$; the dashed and solid vertical lines show the corresponding ordinary profile-likelihood bound $g_{13}=0.607$ and the adopted power-constrained bound $g_{13}=0.640$, respectively. Right: profile-likelihood consistency diagnostic obtained by formally extending the small-coupling response to a signed amplitude $q$. The physical boundary is at zero; the best fit lies mildly in the formally negative, unphysical region, as discussed in the text.}
\label{fig:L}
\end{figure*}

The corresponding profile-likelihood diagnostics are shown in Fig.~\ref{fig:L}.

The goodness of fit is evaluated with the same full non-Gaussian likelihood. The observed generalized deviance is $D_{\rm obs}=34.190$. In 3000 parametric-bootstrap pseudoexperiments generated at the fitted physical model and refitted in the same way, the fraction with a deviance at least this large is
\begin{equation}
 p_{\rm GOF}=0.345\pm0.009.
\end{equation}
The global fit is therefore entirely acceptable. A Gaussian $\chi^2$ calibration is not appropriate for the heavy-tailed finite-sampling kernels. Because each pseudoexperiment contains the same shared nuisance structure as the data fit, this test also checks whether the remaining cluster-to-cluster dispersion is compatible with the genuinely cluster-specific uncertainties after the common modes are accounted for.

We performed several robustness checks using the ordinary asymptotic profile limit. Restoring the three hot-flasher clusters gives $g_{13}<0.634$, compared with $0.607$ for the preferred 24-cluster sample. 
As a population-selection control, the non-variable 24-cluster sample gives the ordinary asymptotic limit $g_{13}<0.585$, compared with $g_{13}<0.607$ for the corresponding variable-inclusive sample.
These variants are used only to test the stability of the result; the PCL calibration and the quoted primary bound refer to the 24-cluster variable-inclusive analysis with the correlated optical-only method model.

\begin{figure*}[t]
 \centering
 \includegraphics[width=0.95\textwidth]{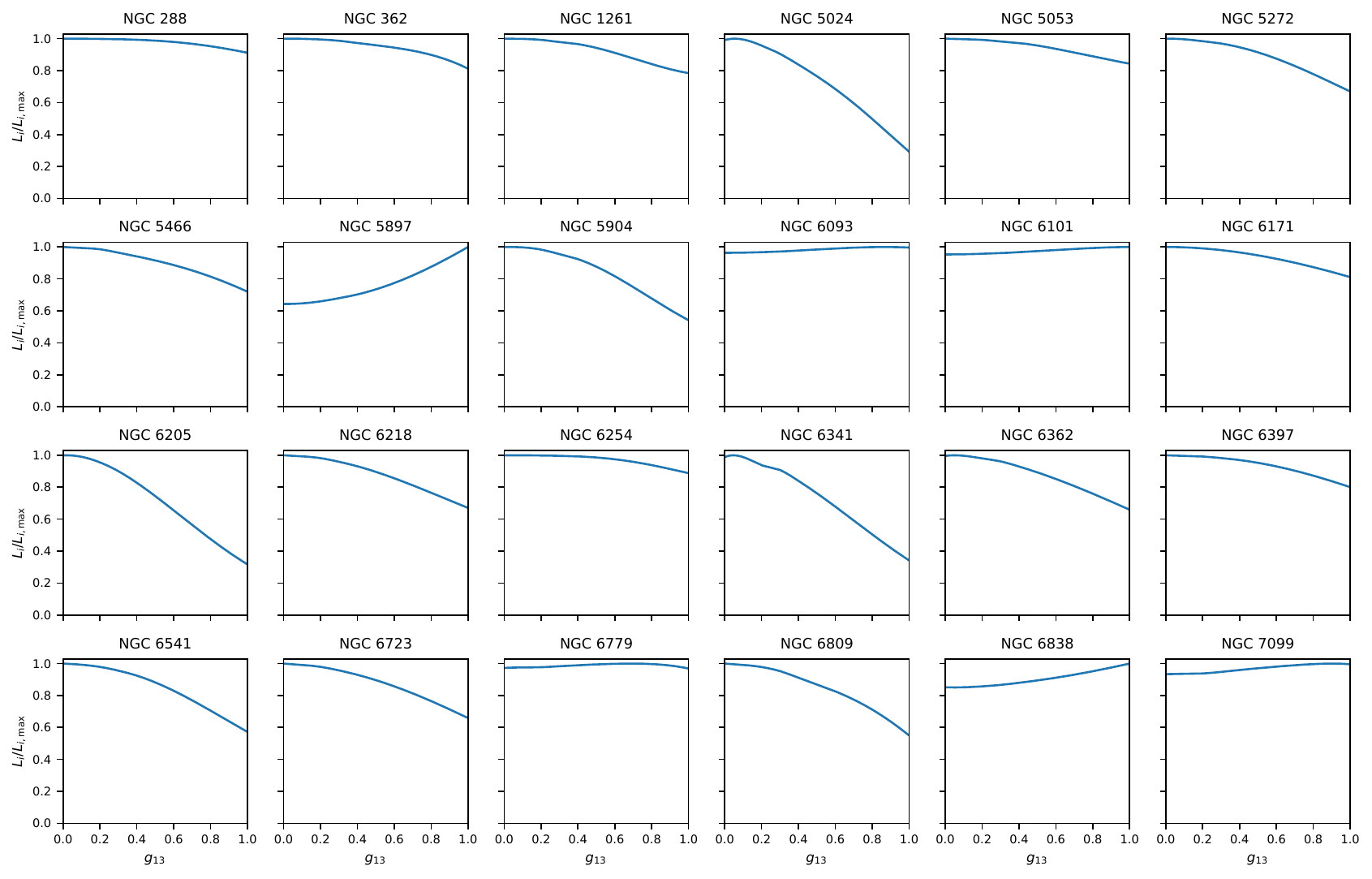}
 \caption{ Normalized individual-cluster likelihoods for the primary 24-cluster variable-inclusive sample, restricted to the MESA-calculated range $0\le g_{13}\le1$. For this diagnostic only, the nuisance variables relevant to each cluster are integrated over their constraint distributions. The combined confidence bound instead uses the correlated profile likelihood of Eq.~\eqref{eq:correlated-likelihood}.}
 \label{fig:individual-likelihoods}
\end{figure*}

Figure~\ref{fig:individual-likelihoods} shows the individual-cluster likelihoods for the primary sample over $0\le g_{13}\le1$. They are individually broad and in several cases remain appreciable at the upper edge of the MESA grid. We therefore use them only as leverage diagnostics and do not quote individual upper limits, which would require assumptions about the uncalculated $g_{13}>1$ tail.

We also test whether the data prefer a response in the direction opposite to additional cooling. For this purpose only, we formally continue the small-coupling response, which is approximately linear in $g_{13}^2$, to a signed amplitude $q$. A linear fit over $g_{13}\le0.6$ reproduces the interpolated MESA response of every cluster in the primary sample to better than $1.3\times10^{-3}$~mag. The continuation to negative amplitude has no physical axion interpretation and is used solely as a consistency diagnostic. Its best fit lies in the unphysical region and gives
\begin{equation}
 2\Delta\ln{\cal L}(0)=3.60,
 \qquad
 \sqrt{2\Delta\ln{\cal L}}=1.90.
\end{equation}
Thus the preference for a response opposite to additional cooling is only $1.90\sigma$ in the Gaussian approximation. Together with $p_{\rm GOF}=0.345\pm0.009$, this provides no significant evidence for an inconsistency of the uncertainty model or for an unaccounted coherent offset.

As a separate Bayesian prior-sensitivity diagnostic, we normalize the nuisance-integrated likelihood over the calculated interval $0\le g_{13}\le1$. The 95\% upper quantile is $g_{13}=0.694$ for a prior flat in $g_{13}$ and $g_{13}=0.824$ for a prior flat in $g_{13}^2$. These values are not used as confidence limits: unlike the frequentist result in Eq.~\eqref{eq:main-gae-limit}, they depend both on the chosen prior measure and on the endpoint of the calculated interval.

The stellar-evolution calculations used in this work assume that the ALP mass is negligible in comparison with the characteristic temperature of the degenerate helium core. The massless-emission approximation is reliable for \(m_a\lesssim1~{\rm keV}\).
At masses of several keV, phase-space and Boltzmann suppression become relevant, and a quantitative constraint requires finite-mass emission rates and a new set of stellar-evolution calculations \cite{Carenza:2021qti}. 

\subsection{Comparison with previous constraints on \texorpdfstring{$g_{ae}$}{gae}}
\label{sec:constraints:gae-comparison}
More recently, Ref.~\cite{Troitsky:2024qea} used Gaia DR3 membership information and distances for seven globular clusters and reported the limit $g_{ae}<5.2\times10^{-14}$ (95\%~{\rm CL}). While this result illustrates the potential of Gaia data, it relies on previously derived parametrizations of the dependence of stellar observables on the ALP couplings, rather than on new stellar-evolution calculations tailored to the individual clusters. The present study implements the more complete analysis identified in Ref.~\cite{Troitsky:2024qea} as an important direction for future work. It extends the observational sample to 27 clusters, combines Gaia data with \textit{HST} and ground-based photometry, determines the relevant stellar-population parameters simultaneously for each cluster, and incorporates these parameters into cluster-specific stellar-evolution calculations and uncertainty estimates. It therefore replaces the interpolating approach of Ref.~\cite{Troitsky:2024qea} with a self-consistent, cluster-by-cluster determination of both the observational TRGB luminosities and their theoretical predictions.  \redd
Ref.~\cite{Troitsky:2024qea} also neglected correlations
among clusters arising from common systematic uncertainties,
which artificially strengthened the limit. In the present work,
these uncertainties are instead retained as correlated nuisance
parameters.
\black
Despite being numerically  weaker than the result of Ref.~\cite{Troitsky:2024qea}, the limit reported here thus supersedes that previous seven-cluster result. 

\redd
A recent preprint \cite{Levasseur:2026rgb} used the
distance-independent bolometric TRGB--RGB-bump magnitude
difference in three globular clusters to obtain
$g_{ae}<1.49\times10^{-13}$ at 95\% CL, providing
a complementary constraint with different systematic
uncertainties.
\black

Ref.~\cite{Dennis:2023czl} addressed the effect of varying stellar nuisance parameters simultaneously rather than one at a time, with the result that couplings up to $g_{ae}\approx4.5\times10^{-13}$ may be allowed from TRGB studies, because correlations between stellar parameters and anomalous cooling can weaken constraints inferred from the theoretical $I$-band tip magnitude. We stress, however, that their analysis addresses a statistical problem which differs from the one considered here. Ref.~\cite{Dennis:2023czl} employed a common set of M5-inspired priors on stellar mass, helium abundance and metallicity when analyzing several empirical TRGB calibrations. In the present work, each cluster is treated individually, and its age, composition and RGB-star mass are constrained by a cluster-specific isochrone fit. The stellar masses are thus confined to the narrow range appropriate for the observed old population, rather than described by a common population model. Moreover, we use the bolometric TRGB luminosity which, as explicitly noted in Ref.~\cite{Dennis:2023czl}, is less sensitive than $M_I$ to uncertainties in the predicted effective temperature and in the corresponding bolometric corrections.  The shared nuisance parameters profiled in our primary likelihood are of a different kind: they encode the coherent effect of shared calibrations and common stellar-physics inputs across the cluster sample. We do not introduce an additional broad simultaneous marginalization over the internally fitted age, distance, reddening, and composition of each individual cluster. 

Constraints on the axion-electron coupling come from laboratory searches, searches for ALPs produced in the Sun, and from the influence of ALP emission on stellar evolution at its various stages; e.g.\ see Ref.~\cite{WISPers} for a comprehensive review. Purely laboratory limits are much weaker than stellar bounds. Searches for spin-dependent forces with a spin-polarized torsion pendulum constrain the coupling at approximately
$g_{ae}\lesssim7.5\times10^{-9}$ for very light ALPs \cite{Terrano:2015sna}. Direct searches for ALPs produced in the Sun improve substantially on this result. In particular, the XENONnT electronic-recoil data imply $g_{ae}<1.9\times10^{-12}$ (90\% CL) in the low-mass regime \cite{XENON:2022ltv}. These limits are direct in the sense that ALP production and detection are both controlled by the electron coupling, but they remain approximately one order of magnitude weaker than the leading stellar-cooling constraints.

White-dwarf cooling is also sensitive to anomalous energy losses \cite{BlinnikovVysotsky}. Analyses of the white-dwarf luminosity function and pulsation periods typically probe couplings of order a few $10^{-13}$, although the result depends on the modeling of the star-formation history, white-dwarf envelopes and crystallization physics \cite{MillerBertolami:2014rka}. A recent analysis  \cite{Alberino:2026wdlf} of the Gaia DR3 100-pc white-dwarf luminosity function reports $g_{ae}<1.68\times10^{-13}$ (95\%~{\rm CL}).  In 47~Tuc, the recent unbinned white-dwarf cooling analysis of Ref.~\cite{Fleury:2025wd47tuc} obtains $g_{ae}\le0.81\times10^{-13}$ at 95\% confidence.  An earlier study of the hot white-dwarf luminosity function in the same cluster found a constraint at approximately the same scale while also testing the normalization of plasmon-neutrino cooling \cite{Hansen:2015wd47tuc}.  These 47~Tuc results are numerically comparable to Eq.~\eqref{eq:main-gae-limit}, though they rely on different stellar populations and systematics.  The agreement in scale between white-dwarf and red-giant constraints is remarkable because the two probes involve different evolutionary stages and different dominant systematic uncertainties.

Searches for ALPs produced by electron bremsstrahlung in magnetic white dwarfs and subsequently converted into X rays constrain the product $\left|g_{ae}g_{a\gamma}\right|$, rather than either coupling separately \cite{Dessert:2021bkv}. Translating such results into a limit on $g_{ae}$ would require an assumption about $g_{a\gamma}$ or about a model-dependent relation between the two couplings. Likewise, limits from absorption of ALP dark matter depend on the assumption that the ALP constitutes a specified fraction of the local dark-matter density and address a different physical scenario.

 The limit obtained in the present work is therefore among the strongest available direct constraints on $g_{ae}$ for ALP masses below $\sim1$~keV. 

\subsection{Indicative sensitivity to other cooling mechanisms}
\label{sec:constraints:other}
\red
The main quantitative new-physics result of this work is the constraint on $g_{ae}$, for which the cooling rate is implemented directly in the same cluster-specific MESA calculations used to predict the standard TRGB.  The present observational sample can nevertheless be used to indicate the parameter scales that could be probed for other cooling mechanisms.  For this purpose, we write the theoretical response as
\black
\begin{equation}
    M_{{\rm bol},i}^{\rm th}(\xi)
    =
    M_{{\rm bol},i}^{\rm MESA,0}
    -
    B_i(\xi),
    \label{eq:Bdifference}
\end{equation}
\red
where $B_i(\xi)>0$ denotes the TRGB brightening produced by a new-physics parameter $\xi$.  In the examples below, $B_i$ is taken from published calculations rather than recomputed with our cluster-specific input physics.  These translations should therefore be regarded as order-of-magnitude sensitivity estimates, not as new confidence limits.  In particular, they do not include the cluster-by-cluster response and the decomposition into shared and cluster-specific theoretical uncertainties used in the primary $g_{ae}$ likelihood.  A dedicated calculation of these responses within the present MESA framework is left for future work.
\black

\subsubsection{Neutrino dipole magnetic moment}
\label{sec:constraints:other:munu}
\red
A nonzero neutrino magnetic moment, $\mu_\nu$, adds an electromagnetic contribution to plasmon decay, $\gamma^\ast\to\nu\bar\nu$, and hence increases the cooling rate of red-giant cores \cite{Bernstein:1963qh,Raffelt:1990pj}.  The corresponding emissivity scales as $\mu_\nu^2$.  To estimate the sensitivity of the present TRGB sample, we use the fitting formulae of Ref.~\cite{Diaz:magnetic} only differentially, i.e.\ to translate a change in $\mu_\nu$ into a TRGB brightening relative to the standard prediction in Eq.~\eqref{eq:Bdifference}.  This exercise indicates sensitivity at the characteristic scale
\(
   \mu_\nu \sim 5\times10^{-13}\,\mu_B,
\)
where $\mu_B$ is the Bohr magneton.  We do not attach a confidence level to this number, because the magnetic-moment response has not been recalculated with the same cluster-specific stellar models and correlated theoretical nuisance structure as the axion analysis.

For orientation, published TRGB analyses obtained $\mu_\nu<4.5\times10^{-12}\,\mu_B$ from M5 \cite{Viaux:magmom} and $\mu_\nu<1.5\times10^{-12}\,\mu_B$ and $1.2\times10^{-12}\,\mu_B$ from the geometrically calibrated TRGBs of NGC~4258 and $\omega$~Cen, respectively \cite{CapozziRaffelt2020}.  The scale inferred above therefore suggests that the present homogeneous GC sample could be useful for a more sensitive dedicated study.  Different laboratory, solar, stellar, and cosmological measurements generally constrain different effective combinations of neutrino dipole moments, so a detailed comparison would require specifying the underlying scenario \cite{TernesTortola:2025effective}.
\black

\subsubsection{Millicharged particles}
\label{sec:constraints:other:MCP}
\red
Light millicharged particles (MCPs), $\chi$, with electric charge $qe$ can also provide an additional cooling channel through plasmon decay, $\gamma^\ast\to\chi\bar\chi$ \cite{Holdom:1986mixing,Davidson:2000MCP}.  We estimate the corresponding reach using the MESA response calculated by Fung et al.~\cite{Fung:2024MCPTRGB}.  Specifically, we digitize their TRGB response for NGC~6341 and use it as a representative proxy for clusters in our sample.  In the low-mass regime, this translation gives a characteristic sensitivity of order
\(
   q\sim 3\times10^{-15}.
\)
The sensitivity degrades rapidly once the MCP mass approaches the range in which plasmon decay becomes kinematically suppressed.  The self-consistent 15-cluster analysis of Ref.~\cite{Fung:2024MCPTRGB} obtained a one-sided 95\% CL limit $q<6.3\times10^{-15}$ in the low-mass regime. A robust MCP constraint based on our cluster sample requires dedicated stellar-evolution calculations analogous to those performed above for axions.
\black

\subsubsection{Other cooling channels}
\label{sec:constraints:other:other}
\red
The same strategy can in principle be applied to other weakly coupled particles, including kinetically mixed dark photons and light scalars or vectors.  In these cases, however, the in-medium production rates can have a nontrivial dependence on plasma properties and may include resonant contributions.  Self-consistent stellar evolution is therefore essential for quantitative limits \cite{2306.13335,1611.05852}.  We restrict the present discussion to the sensitivity estimates above and defer particle-specific stellar calculations to future work.
\black

\subsection{Implications for TRGB distance-scale calibration}

While the goal of this work is to use the bolometric TRGB luminosity as a probe of non-standard energy losses, the same analysis is also relevant for the use of the TRGB in the extragalactic distance ladder.  In distance-scale applications the TRGB is normally employed in optical or near-infrared passbands where part of the intrinsic dependence of the bolometric tip luminosity on stellar population is compensated by bolometric corrections; nevertheless, residual population effects, reddening, RGB sampling, AGB contamination, and the treatment of variable stars remain important sources of systematic uncertainty.  Our cluster-by-cluster determination of $M_{\rm bol}^{\rm TRGB}$  provides a direct test of how the helium-ignition luminosity varies with  stellar-population parameters.  This is particularly relevant in view of current TRGB distance-ladder work, where the transfer of a TRGB zero point from local anchors to SN~Ia host-galaxy halos requires either closely matched stellar populations or well-controlled population corrections; see, e.g., Refs.~\cite{Rizzi:2007ni,Anderson2024,Newman:2024fkx,Li:2024gib,Li:2024pjo}.  

\section{Conclusions}
\label{sec:concl}
We use a new homogeneous observational sample of 27 Galactic globular clusters, whose stellar-population parameters were obtained by fitting BaSTI and DSED isochrones to the ground-based Stetson photometry and the \textit{HST} and \textit{Gaia} data, to constrain anomalous energy losses near the tip of the red-giant branch. The observational and theoretical uncertainties are evaluated individually for every cluster.

We calculate cluster-specific TRGB luminosities with MESA as a function of the axion--electron coupling $g_{ae}$. The primary likelihood contains 24 clusters after excluding three hot-flasher systems and uses well-sampled variable red giants where available. Finite upper-RGB sampling and the correction for removed variables are described by the full conditional density $p(\delta_k\mid N_{\rm obs},k)$, including its long non-Gaussian tail and the exact transformation $M_{\rm tip}=M_\star-\delta_k$. The likelihood distinguishes cluster-specific errors from shared distance, reddening, population, bolometric-correction, and stellar-physics inputs. The method model uses the fitted $\alpha$-enhanced BaSTI and DSED isochrones and separates their family and photometric-system responses; direct optical--infrared SED integration for 22 non-variable stars is retained as an external consistency check.

The physical best fit is very close to zero, $\widehat g_{13}=0.0377$, indicating no preference for additional energy losses. Our primary result is the 95\% CL power-constrained profile-likelihood bound, $g_{ae}<6.4\times10^{-14}$. The parametric-bootstrap goodness of fit is $p_{\rm GOF}=0.345\pm0.009$. The formal signed-response test prefers negative values of energy losses by $\approx 1.9\sigma$; there is therefore no statistically significant preference for the unphysical heating direction.  As a physical scale check, a dedicated NGC~6093 grid with rescaled standard-neutrino cooling maps the direct common TRGB displacement to a representative value $r_\nu\simeq0.77$, while the axion upper-bound response corresponds to $r_\nu\simeq1.08$; we do not interpret this one-cluster translation as a formal neutrino-cooling constraint.

The bound applies to light ALPs with $m_a\lesssim1$~keV and is comparable to the strongest white-dwarf and TRGB constraints in this mass range. Its main advantages include a clean, homogeneous cluster sample, dedicated cluster-specific MESA calculations, an exact treatment of finite upper-RGB sampling, explicit propagation of shared systematics, a prior-independent power-constrained confidence bound, and a toy-calibrated goodness-of-fit test.

\acknowledgments
We are indebted to Konstantin Postnov and Georg Raffelt for encouraging discussions. This work was supported by the Russian Science Foundation, grant 22-12-00253-P.

The authors made use of publicly available large language models to improve the style of some parts of the text of this paper.
\bibliography{trgb}

\end{document}